\documentclass[UTF8,preprint,prd,aps,showpacs,showkeys,nofootinbib]{revtex4}
\usepackage{graphicx}
\usepackage{epstopdf}
\usepackage{dcolumn}
\usepackage{bm}
\usepackage{float}
\usepackage{color}
\usepackage[dvipsnames]{xcolor}
\usepackage{amssymb}
\usepackage{amsmath}
\usepackage{amssymb}
\definecolor{light-gray}{gray}{0.88}
\definecolor{dark-gray}{gray}{0.40}
\begin{document}
\title{The 95 GeV and 125 GeV higgs excesses in the left-right supersymmetric standard
 model }

\author{Zhi-Chuan Wang$^{a,b}$\footnote{wangzc1020@163.com}, Jin-Lei Yang$^{a,b,c}$\footnote{jlyang@hbu.edu.cn}, Qi-Zhen Qin$^{a,b,c}$\footnote{18403586107@163.com}, Wen-Hui Zhang$^{a,b}$\footnote{zwh$\_$0218@163.com}, Tai-Fu Feng$^{a,b,c}$\footnote{fengtf@hbu.edu.cn}}

\affiliation{$^a$ College of Physics Science $\&$ Technology, Hebei University, Baoding, 071002, China\\
$^b$ Hebei Key Laboratory of High-precision Computation and Application of Quantum Field Theory, Baoding, 071002, China\\
$^c$ Research Center for Computational Physics of Hebei Province, Baoding, 071002, China}

\begin{abstract}
This study investigates the excesses observed around 95 GeV in diphoton and $b\bar{b}$ experiments within the framework of the Left-Right Supersymmetric Model (LRSSM). Considering the one-loop and two-loop effective potential corrections to the Higgs masses, the model is able to describe the experimentally observed $\mu(h_{95})_{\gamma\gamma}$ and $\mu(h_{95})_{b\bar{b}}$ signal strengths. In addition, we also present the impacts of the LRSSM-specific parameters $\tan\beta_{R}$, $v_{R}$ and $v_{S}$ on the theoretical predictions of the signal strengths for the 95 GeV and 125 GeV neutral Higgs both.

\keywords{Supersymmetry, 95 GeV Higgs bosons, LRSSM}
\end{abstract}
\maketitle

\section{INTRODUCTION}
In 2012, the Large Hadron Collider (LHC) discovered a Higgs boson with a mass of 125 GeV \cite{ATLAS:2012yve,CMS:2012qbp}, marking a notable success for the Standard Model (SM). However, this discovery did not halt the quest for new physics (NP) at the LHC. Researchers have since focused on searching for lighter scalar particles, conducting experiments at facilities like LEP \cite{OPAL:2002ifx,LEPWorkingGroupforHiggsbosonsearches:2003ing,ALEPH:2006tnd} and LHC. Notably, both the CMS and LEP experiments have reported an excess signal in a similar mass range. CMS, in particular, using a total integrated luminosity of 19.7 $fb^{-1}$ at 8 TeV and 35.9 $fb^{-1}$ at 13 TeV, searched for scalar diphoton resonances at 8 TeV and 13 TeV \cite{CMS:2015ocq,CMS:2018cyk,Biekotter:2023jld}. These investigations reveal a local excess of 2.8$\sigma$ at around 95.3 GeV \cite{CMS:2024yhz}. This unexpected result has generated considerable interest and discussion within the scientific community regarding potential implications for NP models. Analyses in Refs. \cite{Moretti:2006sv,Ellwanger:2010nf,Cao:2011pg,AlbornozVasquez:2012foo,Ellwanger:2012ke,
Boudjema:2012in,Schmidt-Hoberg:2012dba,Badziak:2013bda,Badziak:2013gla,Barbieri:2013nka,Fan:2013gjf} demonstrate that, within the framework of the Next-to-Minimal Supersymmetric Standard Model (NMSSM), the diphoton production rate for a scalar with a given mass can significantly exceed the rate predicted by the SM, sometimes by several multiples. In the Two-Higgs-Doublet Model (THDM) with an additional real singlet (N2HDM), Refs. \cite{Biekotter:2019drs,Biekotter:2019kde,Biekotter:2021ovi,Biekotter:2021qbc,Heinemeyer:2021msz,
Biekotter:2022jyr} have explored the potential to account for the observed excess. In Ref. \cite{Sachdeva:2019hvk}, the authors explore the feasibility of Higgs radion mixing at 125 GeV and the potential of a light radion to explain the CMS diphoton excess within the Higgs-radion mixing model. Considering the one-loop corrections to the neutral scalar mass in the $\mu\nu$SSM, the authors of Refs. \cite{Biekotter:2017xmf,Biekotter:2019gtq} demonstrate how the $\mu\nu$SSM could simultaneously accommodate the two excesses observed at LEP and LHC at the 1$\sigma$ level. Based on the analysis in Ref. \cite{Liu:2024cbr}, an extended scalar sector with an $SU(2)_L$ triplet (hypercharge $Y$=0) can effectively explain the 95 GeV excess. Refs. \cite{Azatov:2012bz,Heinemeyer:2018wzl} reviewed whether certain model implementations could simultaneously explain both excesses while remaining consistent with all other Higgs boson-related constraints and measurements.

In this paper, we investigate two possible scenarios in the LRSSM \cite{Mohapatra:1974hk,Senjanovic:1975rk,Huitu:1997rr,Huitu:1999tg,Alloul:2013fra,Francis:1990pi,Frank:1999sy,Frank:2002nj,Frank:2002pm,Frank:2002es,Frank:2004xt,Frank:2005vd,Frank:2014kma,Chatterjee:2018gca,Frank:2020mqh,Frank:2002hk,Frank:2004gc,Frank:2004hg}. In the first scenario, the lightest Higgs boson is 125 GeV, analogous to the Higgs boson predicted by the SM. In the second scenario, the model contains both of 95 GeV and 125 GeV Higgs bosons, which could provide an explanation for the observed excesses of the 95 GeV in the LEP and LHC experiments. The LRSSM, based on the gauge group $SU(3)\times SU(2)_L \times SU(2)_R \times U(1)_{B-L}$, extends the MSSM by including triplet scalars which interact with leptons to generate the non-zero neutrino masses through the seesaw mechanism \cite{Mohapatra:1979ia}. The observed neutrino flavor oscillations provide definite experimental evidence for physics beyond the SM. Furthermore, the LRSSM resolves some issues faced by the MSSM, such as the strong and electroweak CP problems \cite{Mohapatra:1995xd}. Moreover, in the LRSSM, the gauge structure naturally preserves R-parity, ensuring the stability of the lightest supersymmetric particle (LSP), which is thus a natural dark-matter (DM) candidate. In this framework, there are two typical LSP candidates: a neutralino-like LSP or a right-handed sneutrino LSP. The neutralino-like LSP is a mixture of the gaugino and higgsino states, and its relic density is often constrained by co-annihilation processes with other supersymmetric particles. On the other hand, the right-handed sneutrino, being a Majorana particle, is another viable LSP candidate. The stability of the LSP in both cases is crucial for its role as dark matter. Furthermore, the relic density of these LSPs must comply with observational limits, and direct detection experiments place additional constraints on their properties. The interplay between relic-density constraints and direct-detection limits determines the allowed parameter space of the LRSSM, which is crucial for understanding its implications for dark matter and particle physics \cite{Huitu:2020qxm,Chatterjee:2018gca}. This work is motivated by the rich Higgs sector in the LRSSM, in which a large number of Higgs bosons are introduced, including bidoublets, triplets, and the singlet Higgs bosons \cite{Babu:2014vba}. We propose that a real neutral scalar boson from this model, with the right properties, can explain the 95 GeV excess observed in the LHC and LEP data.

The structure of this paper is as follows. Section II provides a brief introduction to the LRSSM. In Section III, we investigate the 125 GeV and 95 GeV excesses within the LRSSM. Section IV presents the numerical analysis, and Section V concludes with our findings.

\section{The LRSSM model}
The LRSSM is based on the $SU(3)\times SU(2)_L\times SU(2)_R\times U(1)_{B-L}$ gauge group. The model includes the additional local gauge group $\mathrm{SU}(2)_R$, resulting in extra $W_R^{\pm}$ gauge bosons. Furthermore, it incorporates a $U(1)_{B-L}$ gauge group, which produces an additional $Z_R^0$ boson upon gauge symmetry breaking. Additionally, multiple types of Higgs bosons are introduced into the LRSSM, including bidoublets, triplets, and singlet Higgs bosons. The superpotential of the LRSSM is given by \cite{Frank:2017tsm}
\begin{align}
W&=Q_L^TY^{i}_Q\Phi_iQ_R+L_L^TY^{i}_L\Phi_iL_R+L_L^Th_{LL}\delta_LL_R \nonumber \\ &+L_R^Th_{RR}\Delta_RL_R+\lambda_LSTr[\Delta_L\delta_L] +\lambda_RSTr[\Delta_R\delta_R] \nonumber  \\ &+\lambda_3STr[\tau_2\Phi_1^T\tau_2\Phi_2]+\lambda_4STr[\tau_2\Phi_1^T\tau_2\Phi_1] \nonumber \\ &+\lambda_5STr[\tau_2\Phi_2^T\tau_2\Phi_2]+\lambda_SS^3+\xi_F S,
\end{align}
where the $Q_{L(R)}$ and $L_{L(R)}$ are quark and lepton superfields,
\begin{eqnarray}
	&&Q_L=\begin{pmatrix}
		u_L \\
		d_L
	\end{pmatrix}\sim (3,2,1,\frac{1}{3}), \quad\quad
	Q_R=\begin{pmatrix}
	d_R \\
	-u_R
\end{pmatrix}\sim (3,1,2,-\frac{1}{3}) \nonumber \\
	&&L_L=\begin{pmatrix}
		\nu_L \\
		e_L
	\end{pmatrix}\sim (1,2,1,-1),\quad\quad
	L_R=\begin{pmatrix}
		e_R \\
		-\nu_R
	\end{pmatrix}\sim (1,1,2,1),
\end{eqnarray}
and the $\Phi_{i(i=1,2)},\Delta_{R(L)},\delta_{R(L)},S$ are Higgs fields \cite{Zhang:2008jm}.
\begin{align}
	&\Phi_1=\begin{pmatrix}
		\Phi^{+}_{11} & H_{11}^0 \\
		H_{12}^0 &	\Phi^{-}_{12}
	\end{pmatrix}\thicksim{(1,2,2,0)} , \quad
&&\Phi_2=\begin{pmatrix}
		\Phi^{+}_{21} &H_{21}^0 \\
		H_{22}^0 &	\Phi^{-}_{22}
	\end{pmatrix}\thicksim{(1,2,2,0)}, \nonumber \\
	&\Delta_L=\begin{pmatrix}
		\frac{\Delta^{-}_{L}}{\sqrt{2}} & H_{\Delta_{L}}^0 \\
		\Delta^{--}_{L} &	\frac{-\Delta^{-}_{L}}{\sqrt{2}}
	\end{pmatrix}\thicksim{(1,3,1,-2)},\quad\quad
     &&\delta_L=\begin{pmatrix}
		\frac{\delta^{+}_{L}}{\sqrt{2}} & \delta^{++}_{L} \\
		H_{\delta_L}^0 &	\frac{-\delta^{+}_{L}}{\sqrt{2}}
	\end{pmatrix}\thicksim{(1,3,1,2)}, \nonumber \\
    &\delta_R=\begin{pmatrix}
		\frac{\delta^{+}_{R}}{\sqrt{2}} & \delta^{++}_{R} \\
		H_{\delta_{R}}^0 &	\frac{-\delta^{+}_{R}}{\sqrt{2}}
	\end{pmatrix} \thicksim{(1,1,3,2)}, \quad\quad\quad
	&&\Delta_R=\begin{pmatrix}
		\frac{\Delta^{-}_{R}}{\sqrt{2}} & H_{\Delta_{R}}^0 \\
		\Delta^{--}_{R} &	\frac{-\Delta^{-}_{R}}{\sqrt{2}}
	\end{pmatrix}\thicksim{(1,1,3,-2)}, \nonumber \\
	&S=H_{S}^0\thicksim(1,1,1,0), \label{5h}
\end{align}
where we assume the $SU(2)_L$ triplets $\Delta_L$ and $\delta_L$ to be inert. The electroweak symmetry is broken when those Higgs fields acquire some nonzero vacuum expectation values space(VEVs).
\begin{align}
&\Phi_1=\begin{pmatrix}
		0 & \frac{1}{\sqrt{2}}v'_1e^{i\alpha_1} \\
		\frac{1}{\sqrt{2}}v_1 &	0
	\end{pmatrix},
\Phi_2=\begin{pmatrix}
		0 & \frac{1}{\sqrt{2}}v_2 \\
		\frac{1}{\sqrt{2}}v'_2e^{i\alpha_2} &	0
	\end{pmatrix},
\nonumber \\
&\Delta_R=\begin{pmatrix}
		0 & \frac{1}{\sqrt{2}}v_{R1} \\
		0 &	0
	\end{pmatrix},
\delta_R=\begin{pmatrix}
		0 & 0 \\
		\frac{1}{\sqrt{2}}v_{R2} &	0
	\end{pmatrix},
S=\frac{1}{\sqrt{2}}v_Se^{i\alpha_S}.
\end{align}
Within the framework of the LRSSM, the two Higgs doublets $ H_u $ and $ H_d $ from the Minimal Supersymmetric Standard Model (MSSM) are embedded into the bidoublet fields. Specifically, $ H_u $ can be expressed as a superposition of $ H^0_{12} $ and $ H^0_{22}$, while $H_d $ is a superposition of $H^0_{11} $ and $H^0_{21}$. Furthermore, we introduce two parameters $\tan\beta=v_2/v_1,\tan\beta_R=v_{R2}/v_{R1}$. Where $v_1^2+v_2^2=246\ \text{GeV}^2$. The CP-violating mixing between \( W_L^\pm \) and \( W_R^\pm \) depends on terms proportional to
\( v_1 v_1' e^{i\alpha_1} \) and \( v_2 v_2' e^{i\alpha_2} \). This mixing is tightly constrained to be small
due to experimental limits from \( K^0\)-\(\overline{K^0} \) mixing data. To simplify the analysis and reduce
the complexity of the parameter space, we adopt an assumed hierarchy.
\begin{align}
 v_{R1},v_{R2},v_S\gg v_1,v_2,v'_1,v'_2\ \ \text{and}\ \ v'_1=v'_2=\alpha_1=\alpha_2=\alpha_S=0
\end{align}
In the supersymmetric limit, the $F$-terms and $D$-terms vanish, when $\lambda_Rv_{R1}v_{R2}$ = $\xi_F$ and $v_{R1}=v_{R2}$ \cite{Frank:2011jia}. On the other hand, the singlet VEV $v_S$ is induced by the SUSY-breaking linear term $\xi_S$ so that its natural scale is the
supersymmetry-breaking scale. The Right-Hand symmetry
breaking scale is chosen larger than the SUSY breaking scale, so the superpartner masses are lighter than the heavy $W_R^{\pm}$ and $Z_R^0$ gauge bosons \cite{Beall:1981ze,Langacker:1989xa,Colangelo:1990bb}. For simplicity, the left-handed $\Delta_{L}$ and $\delta_{L}$ fields do not develop non-zero VEVs, and hence the neutrino masses do not get any contribution from Type II seesaw. Then the soft supersymmetry-breaking Lagrangian of the LRSSM is generally given as
\begin{align}
	L_{soft}=&-\frac{1}{2}\{
	M_1\tilde{B}\cdot\tilde{B}+M_{2L}\tilde{W}_L^k\cdot\tilde{W}_{Lk}+M_{2R}\tilde{W}_R^k\cdot\tilde{W}_{Rk}+M_3\tilde{g}^a\cdot\tilde{g}_a+h.c.\} \nonumber \\
	&-\tilde{Q}^{\dagger}_Lm_{Q_L}^2\tilde{Q}_L-\tilde{Q}_Rm_{Q_R}^2\tilde{Q}_R^{\dagger}-\tilde{L}^{\dagger}_Lm_{L_L}^2\tilde{L}_L-\tilde{L}_Rm_{L_R}^2\tilde{L}_R^{\dagger} \nonumber \\
	&-(m_\Phi^2)^{f`f}Tr(\Phi_{f}^{\dagger}\Phi_{f`})-m_{\Delta_{L}}^2Tr(\Delta_{L}^{\dagger}\Delta_{L})-m_{\delta_{L}}^2Tr(\delta_{L}^{\dagger}\delta_{L}) \nonumber \\
	&-m_{\Delta_{R}}^2Tr(\Delta_{R}^{\dagger}\Delta_{R})-m_{\delta_{R}}^2Tr(\delta_{R}^{\dagger}\delta_{R})-m_{S}^2S^{\dagger}S \nonumber \\
	&-\{S[T_LTr(\Delta_{L}\delta_{L})+T_RTr(\Delta_{R}\delta_{R})+T_3Tr(\Phi_1^T\tau_2\Phi_2\tau_2)+T_4Tr(\Phi_1^T\tau_2\Phi_1\tau_2) \nonumber \\
	&+T_5Tr(\Phi_2^T\tau_2\Phi_2\tau_2)+T_SS^2+\xi_S]+h.c.\}+\{T^1_Q(\widetilde{Q}_L)^T\Phi_1\widetilde{Q}_R+T^2_Q(\widetilde{Q}_L)^T\Phi_2\widetilde{Q}_R \nonumber \\
	&+T^1_L(\widetilde{L}_L)^T\Phi_1\widetilde{L}_R+T^2_L(\widetilde{L}_L)^T\Phi_2\widetilde{L}_R+T^3_L(\widetilde{L}_L)^T\delta_{L}\widetilde{L}_L+T^4_L(\widetilde{L}_R)^T\Delta_{R}\widetilde{L}_R+h.c.
	\}.
\end{align}
We obtain the hierarchy $v_1, v_2 \gg v'_1, v'_2 \approx 0$ by setting small values for $\lambda_4$, $\lambda_5$ and the corresponding SUSY-breaking terms, which helps to reduce the mixing in the Higgs mass matrix and facilitates the calculation \cite{Frank:2017tsm}. The D-term contribution to the scalar potential (neglecting the squark pieces) is given by
\begin{align}
	V_{D}=&\sum_i [\frac{g_L^2}{8}\vert\ Tr(2\Delta_L^{\dagger}\tau_i\Delta_L+2\delta_L^{\dagger}\tau_i\delta_L+\Phi_1^{\dagger}\tau_i\Phi_1+\Phi_2^{\dagger}\tau_i\Phi_2)+
\widetilde{L}_L^{\dagger}\tau_i\widetilde{L}_L \vert^2  \nonumber \\
&+\frac{g_R^2}{8}\vert\ Tr(2\Delta_R^{\dagger}\tau_i\Delta_R+2\delta_R^{\dagger}\tau_i\delta_R+\Phi_1^{\dagger}\tau_i\Phi_1+\Phi_2^{\dagger}\tau_i\Phi_2)+
\widetilde{L}_R^{\dagger}\tau_i\widetilde{R}_L \vert^2] \nonumber \\
&+\frac{g_{B-L}^2}{2}[Tr(-\Delta_L^{\dagger}\Delta_L+\delta_L^{\dagger}\delta_L-\Delta_R^{\dagger}\Delta_R
+\delta_R^{\dagger}\delta_R)-\widetilde{L}_L^{\dagger}\widetilde{L}_L+\widetilde{L}_R^{\dagger}\widetilde{L}_R].
\end{align}
The F-term contribution to the scalar potential is given by
\begin{align}
V_F=&\vert \lambda_L Tr(\Delta_L\delta_L)+\lambda_R Tr(\Delta_R\delta_R)+\lambda_3Tr(\tau_2\Phi_2^\dagger\tau_2\Phi_2)
   \nonumber \\ & +\lambda_4Tr(\tau_2\Phi_1^\dagger\tau_2\Phi_1)+\lambda_5Tr(\tau_2\Phi_2^\dagger\tau_2\Phi_2)+3\lambda_SS^2+\xi_F\vert^2
   \nonumber \\ &
   +Tr(\vert \lambda_LS\delta_L\vert^2+\vert \lambda_LS\Delta_L\vert^2+\vert \lambda_RS\Delta_R\vert^2+\vert \lambda_RS\delta_R\vert^2) \nonumber \\&
   +Tr(\vert\lambda_3S\Phi_2+\lambda_4S\Phi_1\vert^2+\vert\lambda_3S\Phi_1+\lambda_5S\Phi_2\vert^2).
\end{align}
The effective potential is
\begin{align}
	V_{eff}=V_0+\Delta V ,
\end{align}
where
\begin{align}
&V_0=-L_{soft}+V_D+V_F,  \nonumber \\
&\Delta V=V_1+V_2.
\end{align}
The $V_1$ is the one-loop level
effective potential and $V_2$ is the two-loop level effective potential \cite{Zhang:1998bm,Espinosa:1999zm,Degrassi:2001yf,Degrassi:2009yq}. With the adoption of the Dimensional Reduction
renormalization scheme, the effective Higgs potential up to the one-loop correction is
presented in the Landau gauge, and the specific form of $\Delta V$ is given in~\cite{Quiros:1999jp,Gao:2024ljl}
\begin{align}
\Delta V \;=\; \sum_{i} \frac{n_i}{64\pi^{2}}\, m_i^{4}\!\big(\phi_1,\phi_2,\phi_{\Delta_R},\phi_{\delta_R},\phi_s\big)\,
\left( \log\!\frac{m_i^{2}\!\big(\phi_1,\phi_2,\phi_{\Delta_R},\phi_{\delta_R},\phi_s\big)}{Q^{2}}
- \frac{3}{2} \right).
\end{align}
We denote the degrees of freedom for each mass eigenstate by $n_i$ ($-12$ for quarks, $-4$ for leptons and charginos, $-2$ for neutralinos and neutrinos, $6$ for squarks, $2$ for sleptons and charged Higgs bosons, and $3$ and $6$ for $Z\,(Z')$ and $W_1\,(W_2)$ bosons, $1$ for sneutrinos and the neutral Higgs scalars). The specific potential for
the one-loop correction is as follows
{\small
\begin{align}
\Delta V \;=\; V_{t} + V_{b} + V_{\tau}+V_{\nu_R}
+ V_{\tilde u_i} + V_{\tilde d_i} + V_{\tilde e_i} + V_{\tilde \nu_i}
+ V_{W_1,W_2} + V_{Z,Z'} + V_{\chi_i^{0}} + V_{H_i^{\pm}} + V_{\chi_i^{\pm}} .
\end{align}
}
Here, $V_t$, $V_b$ and $V_\tau$ represent the one-loop effective-potential corrections from $t$, $b$ and $\tau$. $V_{\tilde u_i}$ and $V_{\tilde d_i}$ represent the corrections from scalar up-type quarks and scalar down-type quarks. $V_{\tilde e_i}$ represents the corrections from sleptons. $V_{\tilde \nu_i}$ represents the corrections from sneutrinos. $V_{W_L,W_R}$, $V_{Z_L,Z_R}$, $V_{\chi^{0}}$, $V_{H^{\pm}}$ and $V_{\chi^{\pm}}$ represent the corrections from the $W_1(W_2)$ boson, $Z(Z')$ boson, neutralinos, charged Higgs bosons and charginos, respectively. The two-loop effective potential is mainly contributed by top quarks, scalar top quarks, gluon and gluino. In units of $\alpha_s C_F N_c^f /(4\pi)^3$ where $C_F=4/3$ and $N_c^f=3$ are colour factors. The $V_2$ reads \cite{Degrassi:2009yq}
\begin{align}
V_2=&2J(m_t^2,m_t^2)-4m_t^2I(m_t^2,m_t^2,0)  \nonumber \\
& +\{2m_{\tilde{t}1}^2I(m_{\tilde{t}1}^2,m_{\tilde{t}1}^2,0)+2L(m_{\tilde{t}1}^2,m_{\tilde{g}}^2,m_t^2)- 4m_tm_{\tilde{g}}s_{2\tilde{\theta}}c_{\varphi-\tilde{\varphi}}I(m_{\tilde{t}1}^2,m_{\tilde{g}}^2,m_t^2) \nonumber \\
&+ \frac{1}{2}(1+c_{2\tilde{\theta}}^2)J(m_{\tilde{t}1}^2,m_{\tilde{t}1}^2)+\frac{s_{2\tilde{\theta}}^2}{2}J(m_{\tilde{t}1}^2,m_{\tilde{t}2}^2) +[m_{\tilde{t}1}\leftrightarrow m_{\tilde{t}2},s_{2\tilde{\theta}} \rightarrow -s_{2\tilde{\theta}}]
\}, \label{v2}
\end{align}
where $m_t$ and $m_b$ masses of the top and bottom quarks, $m_{\widetilde{t}_i}$ and $m_{\widetilde{b}_i}$ are the masses of the scalar top quarks and scalar bottom quarks, and $m_{\widetilde{g}}$ is the gluino mass, respectively. For the two-loop functions $I, J, s_{2\theta}, c_{\varphi-\tilde{\varphi}}$ in Eq. (\ref{v2}) can be found in Appendix D of Ref. \cite{Degrassi:2009yq}. The squared pole masses of a set of $n$ scalar bosons that mix with each other are the real parts of the solutions for $p^2$ of the equation \cite{OLeary:2011vlq,Degrassi:2009yq,Pierce:1996zz,Slavich:2020zjv,R:2021bml,Quiros:1997vk}
\begin{align}
\text{det}[\Gamma_{ij}(p^2)]=0 \label{pm},
\end{align}
where $\Gamma_{ij}(p^2)$ is the $n \times n$ inverse propagator matrix and $p$ denotes the incoming external momentum of the self-energies. Generally $\Gamma_{ij}(p^2)$ can be decomposed as
\begin{align}
-i\Gamma_{ij}(p^2)=p^2\delta_{ij}-M^2_{ij,0}-\Delta M^2_{ij}(p^2)
\end{align}
where $M_{ij,0}^2$ is the tree-level squared mass matrix, and $\Delta M_{ij}^2(p^2)$ are the radiative corrections to the squared matrix, respectively.

In the limit of vanishing external momentum, tadpole and self-energy diagrams can be obtained through the derivatives of the effective potential, thus one has \begin{align}
&T_i=-\frac{\partial V_{eff}}{\partial_{\phi_i}}\vert_{min},   \ \ \Pi_{ij}(0)=-\frac{\partial^2 \Delta V}{\partial\phi_i\partial\phi_j}\vert_{min}, \nonumber \\
&\Delta M^2_{ij}(p^2)=\Pi(p^2)-\Pi_{ij}(0)-\frac{\partial^2 \Delta V}{\partial\phi_i\partial\phi_j},\label{xz}
\end{align}
where the detailed calculations of the contributions from $\Delta M^2_{ij}(p^2)$ in Eq. (\ref{xz}) are collected in Appendix \ref{app4}.
In the basis \( (\text{Re}[H_{12}^0], \text{Re}[H_{21}^0], \text{Re}[H_{\Delta_{R}}^0], \text{Re}[H_{\delta_{R}}^0], \text{Re}[H_{S}^0]) \), the tree-level squared mass matrix of the CP-even Higgs fields is written as (refer to Eq. (\ref{5h}) for the notation)

\begin{eqnarray}
	m_{H_3
}^2=\begin{pmatrix}
		m^2_{H_{12}^0H_{12}^0} & m^2_{H_{12}^0H_{21}^0} & m^2_{H_{12}^0H_{\Delta_{R}}^0} & m^2_{H_{12}^0H_{\delta_{R}}^0} & m^2_{H_{12}^0H_{S}^0} \\
		m^2_{H_{21}^0H_{12}^0} & m^2_{H_{21}^0H_{21}^0} & m^2_{H_{21}^0H_{\Delta_{R}}^0} & m^2_{H_{21}^0H_{\delta_{R}}^0} & m^2_{H_{21}^0H_{S}^0} \\
m^2_{H_{\Delta_{R}}^0H_{12}^0} & m^2_{H_{\Delta_{R}}^0H_{21}^0} & m^2_{H_{\Delta_{R}}^0H_{\Delta_{R}}^0} & m^2_{H_{\Delta_{R}}^0H_{\delta_{R}}^0} & m^2_{H_{\Delta_{R}}^0H_{S}^0} \\
m^2_{H_{\delta_{R}}^0H_{12}^0} & m^2_{H_{\delta_{R}}^0H_{21}^0} & m^2_{H_{\delta_{R}}^0H_{\Delta_{R}}^0} & m^2_{H_{\delta_{R}}^0H_{\delta_{R}}^0} & m^2_{H_{\delta_{R}}^0H_{S}^0} \\
m^2_{H_{S}^0H_{12}^0} & m^2_{H_{S}^0H_{21}^0} & m^2_{H_{S}^0H_{\Delta_{R}}^0} & m^2_{H_{S}^0H_{\delta_{R}}^0} & m^2_{H_{S}^0H_{S}^0} \\
	\end{pmatrix}.
\end{eqnarray}
The matrix elements are defined as
\begin{align}
&m^2_{H_{12}^0H_{12}^0}=\frac{1}{8}(8m_1^2+3g_R^2 v_1^2-g_R^2 v_2^2+g_L^2(3v_1^2-v_2^2)+2g_R^2v_{R2}^2+4\lambda_3^2 v_2^2+4\lambda_3^2v_S^2),  \nonumber \\
&m^2_{H_{12}^0H_{21}^0}=\frac{1}{4}(-g_L^2v_1v_2-g_{R}^2v_1v_2-2(\sqrt{2}T_3 v_S+\lambda_3(-2\lambda_3v_1v_2+\lambda_Rv_{R1}v_{R2}+3\lambda_Sv_S^2+2\xi_F))),  \nonumber \\
&m^2_{H_{12}^0H_{\Delta_{R}}^0}=\frac{1}{2}(g_R^2v_1v_{R1}-\lambda_3\lambda_Rv_2v_{R2}),\quad
m^2_{H_{12}^0H_{\delta_{R}}^0}=\frac{1}{2}(-g_R^2v_1v_{R2}-\lambda_3\lambda_Rv_2v_{R1}),
 \nonumber \\
&m^2_{H_{12}^0H_{S}^0}=-\frac{T_3v_2}{\sqrt{2}}+\lambda_3 v_S(\lambda_3 v_1-3\lambda_Sv_2),\quad m_{H_{21}^0H_{S}^0}=\frac{1}{2}(g_R^2  v_2v_{R2}-\lambda_3\lambda_Rv_1v_{R1}),  \nonumber \\
&m^2_{H_{21}^0H_{21}^0}=\frac{1}{8}(8m_2^2-g_R^2v_1^2+3g_R^2v_2^2-g_L^2(v_1^2-3v_2^2)-2g_R^2v_{R1}^2
+2g_R^2v_{R2}^2+4\lambda_3^2v_1^2+4\lambda_3^2v_S^2),  \nonumber \\
&m^2_{H_{21}^0H_{\Delta_{R}}^0}=\frac{1}{2}(-g_R^2v_2v_{R1}-\lambda_3\lambda_Rv_1v_{R2}),\quad
m^2_{H_{21}^0H_{\delta_{R}}^0}=\frac{1}{2}(g_R^2v_2v_{R2}-\lambda_3\lambda_Rv_1v_{R1}),
\nonumber \\
&m^2_{H_{\Delta_{R}}^0H_{\Delta_{R}}^0}=\frac{1}{4}(4m_{\Delta_{R}}^2+g_R^2(v_1^2-v_2^2+6v_{R1}^2-2v_{R2}^2)
+2g_B^2(3v_{R1}^2-v_{R2}^2)+2\lambda_R^2(v_{R2}^2+v_S^2)),  \nonumber \\
&m^2_{H_{\Delta_{R}}^0H_{\delta_{R}}^0}=-g_B^2v_{R1}v_{R2}-g_R^2v_{R1}v_{R2}+\frac{T_Rv_S}{\sqrt{2}}
-\frac{1}{2}\lambda_3\lambda_Rv_1v_2+\lambda_R^2v_{R1}v_{R2}+\frac{3}{2}\lambda_S\lambda_Rv_S^2+\lambda_R\xi_F,
 \nonumber \\
&m^2_{H_{\Delta_{R}}^0H_{S}^0}=\frac{T_Rv_{R2}}{\sqrt{2}}+\lambda_Rv_S(\lambda_Rv_{R1}+3\lambda_Sv_{R2}),\quad
m^2_{H_{\delta_{R}}^0H_{S}^0}=\frac{T_Rv_{R1}}{\sqrt{2}}+\lambda_Rv_S(\lambda_Rv_{R2}+3\lambda_Sv_{R1}) , \nonumber \\
&m^2_{H_{\delta_{R}}^0H_{\delta_{R}}^0}=\frac{1}{4}(4m_{\delta_{R}}^2-2g_{B}^2v_{R1}^2+6g_B^2v_{R2}^2
+g_R^2(-v_1^2+v_2^2-2v_{R1}^2+6v_{R2}^2)+2\lambda_R^2v_{R1}^2+2\lambda_R^2v_S^2)  , \nonumber \\
&m^2_{H_{S}^0H_{S}^0}=\frac{1}{2}(2m_S^2+6\sqrt{2}T_Sv_S+\lambda_3^2(v_1^2+v_2^2
+v_{R1}^2+v_{R2}^2)-6\lambda_3\lambda_S(v_1v_2+v_{R1}v_{R2}) \nonumber \\
&\ \ \ \ \ \ \ \ \ \ \ \ +54\lambda_S^2v_S^2+12\lambda_S\xi_F).
\end{align}
This mass matrix can be diagonalized by the rotation matrix $Z^{H}$ as $Z^{H}m_H^2Z^{H\dagger}$=$m_{H,diag}^2$. And the relationship between the Higgs bosons $H_i$ (with $i=1-5$) in the mass eigenstates and the gauge eigenstates $\phi_{a(a=1,2,\Delta_R,\delta_R,S)}$ are given by $H_{i}=\sum_{a}Z^H_{ia}\phi_a$. The squared mass matrix of other Higgs fields is given in Appendix \ref{app1}. In our calculation we consider the radiative corrections originating from the third generation quarks and scalar quarks up to two-loop level \cite{Babu:2014vba}. Although we have considered the one-loop effective potential and partial contributions from the two-loop effective potential, the portions of the two-loop effective potential not included may introduce some uncertainty. And, the experimental uncertainty in the top quark mass, $m_t = 173.2 \pm 0.9$ GeV, has a direct impact on the Higgs mass calculation; therefore, this error range has been included in our results. Furthermore, QCD effects and the choice of the renormalization scale may also lead to variations in the calculated results, making these scale uncertainties another source of error. So, the theoretical error in the SM-like Higgs boson mass calculation allows for a mass range of 122--128 GeV \cite{Degrassi:2002fi,Martin:2007pg,Harlander:2008ju,Heinemeyer:2011aa,Arbey:2012dq}.
\section{Higgs-boson Signals in the LRSSM}
At the LHC, $h^0$ is produced chiefly from the gluon fusion $(gg\rightarrow h^0)$. The one-loop diagrams with virtual top quarks have the most significant contribution to the leading order (LO). In the LRSSM, the LO contributions need
to be added by the Higgs-new particle couplings, whose effects are significant. So the decay width of $h^0\to gg$ can be shown as\cite{Bae:2022dgj,Shifman:1979eb,Bergstrom:1985hp}
\begin{align}
\Gamma_{NP}(h^0\rightarrow gg)=\frac{G_F\alpha_s^2m_{h^0}^3}{64\sqrt{2}\pi^3}|\sum_{q}g_{h^0qq}A_{\frac{1}{2}}^H(x_q)+
\sum_{\widetilde{q}}g_{h^0\widetilde{q}\widetilde{q}}\frac{m_Z^2}{m_{\widetilde{q}}^2}A_{0}^H(x_{\widetilde{q}})|^2,
\end{align}
with $x_a = m_{h^0}^2/(4 m_a^2)$. Here, $q$ and $\tilde q$ are quarks and squarks. The LO contributions for decay $h^0 \to \gamma\gamma$ also originate from one-loop diagrams. In the SM, the concrete contributions are mainly derived from top quark and charged gauge boson $W_{1}^{\pm}$. Due to the Higgs-new particle couplings in the LRSSM, the decay width of $h^0 \to \gamma\gamma$ can be expressed as\cite{Bae:2022dgj,Shifman:1979eb,Bergstrom:1985hp,Djouadi:2005gj,Cacciapaglia:2009ky}
\begin{align}
\Gamma_{NP}(h^0\rightarrow \gamma\gamma)=&\frac{G_F\alpha_s^2m_{h^0}^3}{128\sqrt{2}\pi^3}|\sum_fN_cQ_f^2g_{h^0ff}A_{\frac{1}{2}}^H(x_f)
+\sum_fg_{h^0\widetilde{f}\widetilde{f}}\frac{m_Z^2}{m_{\widetilde{f}}^2}A_0^H(x_{\widetilde{f}})  \nonumber \\
&+\sum_{i=1}^2g_{h^0H_i^{+}H_i^{-}}\frac{m_Z^2}{m_{h_i^{\pm}}^2}A_0^H(x_{H_i^{\pm}})
+\sum_{j=1}^2g_{h^0H_j^{++}H_j^{--}}\frac{m_Z^2}{m_{h_j^{\pm\pm}}^2}A_0^H(x_{H_j^{\pm\pm}})  \nonumber \\
&+\sum_{k=1}^2g_{h^0\chi_k^{+}\chi_k^{-}}\frac{m_Z^2}{m_{\chi_k^{\pm}}^2}A_{\frac{1}{2}}^H(x_{\chi_k^{\pm}})
+g_{h\chi^{++}\chi^{--}}\frac{m_Z^2}{m_{\chi^{\pm\pm}}^2}A_{\frac{1}{2}}^H(x_{\chi^{\pm\pm}}) \nonumber \\
&+g_{h^0WW}A_1^H(x_W)|^2,
\end{align}
with $\chi_i^{\pm}$ denoting the charginos and $\chi_i^{\pm\pm}$ denoting the doubly-charged charginos. The functions $A_1^H(x)$, $A_{1/2}^H(x)$, $A_0^H(x)$ and $g(x)$ are given by
\begin{align}
A_{1}^H(x)&=-(2x^2+3x+3(2x-1)g(x))/x^2  \nonumber \\
A_{\frac{1}{2}}^H(x)&=2(x+(x-1)g(x))/x^2  \nonumber \\
A_{0}^H(x)&=-(x-g(x))/x^2   \nonumber \\
g(x)=&\begin{cases}\arcsin^{2}\sqrt{x} ,&x\le1
		\\
		-\frac{1}{4}[ln\frac{1+\sqrt{1-1/x}}{1-\sqrt{1-1/x}}-i\pi]^{2},&x> 1
	\end{cases}
\end{align}
The decay width for $h^0 \rightarrow  VV^* \ (V=W_1,\ Z)$ is given by \cite{Bernreuther:2010uw,Gonzalez:2012mq}
\begin{align}
&\Gamma_{NP}(h^0\rightarrow W_1W_1^{*})=\frac{g_2^4 m_{h^0}}{3072 \pi^3}|g_{h^0W_1W_1}|^2F(\epsilon), \\
&\Gamma_{NP}(h^0\rightarrow ZZ^{*})=\frac{g_2^4 m_{h^0}}{2048 \pi^3 c_W^4}(7-\frac{40s_W^2}{3}+\frac{160 s_W^4}{9})|g_{h^0ZZ}|^2F(\epsilon'),
\end{align}
where
\begin{align}		F(x)=&\frac{3(1-8x^2+20x^4)}{(4x^2-1)^2}\arccos(\frac{3x^2-1}{2x^3})-(1-x^2)(\frac{47x^2}{2}-\frac{13}{2}+\frac{1}{x^2})\\ \nonumber
	&-3(1-6x^2+4x^4)\ln x
\end{align}
 and $\epsilon= m_W/m_h$, $\epsilon'= m_Z/m_h$. In the Born approximation, the partial decay widths of CP-even  Higgs bosons decaying into fermion pairs are given by \cite{Resnick:1973vg}
\begin{align}
\Gamma_{NP}(h^0\rightarrow f \bar{f})&=\frac{N_cG_Fm_f^2m_{h^0}}{4\sqrt{2}\pi}|g_{h^0ff}|^2(1-\frac{4m_f^2}{m_{h^0}^2})^\frac{3}{2} ,
\end{align}
The total decay width of the light CP-even Higgs boson in the LRSSM can be written as
\begin{align}
\Gamma_{tot}^{NP}=&\Gamma_{NP}(h^0\rightarrow gg)+\Gamma_{NP}(h^0\rightarrow \gamma\gamma)+\sum_{V=W,Z}\Gamma_{NP}(h^0\rightarrow VV^*)+\sum_{f=\tau,b,c}\Gamma_{NP}(h^0\rightarrow f \bar{f}).
\end{align}
Normalized to the SM expectation, the signal strengths for the Higgs decay channels are quantified by the ratios~\cite{Arbey:2013fqa}
\begin{align}
\mu_{\gamma\gamma, VV^*}^{\text{ggF}} &=
\frac{\sigma_{\text{NP}}(\text{ggF})}{\sigma_{\text{SM}}(\text{ggF})}
\frac{\text{BR}_{\text{NP}}(h \to \gamma\gamma, VV^*)}{\text{BR}_{\text{SM}}(h \to \gamma\gamma, VV^*)},
\qquad (V=Z,W), \\ \nonumber
\mu_{ff}^{\text{VBF}} &=
\frac{\sigma_{\text{NP}}(\text{VBF})}{\sigma_{\text{SM}}(\text{VBF})}
\frac{\text{BR}_{\text{NP}}(h \to f\bar{f})}{\text{BR}_{\text{SM}}(h \to f\bar{f})},
\qquad (f=b,\tau), \label{xh1}
\end{align}
where ggF and VBF stand for gluon-gluon fusion and vector boson fusion, respectively.
Normalized to the SM values, one can evaluate the Higgs production cross sections
\begin{align}
\frac{\sigma_{\text{NP}}(\text{ggF})}{\sigma_{\text{SM}}(\text{ggF})}
&\simeq
\frac{\Gamma_{\text{NP}}(h \to gg)}{\Gamma_{\text{SM}}(h \to gg)}
= \frac{\Gamma^h_{\text{NP}}}{\Gamma^h_{\text{SM}}}
\frac{\text{BR}_{\text{NP}}(h \to gg)}{\text{BR}_{\text{SM}}(h \to gg)}, \\ \nonumber
\frac{\sigma_{\text{NP}}(\text{VBF})}{\sigma_{\text{SM}}(\text{VBF})}
&\simeq
\frac{\Gamma_{\text{NP}}(h \to VV^*)}{\Gamma_{\text{SM}}(h \to VV^*)}
= \frac{\Gamma^h_{\text{NP}}}{\Gamma^h_{\text{SM}}}
\frac{\text{BR}_{\text{NP}}(h \to VV^*)}{\text{BR}_{\text{SM}}(h \to VV^*)},\label{xh2}
\end{align}
where we have neglected the contributions from the rare or invisible decays,
and $\Gamma^{h}_{\text{SM}}$ denotes the SM Higgs total decay width.
Through Eqs.~(\ref{xh1}) and~(\ref{xh2}), we can quantify the signal strengths
for the Higgs decay channels in the LRSSM
\begin{align}
\mu^{\text{ggF}}_{\gamma\gamma} &\approx
\frac{\Gamma_{\text{NP}}(h \to gg)}{\Gamma_{\text{SM}}(h \to gg)}
\frac{\Gamma_{\text{NP}}(h \to \gamma\gamma)/\Gamma^{h}_{\text{NP}}}
{\Gamma_{\text{SM}}(h \to \gamma\gamma)/\Gamma^{h}_{\text{SM}}} \\ \nonumber
&= \frac{\Gamma^{h}_{\text{SM}}}{\Gamma^{h}_{\text{NP}}}
   \frac{\Gamma_{\text{NP}}(h \to gg)}{\Gamma_{\text{SM}}(h \to gg)}
   \frac{\Gamma_{\text{NP}}(h \to \gamma\gamma)}{\Gamma_{\text{SM}}(h \to \gamma\gamma)}, \\ \nonumber
\mu^{\text{ggF}}_{VV^*} &\approx
   \frac{\Gamma_{\text{NP}}(h \to gg)}{\Gamma_{\text{SM}}(h \to gg)}
   \frac{\Gamma_{\text{NP}}(h \to VV^*)/\Gamma^{h}_{\text{NP}}}{\Gamma_{\text{SM}}(h \to VV^*)/\Gamma^{h}_{\text{SM}}} \\ \nonumber
&= \frac{\Gamma^{h}_{\text{SM}}}{\Gamma^{h}_{\text{NP}}}
   \frac{\Gamma_{\text{NP}}(h \to gg)}{\Gamma_{\text{SM}}(h \to gg)}
   |g_{hVV}|^2, \\ \nonumber
\mu^{\text{VBF}}_{ff} &\approx
   \frac{\Gamma_{\text{NP}}(h \to VV^*)}{\Gamma_{\text{SM}}(h \to VV^*)}
   \frac{\Gamma_{\text{NP}}(h \to f\bar{f})/\Gamma^{h}_{\text{NP}}}{\Gamma_{\text{SM}}(h \to f\bar{f})/\Gamma^{h}_{\text{SM}}} \\ \nonumber
&= \frac{\Gamma^{h}_{\text{SM}}}{\Gamma^{h}_{\text{NP}}}
   |g_{hVV}|^2 \, |g_{hff}|^2
   \qquad (V=Z,W_1; \; f=b,\tau),
\end{align}
with
\begin{align}
\frac{\Gamma_{\text{NP}}(h \to VV^*)}{\Gamma_{\text{SM}}(h \to VV^*)}
&= |g_{hVV}|^2 = |g_{hZZ}|^2 = |g_{hW_1W_1}|^2 \\ \nonumber
\frac{\Gamma_{\text{NP}}(h \to f\bar{f})}{\Gamma_{\text{SM}}(h \to f\bar{f})}
&= |g_{hff}|^2 = |g_{hbb}|^2 \simeq |g_{h\tau\tau}|^2.
\end{align}
Therefore, we could just analyze the signal strengths
$\mu^{\text{ggF}}_{\gamma\gamma}$,
$\mu^{\text{ggF}}_{VV^*}$
and
$\mu^{\text{VBF}}_{ff}$ in the following.
In the LRSSM, the concrete expressions for $g_{h^0qq},\ g_{h^0\widetilde{q}\widetilde{q}},\ g_{h^0ff},\ g_{h^0\widetilde{f}\widetilde{f}}
,\ g_{h^0H^+H^-},\ g_{h^0\chi^+\chi^-},\ g_{h^0WW}$ and $g_{h^0ZZ}$ have been discussed in Appendix \ref{app3}. It is intriguing to note that the CMS Collaboration has reported a 2.9$\sigma$ global (1.9$\sigma$ local) excess in the diphoton channel with an invariant mass of 95.4 GeV \cite{CMS:2015ocq,CMS:2018cyk}. A similar analysis by the ATLAS Collaboration has found a 1.7$\sigma$ local excess at the same mass \cite{ATLAS:2023jzc}. In addition, CMS has reported another excess in the $\tau^{+}\tau^{-}$ channel around 100 GeV with 2.6$\sigma$ local
(2.3$\sigma$ global) \cite{CMS:2022goy,Cao:2023gkc,Richard:2017kot,Haisch:2017gql,Liu:2018xsw,Cline:2019okt,
Aguilar-Saavedra:2020wrj,Bhattacharya:2023lmu,Banik:2023vxa,Coloretti:2023yyq} significance. Observed signal rates and the local (global) significance for the 95 GeV excess in different channels are presented in Tab. \ref{tabel1} \cite{Azevedo:2023zkg}. As analyzed in Ref. \cite{CMS:2022arx}, the light CP-even Higgs does suffer strict constraints from the LHC search
 for the top-quark associated production of the SM Higgs boson that decays into $\tau \bar{\tau}$, and the possibility to
 explain the ditau excess by the CP-even scalar is excluded according to the analysis in Ref. \cite{Iguro:2022dok}. However,
 they consider the 1$\sigma$ range of $\mu(\Phi_{95})_{\tau \tau}$ in the analysis, it is found that this constraint can be relaxed if the 2$\sigma$ of $\mu(\Phi_{95})_{\tau\tau}$ is considered. We have therefore decided to focus on the $\gamma\gamma$ and $b\bar{b}$ channels, which are subject to weaker constraints and exhibit cleaner signals, providing a clearer window for probing our model.

\begin{table}[h]
  \centering
  \begin{tabular}{|c|c|c|}
    \hline
    Channel & Signal strength & Local \ (global)\  sig. \\
    \hline
    $gg \rightarrow h_{95} \rightarrow \gamma\gamma$ & $0.33^{+0.19}_{-0.12}$ & $2.9 (1.3)\sigma$ \cite{CMS:2023yay} \\
    $gg \rightarrow h_{95} \rightarrow \tau^{+}\tau^{-}$ & $1.23^{+0.61}_{-0.49}$ & $2.6 (2.3)\sigma$ \cite{CMS:2022goy} \\
    $e^{+}e^{-} \rightarrow Zh_{95} \rightarrow Zb\overline{b}$ & $0.117\pm 0.057$ & $2.3 (<1)\sigma$ \cite{LEPWorkingGroupforHiggsbosonsearches:2003ing} \\
    \hline
  \end{tabular} \label{tabel1}
  \caption{Observed signal rates for a possible new 95 GeV scalar particle.}
\end{table}

For the Higgs boson at about 95 GeV, we have
\begin{align}
\mu(H_{95})_{X}=\frac{\sigma(h_{95})\times Br^{NP}(h_{95} \rightarrow X)}{\sigma^{SM}(h_{95})\times Br^{SM}(h_{95} \rightarrow X)}.
\end{align}
The process measured at LEP was reported a 2.3$\sigma$ local excess in the $b\bar{b}$ final state searches,
with the scalar mass at$\sim$96 GeV. We consider the production of a Higgs boson via Higgstrahlung associated with the Higgs boson decaying to bottom-quarks. In the LRSSM, we use $\mu_{NP}^{bb}$ and $\mu_{NP}^{\gamma\gamma}$ to describe the signal strengths \cite{Cao:2016uwt,Biekotter:2020cjs,Biekotter:2017xmf}
\begin{align}
\mu_{NP}^{bb}&=\frac{\sigma^{NP}(Z^{*}\rightarrow Zh_{1})}{\sigma^{SM}(Z^{*}\rightarrow Zh_{1})}\times \frac{Br^{NP}(h_{1}\rightarrow  b\bar{b})}{Br^{SM}(h_{1}\rightarrow  b\bar{b})} \\ \nonumber
& \approx |C_{h_{1}VV}|^2 \times \frac{\Gamma^{NP}_{h_{1}\rightarrow b\bar{b}}}{\Gamma^{SM}_{h_{1}\rightarrow b\bar{b}}} \times \frac{\Gamma^{SM}_{\mathrm{tot}}}{\Gamma^{NP}_{\mathrm{tot}}}  \\ \nonumber
&\approx \frac{|C_{h_1VV}|^2 \times |C_{h_1dd}|^2}{|C_{h_1dd}|^2 (Br_{h_1 \to b\bar{b}}^{SM} + Br_{h_1 \to \gamma\gamma}^{SM}) + |C_{h_1uu}|^2 (Br_{h_1 \to gg}^{SM} + Br_{h_1 \to c\bar{c}}^{SM})},
\end{align}
\begin{align}
\mu_{NP}^{\gamma\gamma}=&\frac{\sigma^{NP}(gg  \rightarrow h_{1})}{\sigma^{SM}(gg\rightarrow h_{1})}\times \frac{Br^{NP}(h_{1}\rightarrow  \gamma\gamma)}{Br^{SM}(h_{1}\rightarrow  \gamma\gamma)}\\ \nonumber
&  \approx \frac{\Gamma^{NP}_{h_{1}\rightarrow gg}}{\Gamma^{SM}_{h_{1}\rightarrow gg}}\times \frac{\Gamma^{NP}_{h_{1}\rightarrow \gamma\gamma}}{\Gamma^{SM}_{h_{1}\rightarrow \gamma\gamma}}\times \frac{\Gamma^{SM}_{\mathrm{tot}}}{\Gamma^{NP}_{\mathrm{tot}}} \\ \nonumber
&\approx \frac{|C_{h_1uu}|^2 \times |C_{h_1\gamma\gamma}|^2}{|C_{h_1dd}|^2 \left( Br_{h_1 \to b\bar{b}}^{SM} + Br_{h_1 \to \tau\tau}^{SM} \right) + |C_{h_1uu}|^2 \left( Br_{h_1 \to gg}^{SM} + Br_{h_1 \to c\bar{c}}^{SM} \right)},
\end{align}
where \cite{Biekotter:2017xmf}
\begin{align}
|C_{h_1\gamma\gamma}|^2 = \frac{\left|\frac{4}{3}C_{h_1tt}A_{1/2}(\tau_t) + C_{h_1VV}A_1(\tau_W)\right|^2}{\left|\frac{4}{3}A_{1/2}(\tau_t) + A_1(\tau_W)\right|^2}
\end{align}
with $\tau_t = \frac{m_{h_1}^2}{4m_t^2} < 1$ and $\tau_W = \frac{m_{h_1}^2}{4m_W^2} < 1$. The form factors $A_{1/2} $ and $ A_1 $ are given by \cite{Djouadi:2005gi}
\begin{align}
A_{1/2}(x) = 2(x + (x - 1) \arcsin^2\sqrt{x}) x^{-2}, \quad x \leq 1,  \nonumber \\
A_1(x) = -(2x^2 + 3x + 3(2x - 1) \arcsin^2\sqrt{x}) x^{-2}, \quad x \leq 1.
\end{align}
$C_{h_1VV}$ is the ratio of vertices in the SM and LRSSM related to the coupling vertex between the  \( h_1 \) and the gauge boson \( W^{\pm}_L \), while $C_{h_1u\bar{u}(h_1d\bar{d})}$ is the ratio of vertices between $h_1$ and up-type quarks as well as down-type quarks. The specific form of the couplings are given by
\begin{align}
C_{h_1d\bar{d}} = \frac{Z^H_{11}}{\cos\beta}, \quad C_{h_1u\bar{u}} = \frac{Z^H_{12}}{\sin\beta}, \quad C_{h_1VV} = Z^H_{11} \cos\beta + Z^H_{12} \sin\beta. \label{C95}
\end{align}
Here, $Z_H$ is the unitary matrix of the CP-even Higgs field mass matrix.
\section{NUMERICAL ANALYSIS}
In this section, we conduct a detailed exploration of two different scenarios. In the first scenario, we consider only the 125 GeV SM-like Higgs boson, serving as a benchmark model for comparative analysis. The second scenario expands to include two scalar particles with masses of 95 GeV and 125 GeV \cite{ParticleDataGroup:2022pth}. The
 measured SM-like Higgs mass is \cite{ParticleDataGroup:2024cfk}
\begin{align}
  m_{h}=125.09 \pm 0.24\ \text{GeV}\label{mhhh}
\end{align}
The SM input parameters we used are $m_{W}=80.377$ GeV, $m_{Z}=91.1876$ GeV, $\sin^{2}\theta_{W}(m_{Z})=0.2312$, $\alpha_{\mathrm{s}}(m_{Z})=0.118$ and the fermion coupling constant $G_{F}=1.1664\times 10^{-5}$ GeV${}^{-2}$. The masses of SM fermions are taken as $m_{\tau}=1.78$ GeV, $m_{c}=1.28$ GeV, $m_{t}=173.5$ GeV and $m_{b}=4.65$ GeV from the PDG \cite{ParticleDataGroup:2024cfk}. Furthermore, we take $\lambda_4=\lambda_5=T_L=T_4=T_5=0$ for simplicity. It is important to note that while the model framework adopted here is similar to that in Ref. \cite{Frank:2017tsm}, our analysis incorporates one-loop and two-loop effective potential corrections to the Higgs mass matrices and explores a significantly extended parameter space through a comprehensive rescan. The choice to fix the specific parameters mentioned above is a decision informed by the numerical results and feedback obtained from our scanning procedure. This approach simplifies the numerical calculation without loss of generality.
\subsection{Signal Strengths of the 125\,GeV Higgs Bosons}
\begin{figure}[!htb]
\setlength{\unitlength}{5.0mm}
\centering
\includegraphics[width=2.5in]{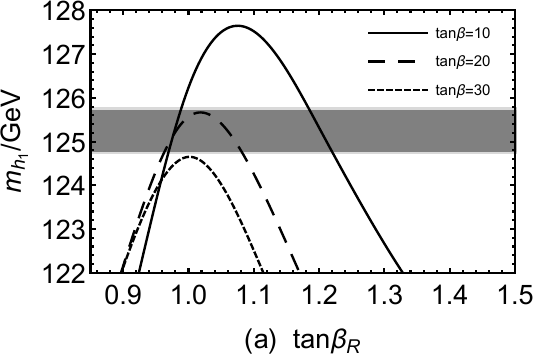}
\vspace{0.2cm}
\setlength{\unitlength}{5.0mm}
\centering
\includegraphics[width=2.5in]{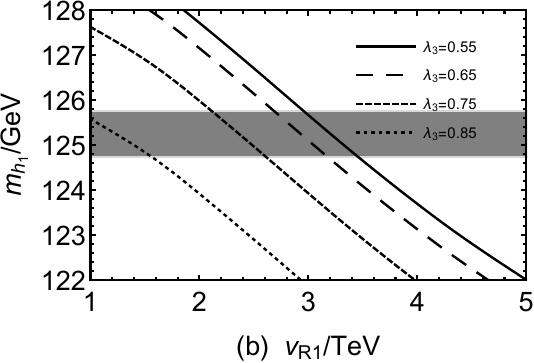}
\vspace{0.2cm} \\
\setlength{\unitlength}{5.0mm}
\centering
\includegraphics[width=2.5in]{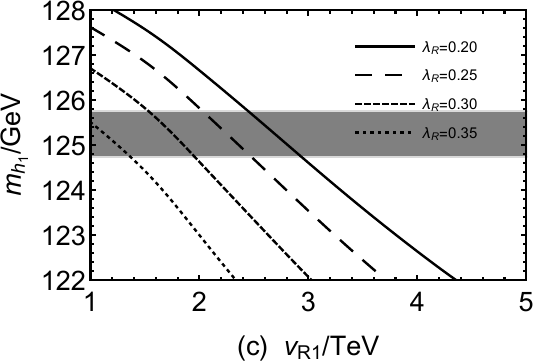}
\vspace{0.2cm}
\setlength{\unitlength}{5.0mm}
\centering
\includegraphics[width=2.5in]{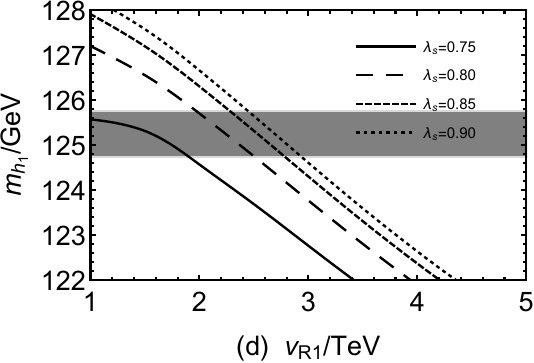}
\vspace{0.2cm}
\caption{(a) displays the mass of the CP-even Higgs boson $ m_{h_1} $ varies with $ \tan \beta_R $ and different values of $\tan \beta$ are taken. (b), (c), (d) display $m_{h_1}$ varies with $v_{R1}$ with taking the different values of $\lambda_3,\lambda_R,\lambda_S$ respectively. The horizontal gray band represents the $3\sigma$ interval of Eq. (\ref{mhhh}).}\label{fig:mh125}
\end{figure}

In order to globally fit the electroweak experimental data\cite{Frank:2017tsm}, we adopt:
\begin{align}
&T_S=-5500\ \text{GeV}, && T_R=-5000\ \text{GeV},\\ \nonumber
&T_3=-2500\ \text{GeV}, && T_Q^1=T_Q^2=-1000\ \text{GeV}, \\ \nonumber
&M_1=1000\ \text{GeV}, && M_{2L}=M_{2R}=5000\ \text{GeV},\\ \nonumber
&m_{Q_R}^2=m_{Q_L}^2=4\times10^{6}\ \text{GeV}^2, && \xi_F=-5500\ \text{GeV}^2, \\ \nonumber
&m_{L_R}^2=m_{L_L}^2=4\times10^{6}\ \text{GeV}^2.
\end{align}
In the numerical analyses we mainly focus on the impact from the following parameters on the Higgs masses and couplings:
\begin{align}
\tan\beta,\tan\beta_R,\lambda_3,\lambda_S,\lambda_R,v_S,v_{R2}
\end{align}
In the sector of the parameter space, the model accommodates a neutral Higgs with mass around 125 GeV.
Taking $v_{R1}=2.5$ $\text{TeV},v_{S}=5$ $\text{TeV},\lambda_3=0.7,\lambda_R=0.2$ and $\lambda_S=0.9$, we plot the theoretical prediction on the mass of the lightest neutral Higgs versus the parameter $\tan\beta_R$ in Fig. \ref{fig:mh125}(a), where the solid line corresponds to $\tan\beta=10$ , the dashed line corresponds to $\tan\beta=20$, and the dotted line corresponds to $\tan\beta=30$, respectively. Fig. \ref{fig:mh125}(b) to (c) also presents the relationship between the mass of the lightest Higgs boson and $v_{R1}$. Wherein the solid, dashed, and dotted lines represent increasing values of the parameters $\lambda_{3}, \lambda_{R}$, and $\lambda_{S}$, respectively. The numerical results indicate that there is parameter space to accommodate the 125 GeV Higgs in this model.

Fig.~\ref{fig:mh125} indicates that by adjusting $\tan\beta,\tan\beta_{R},v_{R1},v_{S},\lambda_3,\lambda_R$ and $\lambda_S$, suitable parameters can be found in the model to align the light Higgs particle mass with the observed value, providing a reference for constructing supersymmetric models consistent with experimental data. Because  the mass of the lightest Higgs boson is influenced significantly in a complicated manner by $\tan\beta,\tan\beta_{R},v_{R1},v_{S},\lambda_3,\lambda_R,\lambda_S$ we scan the following parameter space
 \begin{align}
 &\tan\beta\in(2,50),\tan\beta_{R}\in(0.8,1.5),v_{R1}\in(2,3)\ \text{TeV},v_S\in(4,6)\ \text{TeV},\nonumber \\
 &\lambda_S\in(0.6,1.0),\lambda_3\in(0.6,1.0), \lambda_R\in(0,0.4). \label{cs125}
 \end{align}
 To comprehensively explore the collective influence of $ \tan \beta $, $ \tan \beta_R $, $ v_{R1} $, $ v_S $, $ \lambda_S $, $ \lambda_3 $, and $ \lambda_R $ on the signal strength of the 125 GeV Higgs boson, while maintaining $m_{h_1}$ in the vicinity of 125 GeV during the scan, we specified in Eq. (\ref{cs125}) that $ \tan \beta $ varies between 2 and 50, $ \tan \beta_R $ ranges from 0.8 to 1.5, $\lambda_{3}$ ranges from 0.6 to 1.0, $\lambda_{R}$ from 0 to 0.4 and $\lambda_S$ from 0.6 to 1.0.

\begin{figure}[H]
\setlength{\unitlength}{5.0mm}
\centering
\includegraphics[width=2.5in]{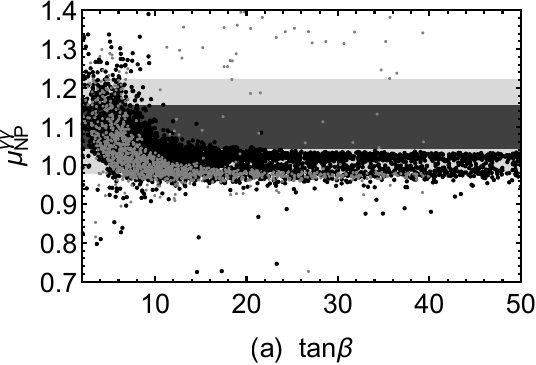}
\vspace{0.2cm}
\setlength{\unitlength}{5.0mm}
\centering
\includegraphics[width=2.5in]{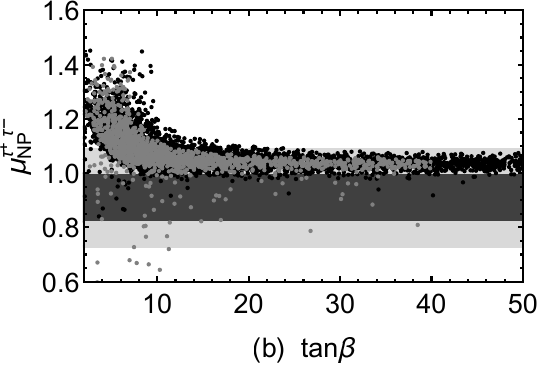}
\vspace{0.2cm}\\ \nonumber
\setlength{\unitlength}{5.0mm}
\centering
\includegraphics[width=2.5in]{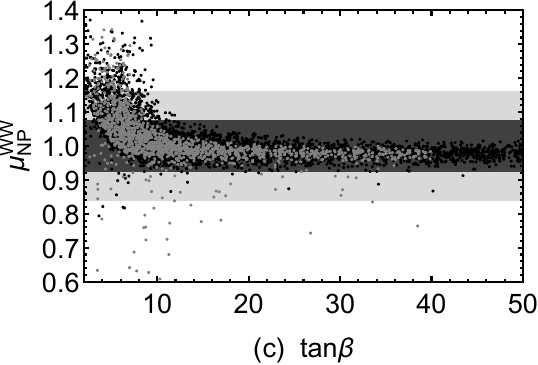}
\vspace{0.2cm}
\setlength{\unitlength}{5.0mm}
\centering
\includegraphics[width=2.5in]{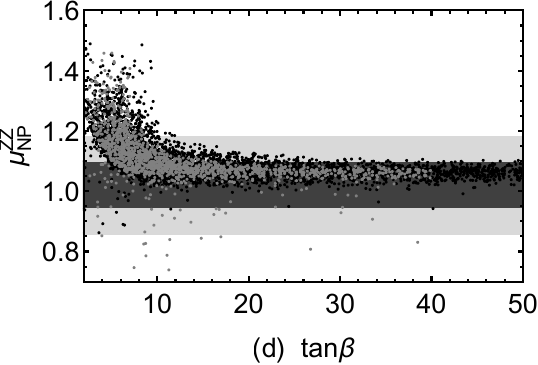}
\vspace{0.2cm}\\ \nonumber
\setlength{\unitlength}{5.0mm}
\centering
\includegraphics[width=2.5in]{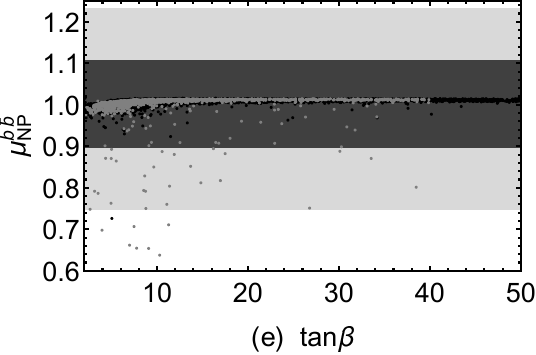}
\vspace{0.2cm}
\setlength{\unitlength}{5.0mm}
\centering
\includegraphics[width=2.7in]{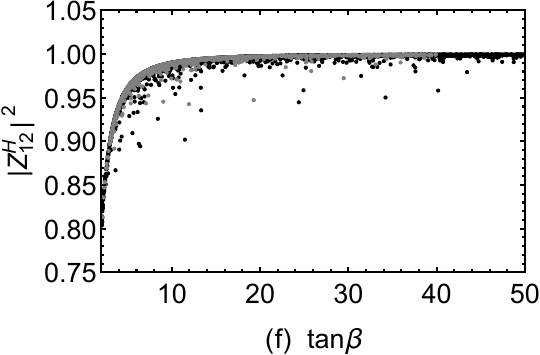}
\vspace{0.2cm}
\caption{\label{fig:125xh}A scan was performed in the parameter space provided by Eq. (\ref{cs125}),  by ensuring that $ m_{h1} $ lies within the range of 124 GeV to 126 GeV. (a) to (e) display the results of signal strengths $\mu_{NP}^{\gamma\gamma},\mu_{NP}^{\tau^{+}\tau^{-}},\mu_{NP}^{WW},\mu_{NP}^{ZZ}$ and $\mu_{NP}^{b\bar{b}}$ respectively, where $ \tan \beta $ on the horizontal axis and the normalized signal strength for various decay channels on the vertical axis. The dark gray shaded areas represent the experimental 1$ \sigma $ range, and the light gray areas represent the experimental 2$ \sigma $ range. (f) displays the relationship between the mixing matrix elements $\tan\beta$ and $|Z^{H}_{12}|^2$ for the two Higgs bosons. The gray points denote values of \(m_{h_1}\) within the range of 124 to 126 GeV, while the black points correspond to masses of \(m_{h_1}\) in the interval of 122 to 128 GeV.}
\end{figure}

With our assumption on the parameter space, it can be readily observed that the distribution of points in the figure is essentially consistent for the case where \(m_{h_1}\) lies within the 122-128 GeV interval and the case where it is constrained to the 124-126 GeV interval. As the scanned mass range of \(m_{h_1}\) becomes narrower, the number of corresponding points in the figure decreases. We present the signal strength $\mu_{NP}^{\gamma\gamma}$ versus $\tan\beta$ in Fig. \ref{fig:125xh}(a), the signal strength $\mu_{NP}^{\tau^{+}\tau^{-}}$ versus $\tan\beta$ in Fig. \ref{fig:125xh}(b), the signal strength $\mu_{NP}^{WW}$ versus $\tan\beta$ in
Fig. \ref{fig:125xh}(d) and the signal strength $ \mu_{NP}^{b\bar{b}}$ versus $\tan\beta$ in Fig. \ref{fig:125xh}(e), respectively. With the increase of $\tan\beta$, the signal strength for each decay channel approaches a certain value, and coincides with the experimental data at large $\tan\beta$.

\begin{figure}
\setlength{\unitlength}{5.0mm}
\centering
\includegraphics[width=2.5in]{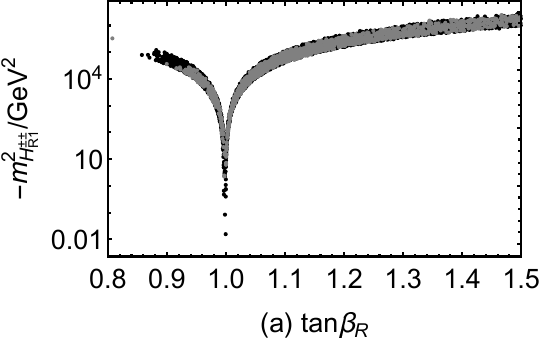}
\vspace{0.2cm}
\setlength{\unitlength}{5.0mm}
\centering
\includegraphics[width=2.4in]{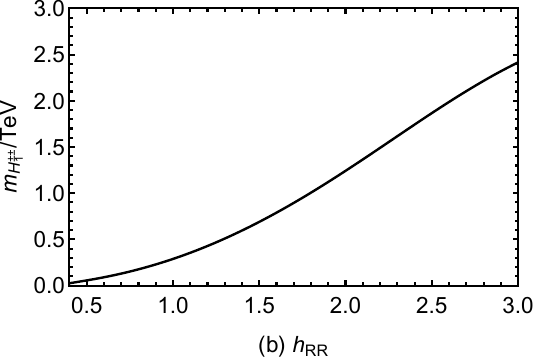}
\vspace{0.2cm}
\caption{\label{fig:hrmm125}The left panel (a) shows the negative doubly charged Higgs mass square as a function of $ \tan \beta_R $ at the tree level, while the right panel (b) illustrates the doubly charged Higgs mass as a function of \(h_{RR}\) after incorporating one-loop effective potential corrections.}
\end{figure}

Fig.~\ref{fig:125xh}(f) shows the relationship between the mixing matrix elements $|Z_{12}^H|^2$ and $\tan\beta$, where $|Z_{12}^H|^2$ represents the contribution of the $ \phi_2 $ component.
 The horizontal axis represents $|Z_{12}^H|^2$, and the vertical axis represents $\tan\beta$. As $\tan\beta$ increases, $|Z_{12}^H|^2$ approaches unity. From Fig. \ref{fig:125xh}(f), it can be observed that $|Z_{12}^H|^2$ is distributed around values close to 1, indicating that the $ \phi_2 $ component dominates the contribution to the 125 GeV Higgs boson. Moreover, the numerical result indicates that when $|Z_{12}^H|^2$ approaches 1, the composition of the Higgs boson is consistent with the expectations of the SM, making the physical properties of the 125 GeV Higgs boson similar in the LRSSM to the SM-like Higgs boson.

 As tan$\beta$ increases, the ratios for both channels approach unity in the LRSSM, but the gray points slightly exceed 1, indicating that the LRSSM predicts a larger signal strength than the Standard Model in this regime.

Fig.~\ref{fig:hrmm125}(a) presents the mass of the lightest doubly charged Higgs boson within the parameter space of the model. In the LRSSM, the tree-level squared mass of the lightest doubly charged Higgs boson, denoted as \(m_{H^{\pm\pm}_{R_1}}^2\), is permitted to be negative and exhibits a strong dependence on \(\tan\beta_R\). As \(\tan\beta_R\) approaches unity, the value of \(m_{H^{\pm\pm}_{R_1}}\) tends toward zero. The explicit functional relationship between \(m_{H^{\pm\pm}_{R_1}}\) and \(\tan\beta_R\) is detailed in Appendix \ref{app5}. As clearly illustrated in Fig. \ref{fig:hrmm125}(a), the correlation between \(m_{H^{\pm\pm}_{R_1}}\) and \(\tan\beta_R\) is readily observable. According to the latest results from the LHC, the lower mass limit for \(m_{H^{\pm\pm}_{R_1}}\) is set at 1.3 TeV \cite{Horii:2024qiw,ATLAS:2022pbd,ATLAS:2021jol}. In our calculation of the one-loop effective potential corrections to \(m_{H^{\pm\pm}_{R_1}}\), the contributions from right-handed neutrinos dominate, while the effects from other particles are minimal and can be largely neglected. Employing the parameters from Fig. 1 with $\tan\beta$=20, $\tan\beta_R$=1.01, $v_{R1}$=2.5 TeV, $v_{S}$=5 TeV, $\lambda_3$=0.7, $\lambda_R$=0.2 and $\lambda_S$=0.9,  the mass of the doubly charged Higgs boson \(m_{H^{\pm\pm}_{R_1}}\) after one-loop effective potential corrections exhibits a positive correlation with the coupling parameter \(h_{RR}\). The expression for the right-handed neutrino mass is $m_{\nu_R}=\frac{1}{\sqrt{2}}\,h_{RR}\,v_{R1}.$ Simultaneously, the mass of the right-handed neutrino increases with larger values of \(h_{RR}\). Consequently, it is straightforward to conclude that higher right-handed neutrino masses facilitate \(m_{H^{\pm\pm}_{R_1}}\) more easily satisfying the experimental lower limit.

\subsection{Signal Strengths of the 95\,GeV and 125\,GeV Higgs Bosons}
\begin{figure}[H]
\setlength{\unitlength}{5.0mm}
\centering
\includegraphics[width=2.4in]{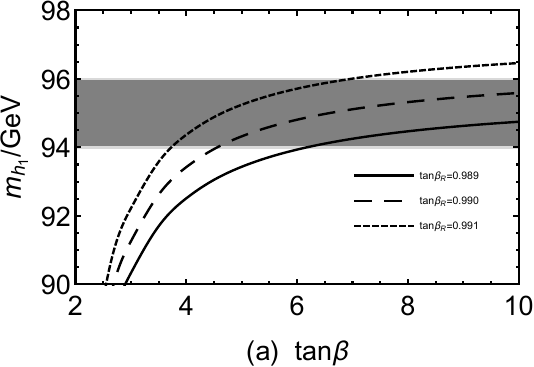}
\vspace{0.2cm}
\setlength{\unitlength}{5.0mm}
\centering
\includegraphics[width=2.4in]{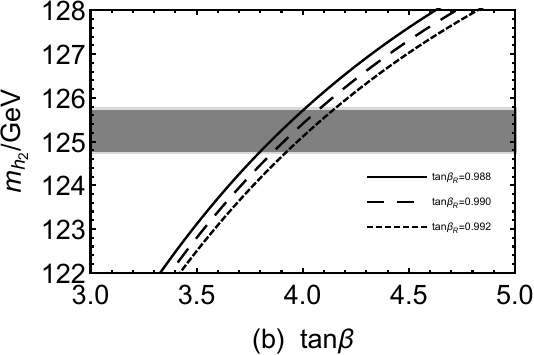}
\vspace{0.2cm} \\
\setlength{\unitlength}{5.0mm}
\centering
\includegraphics[width=2.4in]{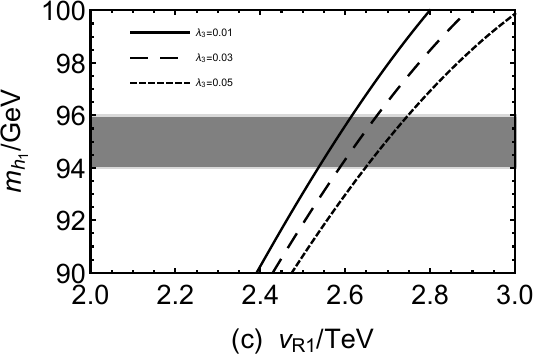}
\vspace{0.2cm}
\setlength{\unitlength}{5.0mm}
\centering
\includegraphics[width=2.5in]{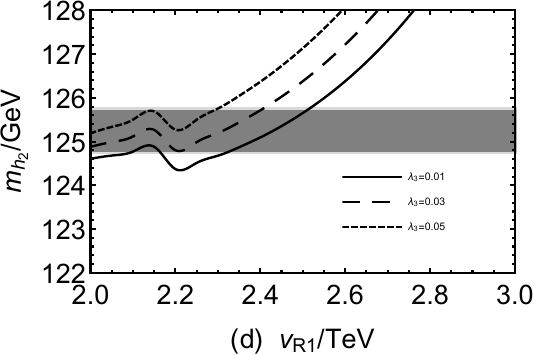}
\vspace{0.2cm} \\
\setlength{\unitlength}{5.0mm}
\centering
\includegraphics[width=2.4in]{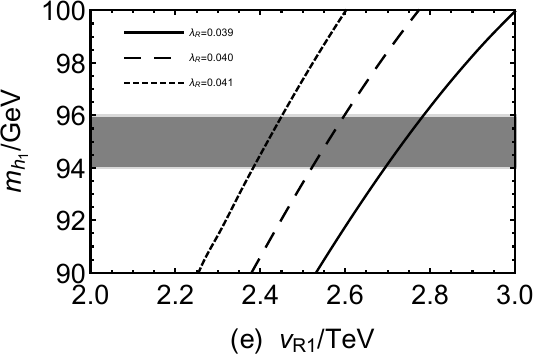}
\vspace{0.2cm}
\setlength{\unitlength}{5.0mm}
\centering
\includegraphics[width=2.4in]{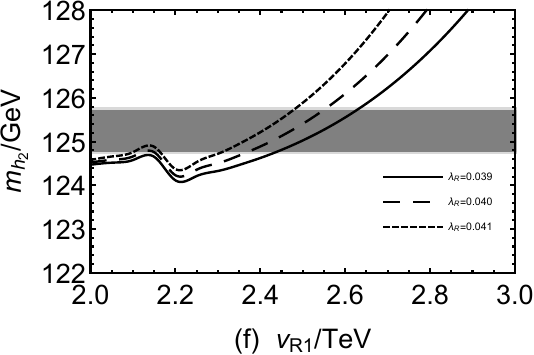}
\vspace{0.2cm} \\
\setlength{\unitlength}{5.0mm}
\centering
\includegraphics[width=2.4in]{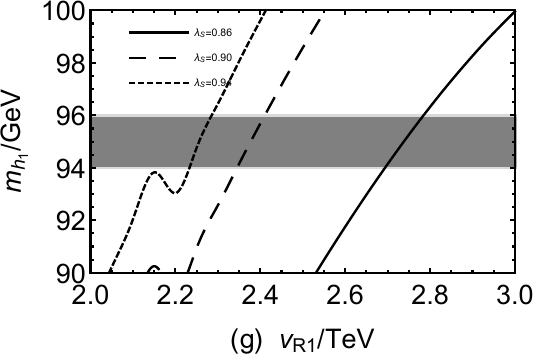}
\vspace{0.2cm}
\setlength{\unitlength}{5.0mm}
\centering
\includegraphics[width=2.4in]{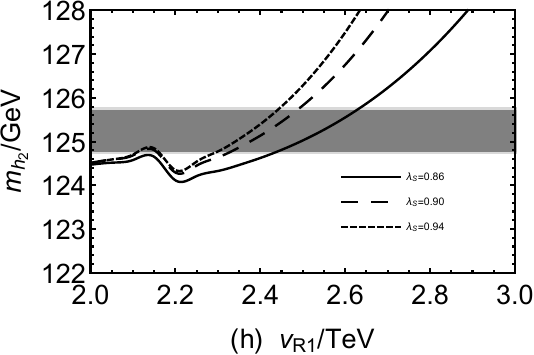}
\vspace{0.2cm}
\caption{
(a) and (b) show the lightest
and second-lightest CP-even Higgs bosons masses versus $\tan\beta$ respectively.
(c), (e) and (g) display the dependence of the lightest CP-even
Higgs-boson mass on $v_{R1}$, while (d), (f) and (h) illustrate the
corresponding variation of the second-lightest CP-even Higgs-boson mass with
$v_{R1}$. In (a),(c), (e) and (g), the horizontal gray band marks the $94$-$96\,\text{GeV}$ mass window,
and in (b),(d), (f) and (h), the horizontal gray band represents the $3\sigma$ interval of Eq. (\ref{mhhh}).
}\label{fig:mh95}
\end{figure}

In our discussion of Higgs bosons where the lightest mass is 95 GeV and the second lightest is 125 GeV, the parameter space differs slightly from that of the previous section, which considered only the existence of a Higgs boson with a mass of 125 GeV. Similarly, we adopt the following assumption on the parameter space:
\begin{align}
&T_S=3000\ \text{GeV}, && T_R=-2500\ \text{GeV},\\ \nonumber
&T_3=4000\ \text{GeV}, && T_Q^1=T_Q^2=-1000\ \text{GeV}, \\ \nonumber
&M_1=1000\ \text{GeV}, && M_{2L}=M_{2R}=5000\ \text{GeV},\\ \nonumber
&m_{Q_R}^2=m_{Q_L}^2=4\times10^{6}\ \text{GeV}^2, && \xi_F=-5500\ \text{GeV}^2, \\ \nonumber
&m_{L_R}^2=m_{L_L}^2=4\times10^{6}\ \text{GeV}^2.
\end{align}
In this parameter space, it is possible to simultaneously have Higgs bosons with masses around 95 GeV and 125 GeV. Taking $v_{R1}=1.5$ $\text{TeV},v_{S}=9.5$ $\text{TeV},\lambda_3=0.1,\lambda_R=0.04$ and $\lambda_S=0.9$, we plot the theoretical prediction of the mass of the lightest and the second lightest neutral Higgs bosons versus the parameter $\tan\beta$, where the solid line corresponds to $\tan\beta_R=0.988$, the dashed line corresponds to $\tan\beta_R=0.990$, and the dotted line corresponds to $\tan\beta_R=0.992$, respectively. We also plot the relationships between the lightest and second-lightest CP-even Higgs bosons and \(v_{R1}\) for different values of \(\lambda_3\), \(\lambda_R\), and \(\lambda_S\). In Fig. \ref{fig:mh95}(b) to (d), the solid and dashed lines represent increasing values of the parameters \(\lambda_3\), \(\lambda_R\), and \(\lambda_S\), respectively. The numerical results shown in Fig. \ref{fig:mh95} indicate that there is parameter space to accommodate the 95 GeV Higgs boson in this model.

We scanned the parameter space with
\begin{align}
 &\tan\beta\in(2,10),\tan\beta_{R}\in(0.98,1.00),v_{R1}\in(2,3)\ \text{TeV},v_S\in(9,11)\ \text{TeV},\nonumber \\
 &\lambda_S\in(0.9,1.0),\lambda_3\in(0,0.5), \lambda_R\in(0.03,0.04).
 \end{align}
With the above assumption on the parameter space, we investigate the signal strengths of 95 GeV and 125 GeV simultaneously. Obviously, the explanation of the 95 GeV and 125 GeV Higgs boson particles and their signal strengths imposes stronger constraints on the parameter space of the model concerned here. For example, the scanning ranges for $ \tan \beta_R $ and $ \lambda_R $ are significantly smaller, and the signal of the Higgs with a mass of 125 GeV must also conform to experimental findings. In the CP-even Higgs sector of the LRSSM, the lightest Higgs boson has a mass of 95 GeV, while the next lightest Higgs boson has a mass of 125 GeV. In the scanned parameter space, we first analyze the signal strength of the 125 GeV Higgs boson.

We selected the signal strength that matches the 125 GeV Higgs boson as provided by the PDG \cite{ParticleDataGroup:2024cfk}. Under the aforementioned conditions, we calculated the signal strengths of the 95 GeV Higgs boson for its final states of diphoton and $b\bar{b}$.
\begin{figure}[h]
\setlength{\unitlength}{5.0mm}
\centering
\includegraphics[width=2.5in]{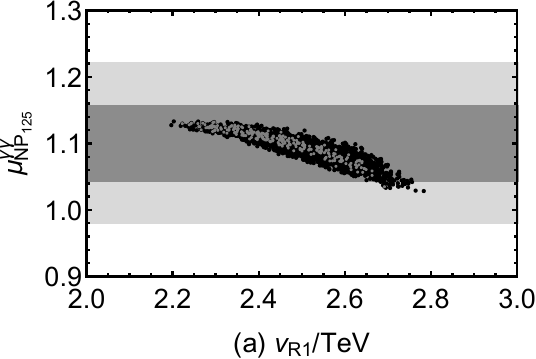}
\vspace{0.2cm}
\setlength{\unitlength}{5.0mm}
\centering
\includegraphics[width=2.5in]{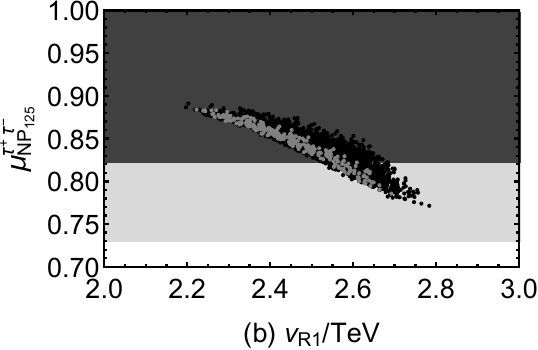}
\vspace{0.2cm}\\ \nonumber
\setlength{\unitlength}{5.0mm}
\centering
\includegraphics[width=2.5in]{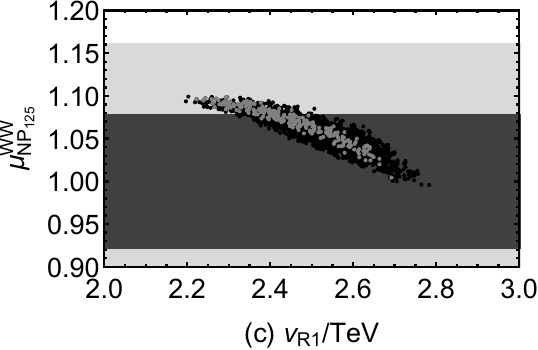}
\vspace{0.2cm}
\setlength{\unitlength}{5.0mm}
\centering
\includegraphics[width=2.5in]{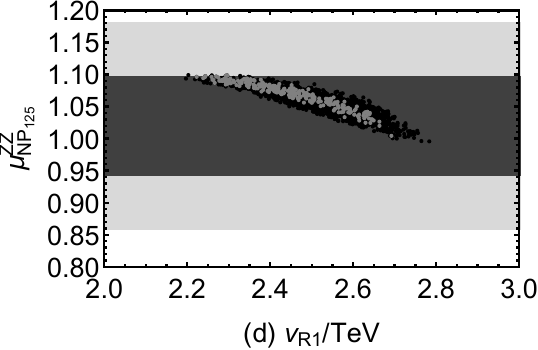}
\vspace{0.2cm}\\ \nonumber
\setlength{\unitlength}{5.0mm}
\centering
\includegraphics[width=2.5in]{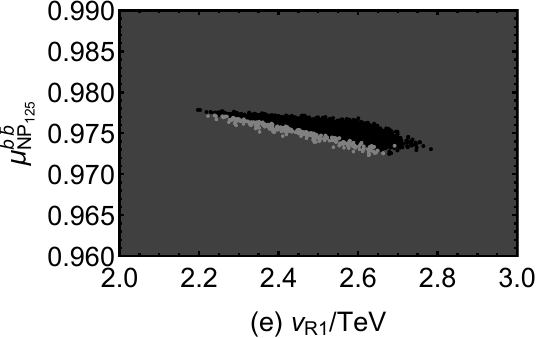}
\vspace{0.2cm}
\setlength{\unitlength}{5.0mm}
\centering
\includegraphics[width=2.55in]{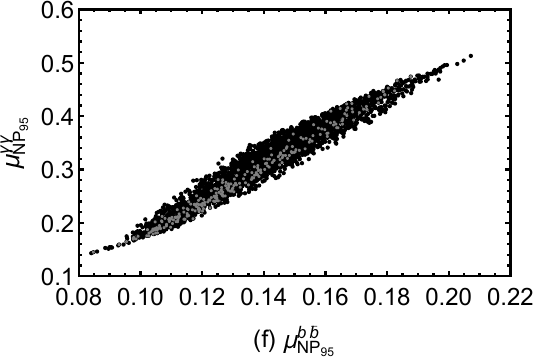}
\vspace{0.2cm}
\caption{(a) to (e) present the signal strength of the 125 GeV Higgs boson in different decay channels as a function of the right-handed Higgs vacuum expectation value $ v_{R1} $, with the dark and light gray regions indicating the $1\sigma$ and $2\sigma$ experimental ranges, respectively. Plot (f) shows the 95\,GeV Higgs boson signal strengths of $\gamma\gamma$ ($\mu_{{\rm NP}_{95}}^{\gamma\gamma}$) and $b\bar{b}$ ($\mu_{{\rm NP}_{95}}^{b\bar b}$) decay channels. Gray and black points denote benchmarks with the next-to-lightest Higgs mass in 124-126 GeV and 122-128 GeV, respectively.}\label{fig:95xh}
\end{figure}

\begin{figure}[h]
\setlength{\unitlength}{5.0mm}
\centering
\includegraphics[width=2.0in]{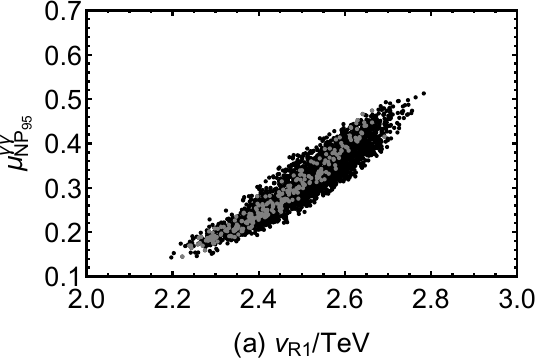}
\vspace{0.2cm}
\setlength{\unitlength}{5.0mm}
\centering
\includegraphics[width=2.0in]{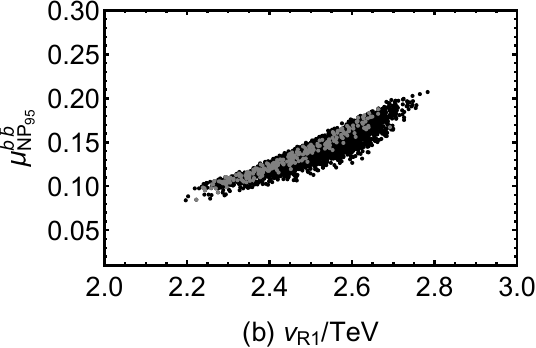}
\vspace{0.2cm}
\caption{$\mu_{{\rm NP}_{95}}^{\gamma\gamma}$ (a) and $\mu_{{\rm NP}_{95}}^{b\bar b}$ (b) as functions of $v_{R1}/\text{TeV}$ are plotted in (a) and (b) respectively. Gray points denote benchmarks with the next-to-lightest Higgs mass in 124-126 GeV, and black points for those in 122-128 GeV. }\label{fig:rrbb95}
\end{figure}

\begin{figure}[h]
\setlength{\unitlength}{5.0mm}
\centering
\includegraphics[width=2.0in]{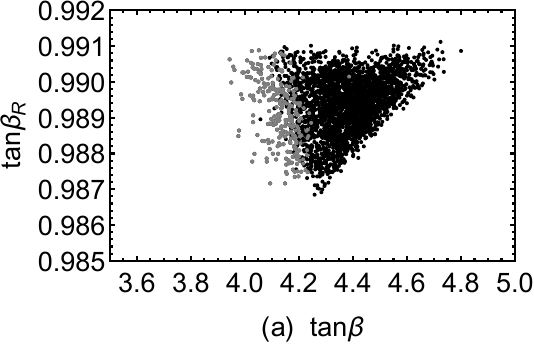}
\vspace{0.2cm}
\setlength{\unitlength}{5.0mm}
\centering
\includegraphics[width=2.0in]{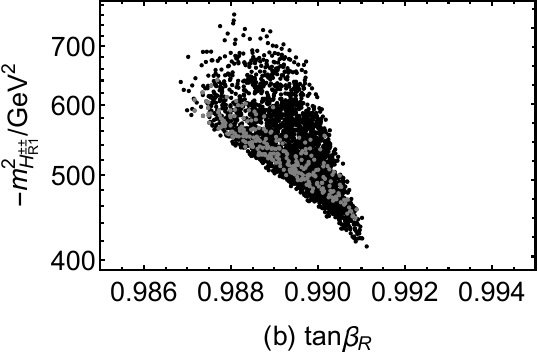}
\vspace{0.2cm}
\setlength{\unitlength}{5.0mm}
\centering
\includegraphics[width=2.0in]{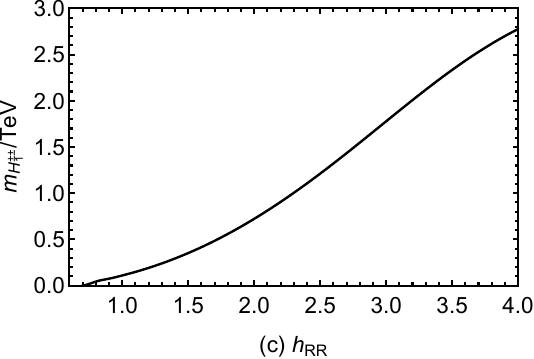}
\vspace{0.2cm}
\caption{(a) shows the relationship between $\tan\beta$ and $\tan\beta_R$. (b) shows the negative doubly charged Higgs mass square as a function of $ \tan \beta_R $. Gray points denote benchmarks with the next-to-lightest Higgs mass in 124-126 GeV, and black points for those in 122-128 GeV. (c) illustrates the relationship between the lightest doubly charged Higgs mass and the parameter \(h_{RR}\) after incorporating the one-loop effective potential corrections.}\label{fig:tbtr}
\end{figure}

With our assumption on the parameter space, we present the signal strength $\mu_{NP}^{\gamma\gamma}$ versus $v_{R2}$ in Fig. \ref{fig:95xh}(a), the signal strength $\mu_{NP}^{\tau^{+}\tau^{-}}$ versus $v_{R1}$ in Fig. \ref{fig:95xh}(b), the signal strength $\mu_{NP}^{WW}$ versus $v_{R1}$ in
Fig. \ref{fig:95xh}(c), the signal strength $ \mu^{ZZ}_{\text{NP}}$ versus $v_{R2}$ in Fig. \ref{fig:95xh}(d), the signal strength $ \mu^{b\bar{b}}_{NP}$ versus $v_{R2}$ in Fig. \ref{fig:95xh}(e), respectively. In Fig. \ref{fig:95xh}(a) to (e), the dark gray regions correspond to the \(1\sigma\) signal strength range of the 125 GeV Higgs boson, while the light gray areas represent the \(2\sigma\) signal strength range. In Figs.~\ref{fig:95xh}, \ref{fig:rrbb95}, and \ref{fig:tbtr}, the gray points denote benchmarks where the next-to-lightest Higgs mass lies in $124\text{-}126$~GeV, while the black points denote those with the next-to-lightest Higgs mass in $122\text{-}128$~GeV. As can be readily observed in Fig. \ref{fig:95xh}, under the condition that both the $\gamma\gamma$ and $b\bar{b}$ signal strengths of the lightest Higgs boson at 95 GeV and those of the second-lightest Higgs boson at 125 GeV largely fall within the 1$\sigma$ confidence level, the majority of the parameter points are concentrated in the range of 2 TeV $<v_{R1}<$ 3 TeV. We also present the correlation between the $b\bar{b}$ and $\gamma\gamma$ signal strengths of the 95 GeV Higgs boson in this parameter space, as shown in Fig. \ref{fig:95xh}(f). We considered the experimental constraints on the charged Higgs boson sector by applying the latest lower mass bounds from LHC searches \cite{CMS:2020imj}. Although certain marginal points in the preliminary scan exhibited enhanced signal strengths due to the presence of a lighter $H_1^\pm$, these have been effectively excluded after enforcing this stringent limit. The plots in Fig. \ref{fig:rrbb95}(a) and (b) show the signal strength distributions of the 95 GeV Higgs boson in the
$\gamma\gamma$ and $b\bar{b}$ decay channels, respectively, as functions of the parameter $v_{R1}$ in the LRSSM. To simultaneously explain the 95 GeV and 125 GeV Higgs bosons, the parameter $v_{R1}$ is constrained to the range of 2 to 3 TeV.

Fig. \ref{fig:tbtr}(a) illustrates the correlation between
$\tan\beta$ and $\tan\beta_R$. As can be seen from Fig. \ref{fig:tbtr}(a), the values of $\tan\beta_R$ approach but do not equal unity, clustering predominantly within the range of 0.986 to 0.992.
 This indicates that for the condition of satisfying the 95 GeV signal strength within the 1$\sigma$ range, the value of $\tan\beta_R$ cannot be strictly equal to 1 but is slightly deviated from it. In the analytical approximation presented in Appendix \ref{app5} for \(m_{H^{\pm}_1}\) and its dependence on \(\tan\beta_R\), it can be seen that when
$\tan\beta_R$=1, the mass of the charged Higgs boson becomes very small, nearly approaching zero, which cannot satisfy the experimental constraints. For the doubly charged Higgs boson, the negative value of its mass squared is also small when $\tan \beta_R$ is close to 1, but because it is negative, it can easily be corrected to yield a positive mass squared. Therefore, the mass of the doubly charged Higgs is more easily adjusted to meet experimental requirements through one-loop corrections in the model when $\tan \beta_R$ is near 1, which is also the reason for choosing $\tan \beta_R$ around 1. As similarly demonstrated for \(m_{H^{\pm\pm}_{R1}}\) in the previous section, the one-loop effective potential correction leads to a consistent conclusion. The corrected mass of \(m_{H^{\pm\pm}_{R1}}\) exhibits a positive correlation with the mass of the right-handed neutrino. Specifically, larger right-handed neutrino masses render it easier for \(m_{H^{\pm\pm}_{R1}}\) to satisfy the experimental lower limit. Under the current parameter space with $\tan\beta$=4.5, $\tan\beta_R$=0.99, $v_{R1}$=2.5 TeV, $v_{S}$=9.5 TeV, $\lambda_3$=0.2, $\lambda_R$=0.04 and $\lambda_S$=0.95. This ensures that the masses of the Higgs bosons in the model comply with the existing experimental limits.

\begin{figure}[h]
\setlength{\unitlength}{5.0mm}
\centering
\includegraphics[width=2.5in]{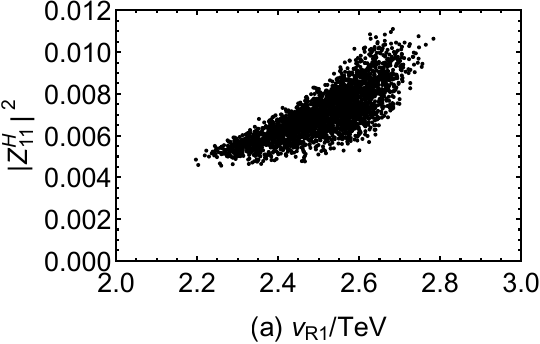}
\vspace{0.2cm}
\setlength{\unitlength}{5.0mm}
\centering
\includegraphics[width=2.3in]{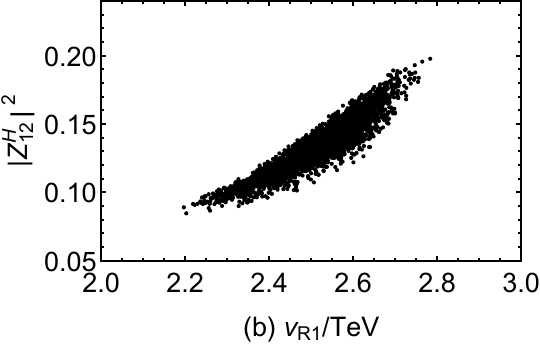}
\vspace{0.2cm} \\
\setlength{\unitlength}{5.0mm}
\centering
\includegraphics[width=2.3in]{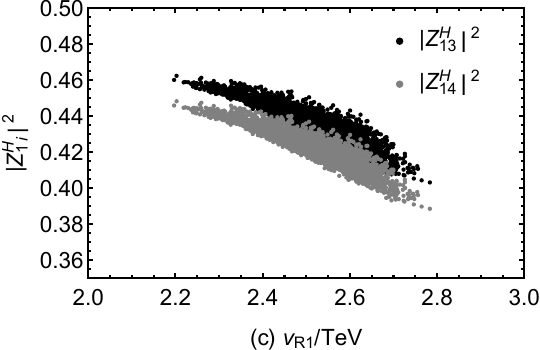}
\vspace{0.2cm}
\setlength{\unitlength}{5.0mm}
\centering
\includegraphics[width=2.3in]{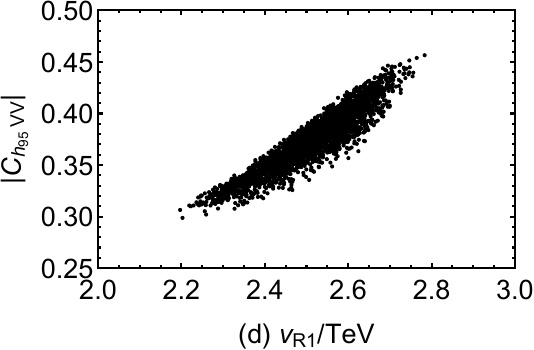}
\vspace{0.2cm}
\caption{$|Z^{H}_{11}|^2$ (a), $|Z^{H}_{12}|^2$ (b), $|Z^{H}_{13}|^2$ (black points in (c)), $|Z^{H}_{14}|^2$ (gray points in (c)) and $|C_{h_95VV}|$ (d) as functions of $v_{R1}$ are plotted. The points represent the parameter benchmarks from the prior signal strength calculation.}\label{fig:z95}
\end{figure}

\begin{figure}[h]
\setlength{\unitlength}{5.0mm}
\centering
\includegraphics[width=2in]{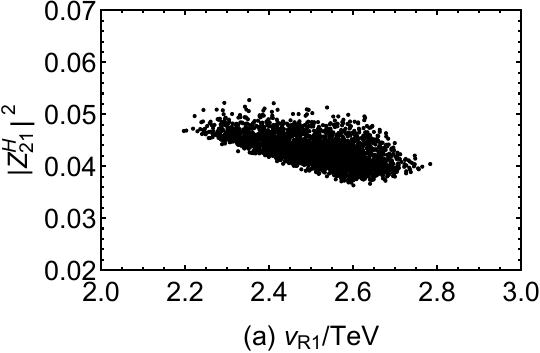}
\vspace{0.2cm}
\setlength{\unitlength}{5.0mm}
\centering
\includegraphics[width=2in]{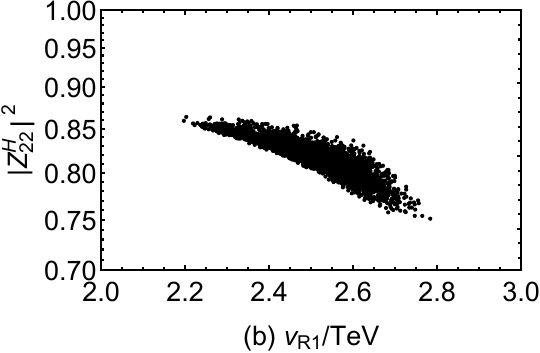}
\vspace{0.2cm}
\setlength{\unitlength}{5.0mm}
\centering
\includegraphics[width=2in]{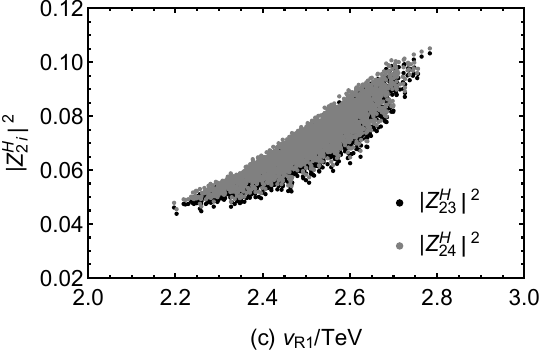}
\vspace{0.2cm}
\setlength{\unitlength}{5.0mm}
\caption{$|Z^{H}_{21}|^2$ (a), $|Z^{H}_{22}|^2$ (b), $|Z^{H}_{23}|^2$ (black points in (c)) and $|Z^{H}_{24}|^2$ (gray points in (c)) as functions of $v_{R1}$ are plotted. The points represent the parameter benchmarks from the prior signal strength calculation.
}\label{fig:zrrgg125}
\end{figure}
We also perform an analysis of the mass squared matrix structure and components of different Higgs states in the LRSSM, and explore the components of various gauge eigenstates in the 95 GeV and 125 GeV Higgs mass eigenstates, as shown in Fig.\ \ref{fig:z95} and Fig.\ \ref{fig:zrrgg125}. Specifically, the 125 GeV Higgs boson is primarily composed of the $\Phi_2$ field. In contrast, the 95 GeV Higgs boson is primarily composed of the
$\delta_R$ and $\Delta_R$ fields. In the LRSSM, these two fields belong to the right-handed Higgs triplet and are associated with the breaking of right-handed gauge symmetry. This association causes the 95 GeV Higgs state to primarily originate from the mixing of $\delta_R$ and $\Delta_R$. From Fig. \ref{fig:z95}(b), one can see that $|Z_{12}^H|^2$ lies in the range $0.1\text{--}0.2$, indicating that the
$95\,\mathrm{GeV}$ Higgs state contains a $10\%\text{--}20\%$ component from $\Phi_{2}$.
From Eq.~(\ref{C95}), it follows straightforwardly that the vertex $|C_{h_{95}VV}|$ depends on $|Z_{12}^H|^2$.
The numerical values of $|C_{h_{95}VV}|$ are shown in Fig. \ref{fig:z95}(d).
 The distribution of the squared values of the matrix elements in the figure allows us to see that different Higgs mass states reflect the complex field contributions and interactions behind them.

\section{SUMMARY}
After considering the two-loop radiative corrections to the effective potential, we analyze the pole mass squared of the CP-even Higgs boson in the LRSSM. The numerical results show that within the framework of this model, there exists a parameter space that can simultaneously account for the signal excesses of 95 GeV and 125 GeV observed in the experiments. The dominant component of the 125 GeV Higgs boson comes from the bidoublet $\Phi_2$, whereas the 95~GeV state is dominantly a mixture of the two triplets $\delta_R$, $\Delta_R$ and a portion of the bidoublet $\Phi_{2}$.

\begin{acknowledgments}
\indent\indent
The work has been supported by the National Natural Science Foundation of China (NNSFC) with Grants No. 12075074, No. 12235008, No. 11535002, No. 11705045, Natural Science Foundation for Distinguished Young Scholars of Hebei Province with Grant No. A2022201017, Natural Science Foundation of Guangxi Autonomous Region with Grant No. 2022GXNSFDA035068, the youth top-notch talent support program of the Hebei Province, and Midwest Universities Comprehensive Strength Promotion project.
\end{acknowledgments}

\appendix

\section{The Higgs Mass Matrix\label{app1}}
The minimization of the scalar potential proceeds as
\begin{align}
\frac{\partial V_{0}}{\partial H_{12}^0}=&\frac{1}{8}(8m_1^2v_1+g_L^2v_1(v_1^2-v_2^2)+g_R^2v_1^2(v_1^2-v_2^2+2v_{R1}^2-2v_{R2}^2)
-4\sqrt{2}T_3v_2v_S  \nonumber \\
&+4\lambda_3^2v_1v_S^2+4\lambda_3v_2(\lambda_3v_1v_2-\lambda_Rv_{R1}v_{R2}
-3\lambda_Sv_S^2-2\xi_F)),\\
\frac{\partial V_{0}}{\partial H_{21}^0}=&\frac{1}{8}(8m_2^2v_2-g_R^2v_1^2v_2+g_R^2v_2^3+g_L^2(-v_1^2v_2+v_2^3)-2g_R^2v_2v_{R1}^2+
2g_R^2v_2v_{R2}^2-4\sqrt{2}T_3v_1v_S \nonumber \\
&+4\lambda_3^2v_1^2v_2+4\lambda_3^2v_S^2v_2-4\lambda_S\lambda_Rv_1v_{R1}v_{R2}
-12\lambda_3\lambda_Sv_1v_S^2-8\lambda_3\xi_Fv_1), \\
\frac{\partial V_{0}}{\partial H_{\Delta_R}^0}=&\frac{1}{4}(4m_{\Delta_R}^2v_{R1}+g_R^2v_{R1}(v1^2-v_2^2+2v_{R2}^2-2v_{R2}^2)+2(g_{B}(v_{R1}^3-v_{R1}^2v_{R2}^2) \nonumber \\
&+\sqrt{2}T_Rv_{R2}v_S+\lambda_R(-\lambda_3v_1v_2v_{R2}+\lambda_Rv_{R1}(v_{R2}^2+v_S^2)+v_{R2}(3\lambda_Sv_S^2+2\xi_F)))), \\
\frac{\partial V_{0}}{\partial H_{\delta_R}^0}=&\frac{1}{4}(4m_{\Delta_R}^2v_{R2}+2g_{B}^2v_{R2}(-v_{R1}^2+v_{R2}^2)
+g_{R}^2v_{R2}(-v_1^2+v_2^2-2v_{R1}^2+2v_{R2}^2)  \nonumber \\
&+2\sqrt{2}T_Rv_{R1}v_S+2\lambda_R^2v_S^2v_{R2}+2\lambda_Rv_{R1}(-\lambda_3v_1v_2+\lambda_Rv_{R1}v_{R2}
+\lambda_Sv_S^2+2\xi_F)),  \\
\frac{\partial V_{0}}{\partial H_{S}^0}=&\frac{1}{2}(-\sqrt{2}T_3v_1v_2+\sqrt{2}T_Rv_{R1}v_{R2}+2m_S^2v_S+3\sqrt{2}T_Sv_{S}^2
+\lambda_3^2v_1^2v_S+\lambda_3^2v_2^2v_S \nonumber \\
&+\lambda_R^2v_{R1}^2v_S +\lambda_R^2v_{R2}^2v_S-6\lambda_S\lambda_3v_1v_2v_S+6\lambda_R\lambda_Sv_{R1}v_{R2}v_S
+18\lambda_Sv_S^3+12\lambda_S\xi_Fv_S).
\end{align}
The $2\times2$ mass squared matrix for these fields in the basis $(\text{Re}[H_{11}^0],\text{Re}[H_{22}^0])$
\begin{eqnarray}
	m_{H_1^0}^2=\begin{pmatrix}
m^2_{H_{11}^0H_{11}^0} & m^2_{H_{11}^0H_{22}^0} \\
m^2_{H_{22}^0H_{11}^0} & m^2_{H_{22}^0H_{22}^0} \\
\end{pmatrix}.
\end{eqnarray}
The matrix elements are defined as
\begin{align}
m^2_{H_{11}^0 H_{11}^0} &= m_1^2 - \frac{g_L^2}{4} \left(\frac{v_1^2}{2} - \frac{v_2^2}{2}\right) - \frac{g_R^2}{4} \left(\frac{v_1^2}{2} - \frac{v_2^2}{2} + v_{R1}^2 - v_{R2}^2\right) + \frac{v_{S}^2 \lambda_3^2}{2} + 2\lambda_4^2v_1^2+\frac{\lambda_4^2v_S^2}{2},  \nonumber \\
m^2_{H_{11}^0 H_{22}^0} &=m_{H_{22}^0 H_{11}^0}^2= -T_3 \frac{v_S}{\sqrt{2}} + 2\lambda_4 \lambda_5 v_1 v_2 - \lambda_3 \left(-\frac{\lambda_3 v_1 v_2}{2} + \frac{\lambda_R v_{R1} v_{R2} }{2} + 3 \frac{\lambda_S v_{S}^2}{2}+8\xi_F \right),    \nonumber \\
m^2_{H_{22}^0 H_{22}^0} &= m_1^2 + \frac{g_L^2}{4} \left(\frac{v_1^2}{2} - \frac{v_2^2}{2}\right) + \frac{g_R^2}{4} \left(\frac{v_1^2}{2} + \frac{v_2^2}{2} + v_{R1}^2 - v_{R2}^2\right) + \frac{v_{S}^2 \lambda_3^2}{2} + 2\lambda_5^2 v_2^2 + \frac{v_S^2 \lambda_5^2}{2}.
\end{align}
The $2\times2$ mass squared matrix for these fields in the basis  $(\text{Re}[\Delta_{L}^0],\text{Re}[\delta_{L}^0])$
\begin{eqnarray}
	m_{H_2^0}^2=\begin{pmatrix}
m^2_{\Delta_{L}^0\Delta_{L}^0} & m^2_{\Delta_{L}^0\delta_{L}^0} \\
m^2_{\delta_{L}^0\Delta_{L}^0} & m^2_{\delta_{L}^0\delta_{L}^0} \\
\end{pmatrix}.
\end{eqnarray}
The matrix elements are defined as
\begin{align}
&m^2_{\Delta_{L}^0\Delta_{L}^0} = m_{\delta_L}^2 + \frac{1}{2}g_L^2 \left(\frac{v_1^2}{2} - \frac{v_2^2}{2}\right) - g_B^2 \left(\frac{v_{R1}^2}{2} + \frac{v_{R2}^2}{2}\right) + \frac{\lambda_5^2 v_S^2}{2}, \nonumber \\
&m^2_{\Delta_{L}^0\delta_{L}^0} =m^2_{\delta_{L}^0\Delta_{L}^0}= \frac{T_L v_S}{\sqrt{2}} + \lambda_L \left(-\frac{1}{2} \lambda_3 v_1 v_2 +\frac{ \lambda_5}{2} v_{R1} v_{R2} + \frac{3 \lambda_5 v_S^2}{2} + \xi_F \right), \nonumber \\
&m^2_{\delta_{L}^0\delta_{L}^0} = m_{\delta_L}^2 - \frac{1}{2}g_L^2 \left(\frac{v_1^2}{2} - \frac{v_2^2}{2}\right) + g_B^2 \left(\frac{v_{R2}^2}{2} - \frac{v_{R1}^2}{2}\right) + \frac{\lambda_L^2 v_S^2}{2}.
\end{align}
Similarly, in the LRSSM, the imaginary parts of the neutral Higgs fields generate pseudo-scalar (CP-odd) states. Their mass-squared matrix is provided in the gauge basis $(\text{Im}[H_{12}^0],\text{Im}[H_{21}^0],\text{Im}[H_{\Delta_{R}}^0],\text{Im}[H_{\delta_{R}}^0],\text{Im}[H_{S}^0])$
\begin{eqnarray}
	m_{A_1
}^2=\begin{pmatrix}
		m^2_{\sigma_{12}^0\sigma_{12}^0} & m^2_{\sigma_{12}^0\sigma_{21}^0} &-\frac{1}{2} \lambda_3 \lambda_R v_2 v_{R2} & -\frac{1}{2} \lambda_3 \lambda_R v_2 v_{R1} & m^2_{\sigma_{12}^0\sigma_{S}^0} \\
		m^2_{\sigma_{21}^0\sigma_{12}^0} & m^2_{\sigma_{21}^0\sigma_{21}^0} & -\frac{1}{2} \lambda_3 \lambda_R v_1 v_{R2} & -\frac{1}{2} \lambda_3 \lambda_R v_1 v_{R1} & m_{\sigma_{21}^0\sigma_{S}^0} \\
-\frac{1}{2} \lambda_3 \lambda_R v_2 v_{R2}& -\frac{1}{2} \lambda_3 \lambda_R v_1 v_{R2} & m^2_{\sigma_{\Delta_{R}}^0\sigma_{\Delta_{R}}^0} & m^2_{\sigma_{\Delta_{R}}^0\sigma_{\delta_{R}}^0} & m^2_{\sigma_{\Delta_{R}}^0\sigma_{S}^0} \\
-\frac{1}{2} \lambda_3 \lambda_R v_2 v_{R1}& -\frac{1}{2} \lambda_3 \lambda_R v_1 v_{R1} & m^2_{\sigma_{\delta_{R}}^0\sigma_{\Delta_{R}}^0} & m^2_{\sigma_{\delta_{R}}^0\sigma_{\delta_{R}}^0} & m^2_{\sigma_{\delta_{R}}^0\sigma_{S}^0} \\
m^2_{\sigma_{S}^0\sigma_{12}^0} & m^2_{\sigma_{S}^0\sigma_{21}^0} & m^2_{\sigma_{S}^0\sigma_{\Delta_{R}}^0} & m^2_{\sigma_{S}^0\sigma_{\delta_{R}}^0} & m^2_{\sigma_{S}^0\sigma_{S}^0} \\
	\end{pmatrix}.
\end{eqnarray}
The matrix elements are defined as
\begin{align}
&m^2_{\sigma_{12}^0\sigma_{21}^0}=m^2_{\sigma_{21}^0\sigma_{12}^0}=\frac{1}{2} \left(\sqrt{2} T_3 v_s + \lambda_3 \left(2 \xi_F + \lambda_R v_{R1} v_{R2} + 3 \lambda_s v_s^2\right)\right), \nonumber \\
&m^2_{\sigma_{12}^0\sigma_{12}^0}=\frac{1}{8}(8m_1^2+g_R^2v_1^2-g_Rv_{2}^2+g_L^2(v_1^2-v_2^2)+2g_R^2v_{R1}^2-
2g_R^2v_{R2}^2+4\lambda_3^2v_2^2+4\lambda_3^2v_S^2+4\lambda_4v_S^2), \nonumber \\
&m^2_{\sigma_{21}^0\sigma_{21}^0}=\frac{1}{8} \left(g_L^2 \left(v_2^2 - v_1^2\right) - g_R^2 v_1^2 + g_R^2 v_2^2 - 2 g_R^2 v_{R1}^2 + 2 g_R^2 v_{R2}^2 + 8 m_2^2 + 4 \lambda_3^2 v_1^2 + 4 \lambda_3^2 v_s^2 + 4 \lambda_5^2 v_s^2\right), \nonumber \\
&m^2_{\sigma_{\Delta_{R}}^0\sigma_{\Delta_{R}}^0}=\frac{1}{4} \left(2 g_B^2 \left(v_{R1}^2 - v_{R2}^2\right) + g_R^2 \left(v_1^2 - v_2^2 + 2 v_{R1}^2 - 2 v_{R2}^2\right) + 4 m_{\Delta_R}^2 + 2 \lambda_R^2 \left(v_{R2}^2 + v_s^2\right)\right), \nonumber \\
&m^2_{\sigma_{\delta_{R}}^0\sigma_{\delta_{R}}^0}=\frac{1}{4} \left(2 g_B^2 \left(v_{R2}^2 - v_{R1}^2\right) + g_R^2 \left(-v_1^2 + v_2^2 - 2 v_{R1}^2 + 2 v_{R2}^2\right) + 4 m_{\delta_R}^2 + 2 \lambda_R^2 \left(v_{R1}^2 + v_s^2\right)\right), \nonumber \\
&m^2_{\sigma_{\Delta_{R}}^0\sigma_{\delta_{R}}^0}=m_{\sigma_{\delta_{R}}^0\sigma_{\Delta_{R}}^0}=\frac{1}{2} \left(\lambda_R \left(-2 \xi_F + \lambda_3 v_1 v_2 - 3 \lambda_s v_s^2\right) - \sqrt{2} T_R v_s\right), \nonumber \\
&m^2_{\sigma_{S}^0\sigma_{S}^0}=\frac{1}{2} (-12 \lambda_s \xi_F + 2 m^2_s - 6 \sqrt{2} T_S v_s + \lambda_3^2 v_1^2 + \lambda_4^2 v_1^2 + 6 \lambda_3 \lambda_s v_1 v_2 + \lambda_3^2 v_2^2 + \lambda_5^2 v_2^2 + \lambda_R^2 v_{R1}^2 \nonumber \\
&\ \ \quad\quad\quad - 6 \lambda_R \lambda_s v_{R1} v_{R2} + \lambda_R^2 v_{R2}^2 + 18 \lambda_s^2 v_s^2), \nonumber \\
&m^2_{\sigma_{S}^0\sigma_{12}^0}=m^2_{\sigma_{12}^0\sigma_{S}^0}=\frac{T_3 v_2}{\sqrt{2}} - 3 \lambda_3 \lambda_s v_2 v_s, \quad m^2_{\sigma_{S}^0\sigma_{21}^0}=m^2_{\sigma_{21}^0\sigma_{S}^0}=\frac{T_3 v_1}{\sqrt{2}} - 3 \lambda_3 \lambda_s v_1 v_s,  \nonumber \\
&m^2_{\sigma_{S}^0\sigma_{\Delta_R}^0}=m^2_{\sigma_{\Delta_R}^0\sigma_{S}^0}=3 \lambda_R \lambda_s v_{R2} v_s - \frac{T_R v_{R2}}{\sqrt{2}}, \quad m^2_{\sigma_{S}^0\sigma_{\delta_R}^0}=m^2_{\sigma_{\delta_R}^0\sigma_{S}^0}=3 \lambda_R \lambda_s v_{R1} v_s - \frac{T_R v_{R1}}{\sqrt{2}}
\end{align}
By diagonalizing the matrix into physical mass eigenstates using the rotation matrix \( Z^A \), three physical particles with definite masses, \( A_1 \), \( A_2 \), and \( A_3 \), are obtained. Meanwhile, the remaining two neutral states, \( G_1^0 \) and \( G_2^0 \), become massless Goldstone bosons, which are absorbed by the corresponding gauge bosons \( Z \) and \( Z_R \), giving them mass. The form of the transpose matrix is given below.
\begin{align}
Z^A=\begin{pmatrix}
c_{\beta} & -s_{\beta} & 0 & 0 & 0 \\
s_{\beta} & c_{\beta} & 0 & 0 & 0 \\
0 & 0 & c_{\beta_R} & -s_{\beta_R}  & 0 \\
0 & 0 & s_{\beta_R} & c_{\beta_R}  & 0 \\
0 & 0 & 0 & 0  & 1
\end{pmatrix},
\end{align}
where $c_{\beta}=\frac{v_1}{\sqrt{v_1^2+v_2^2}}$, $s_{\beta}=\frac{v_2}{\sqrt{v_1^2+v_2^2}}$, $c_{\beta_R}=\frac{v_{R1}}{\sqrt{v_{R1}^2+v_{R2}^2}}$ and $s_{\beta_R}=\frac{v_{R2}}{\sqrt{v_{R1}^2+v_{R2}^2}}$. Similarly, in the LRSSM, there are also two
$2\times2$ CP-odd Higgs boson mass matrices, one of which has a basis of $(\text{Im}[H_{11}^0],\text{Im}[H_{22}^0])$
\begin{align}
m^2_{A_2}=\begin{pmatrix}
m^2_{\sigma_{11}^0\sigma_{11}^0} & m^2_{\sigma_{11}^0\sigma_{22}^0} \\
m^2_{\sigma_{22}^0\sigma_{11}^0} & m^2_{\sigma_{22}^0\sigma_{22}^0}
\end{pmatrix}.
\end{align}
The various forms of the matrix elements are respectively
\begin{align}
m^2_{\sigma_{11}^0\sigma_{22}^0}&=m^2_{\sigma_{22}^0\sigma_{11}^0}=\frac{T_3 v_s}{\sqrt{2}} + 2 \lambda_4 \lambda_5 v_1 v_2 + \lambda_3 \left(\xi_F - \frac{\lambda_3 v_1 v_2}{2} + \frac{\lambda_R v_{R1} v_{R2}}{2} + \frac{3 \lambda_s v_s^2}{2}\right), \nonumber \\
m^2_{\sigma_{11}^0\sigma_{11}^0}&=-\frac{1}{4} g_L^2 \left(\frac{v_1^2}{2} - \frac{v_2^2}{2}\right) - \frac{1}{4} g_R^2 \left(\frac{v_1^2}{2} - \frac{v_2^2}{2} + v_{R1}^2 - v_{R2}^2\right) + m_1^2 + 2 \lambda_4^2 v_1^2 + \frac{\lambda_3^2 v_s^2}{2} + \frac{\lambda_4^2 v_s^2}{2}, \nonumber \\
m^2_{\sigma_{22}^0\sigma_{22}^0}&=\frac{1}{4} g_L^2 \left(\frac{v_1^2}{2} - \frac{v_2^2}{2}\right) + \frac{1}{4} g_R^2 \left(\frac{v_1^2}{2} - \frac{v_2^2}{2} + v_{R1}^2 - v_{R2}^2\right) + m_2^2 + 2 \lambda_5^2 v_2^2 + \frac{\lambda_3^2 v_s^2}{2} + \frac{\lambda_5^2 v_s^2}{2}.
\end{align}
The basis of the other $2\times2$ CP-odd Higgs boson mass matrix is  $(\text{Im}[\Delta_{L}^0],\text{Im}[\delta_{L}^0])$
\begin{align}
m^2_{A_3}=\begin{pmatrix}
m^2_{\sigma_{\Delta_L}^0\sigma_{\Delta_L}^0} & m^2_{\sigma_{\Delta_L}^0\sigma_{\delta_L}^0} \\
m^2_{\sigma_{\delta_L}^0\sigma_{\Delta_L}^0} & m^2_{\sigma_{\delta_L}^0\sigma_{\delta_L}^0}
\end{pmatrix}.
\end{align}
The matrix elements are defined as
\begin{align}
&m^2_{\sigma_{\Delta_L}^0\sigma_{\Delta_L}^0}=-g_B^2 \left(\frac{v_{R2}^2}{2} - \frac{v_{R1}^2}{2}\right) + \frac{1}{2} g_L^2 \left(\frac{v_1^2}{2} - \frac{v_2^2}{2}\right) + m_{\Delta_L}^2 + \frac{\lambda_L^2 v_s^2}{2}, \nonumber \\
&m^2_{\sigma_{\delta_L}^0\sigma_{\delta_L}^0}=g_B^2 \left(\frac{v_{R2}^2}{2} - \frac{v_{R1}^2}{2}\right) - \frac{1}{2} g_L^2 \left(\frac{v_1^2}{2} - \frac{v_2^2}{2}\right) + m_{\delta_L}^2 + \frac{\lambda_L^2 v_s^2}{2}, \nonumber \\
&m^2_{\sigma_{\Delta_L}^0\sigma_{\delta_L}^0}=m_{\sigma_{\delta_L}^0\sigma_{\Delta_L}^0}=-\frac{T_L v_s}{\sqrt{2}} - \lambda_L \left(\xi_F - \frac{\lambda_3 v_1 v_2}{2} + \frac{\lambda_R v_{R1} v_{R2}}{2} + \frac{3 \lambda_s v_s^2}{2}\right).
\end{align}
The mass squared matrix for the singly charged Higgs bosons in the basis $(\Phi_{12}^{\pm}, \Phi_{21}^{\pm}, \Delta_R^{\pm}, \delta_R^{\pm})$ is given as follows:
\begin{align}
m_{H^{\pm}_1}^2 = \left(
\begin{array}{cccc}
m^2_{\phi_{12}^+ \phi_{12}^-} & m^2_{\phi_{12}^+ \phi_{21}^-} & m^2_{\phi_{12}^+ \Delta_R^-} & m^2_{\phi_{12}^+ \delta_R^-} \\
m^2_{\phi_{21}^+ \phi_{12}^-} & m^2_{\phi_{21}^+ \phi_{21}^-} & m^2_{\phi_{21}^+ \Delta_R^-} & m^2_{\phi_{21}^+ \delta_R^-} \\
m^2_{\Delta_R^+ \phi_{12}^-} & m^2_{\Delta_R^+ \phi_{21}^-} & m^2_{\Delta_R^+ \Delta_R^-} & m^2_{\Delta_R^+ \delta_R^-} \\
m^2_{\delta_R^+ \phi_{12}^-} & m^2_{\delta_R^+ \phi_{21}^-} & m^2_{\delta_R^+ \Delta_R^-} & m^2_{\delta_R^+ \delta_R^-} \\
\end{array}
\right).
\end{align}
The matrix elements are defined as
\begin{align}
&m^2_{\phi_{12}^+ \phi_{12}^-} = \frac{1}{8} \left(g_L^2 \left(v_1^2 + v_2^2\right) + g_R^2 v_1^2 + g_R^2 v_2^2 - 2 g_R^2 v_{R1}^2 + 2 g_R^2 v_{R2}^2 + 8 m_1^2 + 4 \lambda_3^2 v_s^2 + 4 \lambda_4^2 v_s^2\right), \nonumber \\
&m^2_{\phi_{12}^+ \phi_{21}^-} = m_{\phi_{21}^+ \phi_{12}^-} = \frac{1}{4} \left(g_L^2 v_1 v_2 + g_R^2 v_1 v_2 + 4 \lambda_3 \xi_F + 2 \sqrt{2} T_3 v_s - 2 \lambda_3^2 v_1 v_2 + 2 \lambda_3 \lambda_R v_{R1} v_{R2} + 6 \lambda_3 \lambda_s v_s^2\right), \nonumber \\
&m^2_{\phi_{21}^+ \phi_{21}^-} = \frac{1}{8} \left(g_L^2 \left(v_1^2 + v_2^2\right) + g_R^2 v_1^2 + g_R^2 v_2^2 + 2 g_R^2 v_{R1}^2 - 2 g_R^2 v_{R2}^2 + 8 m_2^2 + 4 \lambda_3^2 v_s^2 + 4 \lambda_5^2 v_s^2\right), \nonumber \\
&m^2_{\phi_{21}^+ \Delta_R^-} = m_{\Delta_R^+ \phi_{21}^-} = \frac{g_R^2 v_2 v_{R2}}{2 \sqrt{2}},\ \ m_{\phi_{12}^+ \Delta_R^-} = m_{\Delta_R^+ \phi_{12}^-} = -\frac{g_R^2 v_2 v_{R1}}{2 \sqrt{2}}, \nonumber \\
&m^2_{\phi_{21}^+ \delta_R^-} = m_{\delta_R^+ \phi_{21}^-} = \frac{g_R^2 v_1 v_{R2}}{2 \sqrt{2}},\ \ m_{\phi_{12}^+ \delta_R^-} = m_{\delta_R^+ \phi_{12}^-} = -\frac{g_R^2 v_1 v_{R1}}{2 \sqrt{2}}, \nonumber \\
&m^2_{\Delta_R^+ \Delta_R^-} = \frac{1}{2} \left(g_B^2 \left(v_{R1}^2 - v_{R2}^2\right) + g_R^2 v_{R1}^2 + 2 m_{\Delta_R}^2 + \lambda_R^2 v_s^2\right), \nonumber \\
&m^2_{\Delta_R^+ \delta_R^-} =m_{\delta_R^+ \Delta_R^-}= \frac{1}{2} \left(-g_R^2 v_{R1} v_{R2} + \sqrt{2} T_R v_s + \lambda_R \left(2 \xi_F - \lambda_3 v_1 v_2 + \lambda_R v_{R1} v_{R2} + 3 \lambda_s v_s^2\right)\right), \nonumber \\
&m^2_{\delta_R^+ \delta_R^-} =  \frac{1}{2} \left(g_B^2 \left(v_{R2}^2 - v_{R1}^2\right) + g_R^2 v_{R2}^2 + 2 m_{\delta_R}^2 + \lambda_R^2 v_s^2\right).
\end{align}
The basis of the other $2\times2$ charged Higgs boson mass matrix is $(\Phi_{11}^{\pm},\Phi_{22}^{\pm})$.
\begin{align}
m_{H^{\pm}_2}^2 = \begin{pmatrix}
m^2_{\phi_{11}^{+}\phi_{11}^{-}} & m^2_{\phi_{11}^{+}\phi_{22}^{-}} \\
m^2_{\phi_{22}^{+}\phi_{11}^{-}} & m^2_{\phi_{22}^{+}\phi_{22}^{-}} \\
\end{pmatrix}.
\end{align}
The matrix elements are defined as
\begin{align}
&m^2_{\phi_{11}^{+}\phi_{11}^{-}}= \frac{1}{8} \left(g_L^2 \left(v_1^2 - v_2^2\right) + g_R^2 v_1^2 - g_R^2 v_2^2 + 2 g_R^2 v_{R1}^2 - 2 g_R^2 v_{R2}^2 + 8 m_1^2 + 4 \lambda_3^2 v_s^2 + 4 \lambda_4^2 v_s^2\right), \nonumber \\
&m^2_{\phi_{22}^{+}\phi_{22}^{-}}=\frac{1}{8} \left(g_L^2 \left(v_2^2 - v_1^2\right) - g_R^2 v_1^2 + g_R^2 v_2^2 - 2 g_R^2 v_{R1}^2 + 2 g_R^2 v_{R2}^2 + 8 m_2^2 + 4 \lambda_3^2 v_s^2 + 4 \lambda_5^2 v_s^2\right),  \nonumber \\
&m^2_{\phi_{22}^{+}\phi_{11}^{-}}=m^2_{\phi_{11}^{+}\phi_{22}^{-}}=\frac{1}{2} \left(\sqrt{2} T_3 v_s + \lambda_3 \left(2 \xi_F - \lambda_3 v_1 v_2 + \lambda_R v_{R1} v_{R2} + 3 \lambda_s v_s^2\right)\right).
\end{align}
The basis of the other $2\times2$ charged Higgs boson mass matrix is $(\Delta_{L}^{\pm},\delta_{L}^{\pm})$
\begin{align}
m_{H^{\pm}_3}^2 = \begin{pmatrix}
m^2_{\Delta_{L}^{+}\Delta_{L}^{-}} & m^2_{\Delta_{L}^{+}\delta_{L}^{-}} \\
m^2_{\delta_{L}^{+}\Delta_{L}^{-}} & m^2_{\delta_{L}^{+}\delta_{L}^{-}} \\
\end{pmatrix}.
\end{align}
The various forms of the matrix elements are respectively
\begin{align}
&m^2_{\Delta_{L}^{+}\Delta_{L}^{-}}=\frac{1}{2} \left(g_B^2 \left(v_{R1}^2 - v_{R2}^2\right) + 2 m_{\Delta_L}^2 + \lambda_L^2 v_s^2\right), \nonumber \\
&m^2_{\delta_{L}^{+}\delta_{L}^{-}}=\frac{1}{2} \left(g_B^2 \left(v_{R2}^2 - v_{R1}^2\right) + 2 m_{\delta_L}^2 + \lambda_L^2 v_s^2\right),  \nonumber \\
&m^2_{\Delta_{L}^{+}\delta_{L}^{-}}=m^2_{\delta_{L}^{+}\Delta_{L}^{-}}=\frac{1}{2} \left(\sqrt{2} T_L v_s + \lambda_L \left(2 \xi_F - \lambda_3 v_1 v_2 + \lambda_R v_{R1} v_{R2} + 3 \lambda_s v_s^2\right)\right).
\end{align}
In the LRSSM, the mass matrices $m_{H^{\pm}_1}^2$ and $m_{H^{\pm}_2}^2$ each contain a charged Goldstone boson, which, after electroweak symmetry breaking, is absorbed into the corresponding gauge bosons $W^{\pm}$ and $W_R^{\pm}$, imparting mass to these bosons. The mass matrix of the doubly charged Higgs bosons in the LRSSM is relatively simple, consisting of only two $2\times2$ matrices. The basis of the doubly charged Higgs boson mass matrix is $(\Delta_{R}^{\pm\pm},\delta_{R}^{\pm\pm})$ and $(\Delta_{L}^{\pm\pm},\delta_{L}^{\pm\pm})$. We denote them as $m_{H^{\pm\pm}_1}^2$ and $m_{H^{\pm\pm}_2}^2$, and their matrix forms are given below
\begin{align}
m_{H^{\pm\pm}_1}^2=\begin{pmatrix}
m^2_{\Delta_R^{++}\Delta_R^{--}} & m^2_{\Delta_R^{++}\delta_R^{--}} \\
m^2_{\delta_R^{++}\Delta_R^{--}} & m^2_{\delta_R^{++}\delta_R^{--}} \\
\end{pmatrix},
\end{align}
\begin{align}
&m^2_{\Delta_R^{++}\Delta_R^{--}}=\frac{1}{4} \left(2 g_B^2 \left(v_{R1}^2 - v_{R2}^2\right) + g_R^2 \left(-v_1^2 + v_2^2 - 2 v_{R1}^2 + 2 v_{R2}^2\right) + 4 m_{\Delta_R}^2 + 2 \lambda_R^2 v_s^2\right), \nonumber \\
&m^2_{\delta_R^{++}\delta_R^{--}}=\frac{1}{4} \left(-2 g_B^2 \left(v_{R1}^2 - v_{R2}^2\right) + g_R^2 \left(v_1^2 - v_2^2 + 2 v_{R1}^2 - 2 v_{R2}^2\right) + 4 m_{\delta_R}^2 + 2 \lambda_R^2 v_s^2\right), \nonumber \\
&m^2_{\Delta_R^{++}\delta_R^{--}}=m^2_{\delta_R^{++}\Delta_R^{--}}=\frac{1}{2} \left(\sqrt{2} T_R v_s + \lambda_R \left(2 \xi_F - \lambda_3 v_1 v_2 + \lambda_R v_{R1} v_{R2} + 3 \lambda_s v_s^2\right)\right),
\end{align}
\begin{align}
m_{H^{\pm\pm}_2}^2=\begin{pmatrix}
m^2_{\Delta_L^{++}\Delta_L^{--}} & m^2_{\Delta_L^{++}\delta_L^{--}} \\
m^2_{\delta_L^{++}\Delta_L^{--}} & m^2_{\delta_L^{++}\delta_L^{--}} \\
\end{pmatrix},
\end{align}
\begin{align}
&m^2_{\Delta_{L}^{++}\Delta_{L}^{--}}=\frac{1}{4} \left(2 g_B^2 \left(v_{R1}^2 - v_{R2}^2\right) + g_L^2 \left(v_2^2 - v_1^2\right) + 4 m_{\Delta_L}^2 + 2 \lambda_L^2 v_s^2\right), \nonumber \\
&m^2_{\delta_{L}^{++}\delta_{L}^{--}}=\frac{1}{4} \left(-2 g_B^2 v_{R1}^2 + 2 g_B^2 v_{R2}^2 + g_L^2 \left(v_1^2 - v_2^2\right) + 4 m_{\delta_L}^2 + 2 \lambda_L^2 v_s^2\right), \nonumber \\
&m^2_{\Delta_{L}^{++}\delta_{L}^{--}}=m^2_{\delta_{L}^{++}\Delta_{L}^{--}}=\frac{1}{2} \left(\sqrt{2} T_L v_s + \lambda_L \left(2 \xi_F - \lambda_3 v_1 v_2 + \lambda_R v_{R1} v_{R2} + 3 \lambda_s v_s^2\right)\right).
\end{align}
The mass-squared matrix $m_{H^{\pm\pm}_1}^2$ of this doubly charged Higgs boson, when diagonalized, produces a negative eigenvalue, which is physically unreasonable as it would result in a tachyonic state. However, by incorporating radiative corrections into the mass of the doubly charged Higgs boson, this issue can be resolved, turning the eigenvalue positive \cite{Babu:2014vba,Babu:2008ep}.
\section{Sfermion sectors, Neutralinos and charginos\label{app2}}
\subsection{Sfermion sectors}
In the LRSSM, fermions include scalar superpartners of up-type and down-type quarks, as well as charged and neutral leptons. Due to the extended gauge symmetry in this model, the existence of right-handed neutrinos and their superpartners is ensured, making all fermion mass-squared matrices (including the sneutrinos) typically $6\times6$ matrices. We calculated the mass-squared matrices for scalar down-type squarks, up-type squarks, charged sleptons, and sneutrinos, which are represented in the gauge basis of $(\tilde{d}_L, \tilde{d}_R)\), \((\tilde{u}_L, \tilde{u}_R)\), \((\tilde{e}_L, \tilde{e}_R)$ and $(\tilde{\nu}_L, \tilde{\nu}_R)$. Therefore, we can express the mass-squared matrix of the squarks in the following form
\begin{equation}
\mathbf{m}^2_{\tilde{d}} =
\begin{pmatrix}
m^2_{\tilde{d}_L \tilde{d}^*_L} & m^2_{\tilde{d}_L \tilde{d}^*_R} \\
m^2_{\tilde{d}_R \tilde{d}^*_L} & m^2_{\tilde{d}_R \tilde{d}^*_R}
\end{pmatrix},
\qquad
\mathbf{m}^2_{\tilde{u}} =
\begin{pmatrix}
m^2_{\tilde{u}_L \tilde{u}^*_L} & m^2_{\tilde{u}_L \tilde{u}^*_R} \\
m^2_{\tilde{u}_R \tilde{u}^*_L} & m^2_{\tilde{u}_R \tilde{u}^*_R}
\end{pmatrix}.
\end{equation}
Each matrix element itself is a $3\times3$ matrix.
\begin{align}
&m^2_{\tilde{d}_L \tilde{d}^*_L}=\frac{1}{3} g_{B}^2 \left(\frac{v_{R2}^2}{2}-\frac{v_{R1}^2}{2}\right)+\left(\frac{1}{4} g_{L}^2 \left(\frac{v_{2}^2}{2}-\frac{v_{1}^2}{2}\right)+m^2_{Q_L}+\frac{v_{1}^2 (Y_{Q}^1)^2}{2}\right)\delta_{\alpha_1\beta_1}, \nonumber \\
&m^2_{\tilde{d}_R \tilde{d}^*_R}=-\frac{1}{3} g_{B}^2 \left(\frac{v_{R2}^2}{2}-\frac{v_{R1}^2}{2}\right)+\left(\frac{1}{4} g_{R}^2 \left(-\frac{v_{1}^2}{2}+\frac{v_{2}^2}{2}+v_{R1}^2-v_{R2}^2\right)+m^2_{Q_R}+\frac{v_{1}^2
(Y^1_{Q})^2}{2}\right)\delta_{\alpha_2\beta_2}, \nonumber \\
&m^2_{\tilde{d}_L \tilde{d}^*_R}=m^2_{\tilde{d}_R \tilde{d}^*_L}=\left( -\frac{T^1_{Q} v_{1}}{\sqrt{2}}-\frac{1}{2} \lambda_{3} v_{2} vs Y_{Q}^1\right)\delta_{\alpha_1\beta_2}, \nonumber \\
&m^2_{\tilde{u}_L \tilde{u}^*_L}=\frac{1}{3} g_B^2 \left( -\frac{v_{R1}^2}{2} + \frac{v_{R2}^2}{2} \right)+\left(m^2_{Q_L} + \frac{1}{4} g_L^2 \left( \frac{v_1^2}{2} - \frac{v_2^2}{2} \right)  + \frac{v_2^2 (Y^2_{Q})^2}{2}\right)\delta_{\alpha_1\beta_1}, \nonumber \\
&m^2_{\tilde{u}_R \tilde{u}^*_R}= - \frac{1}{3} g_B^2 \left( -\frac{v_{R1}^2}{2} + \frac{v_{R2}^2}{2} \right)+\left(m^2_{Q_R} - \frac{1}{4} g_R^2 \left( -\frac{v_1^2}{2} + \frac{v_2^2}{2} + v_{R1}^2 - v_{R2}^2 \right)+ \frac{v_2^2 (Y^2_{Q})^2}{2}\right)\delta_{\alpha_2\beta_2}, \nonumber \\
&m^2_{\tilde{u}_L \tilde{u}^*_R}=m^2_{\tilde{u}_R \tilde{u}^*_L}=\left(\frac{T^2_{Q} v_2}{\sqrt{2}} + \frac{1}{2} v_1 v_s Y^2_{Q} \lambda_3\right)\delta_{\alpha_1\beta_2}.
\end{align}
Here $\alpha_1, \alpha_2, \beta_1$ and $\beta_2$ denote color indices. In general, these $3 \times 3$ matrices can be non-diagonal, allowing off-diagonal elements to enable mixing between different flavors. In our study, we do not consider flavor-violating processes; therefore, for simplicity, we assume that the matrices are diagonal and their elements are real. Likewise, the mass-squared matrix of the slepton is expressed as follows:
\begin{align}
\mathbf{m}^2_{\tilde{e}} =
\begin{pmatrix}
m^2_{\tilde{e}_L \tilde{e}^*_L} & m^2_{\tilde{e}_L \tilde{e}^*_R} \\
m^2_{\tilde{e}_R \tilde{e}^*_L} & m^2_{\tilde{e}_R \tilde{e}^*_R}
\end{pmatrix},
\qquad
\mathbf{m}^2_{\tilde{\nu}} =
\begin{pmatrix}
m^2_{\tilde{\nu}_L \tilde{\nu}^*_L} & m^2_{\tilde{\nu}_L \tilde{\nu}^*_R} \\
m^2_{\tilde{\nu}_R \tilde{\nu}^*_L} & m^2_{\tilde{\nu}_R \tilde{\nu}^*_R}
\end{pmatrix}.
\end{align}
with the matrix elements in the slepton sector being
\begin{align}
&m^2_{\tilde{e}_L \tilde{e}^*_L}=\frac{g_B^2}{2} (v_{R2}^2 - v_{R1}^2) + \frac{g_L^2}{2} (v_1^2 - v_2^2) + (Y_L^1)^2 v_2^2 + m^2_{L_L}, \nonumber \\
&m^2_{\tilde{e}_R \tilde{e}^*_R}=\frac{g_B^2}{2} (v_{R1}^2 - v_{R2}^2) + \frac{g_R^2}{4} (2v_{R2}^2 - 2v_{R1}^2 - v_2^2 + v_1^2) + v_2^2 (Y_L^1)^2 + m^2_{L_R}, \nonumber \\
&m^2_{\tilde{e}_L \tilde{e}^*_R}=m^2_{\tilde{e}_R \tilde{e}^*_L}=Y_{L}^1 \lambda_3 v_s v_1 - v_2 T_{L}^1, \nonumber \\
&m^2_{\tilde{\nu}_L \tilde{\nu}^*_L} = \frac{g_B^2}{2} (v_{R2}^2 - v_{R1}^2) + \frac{g_L^2}{2} (v_2^2 - v_1^2) + (Y_L^1)^2 v_1^2 + m^2_{L_L}, \nonumber \\
&m^2_{\tilde{\nu}_R \tilde{\nu}^*_R} = v_{R2}^2 h_{RR}^2 + \left( \frac{g_B^2}{2} (v_{R1}^2 - v_{R2}^2) + \frac{g_R^2}{4} (2v_{R1}^2 - 2v_{R2}^2 - v_1^2 + v_1^2) + v_1^2 (Y_L^1)^2 + m^2_{L_R} \right), \nonumber \\
&m^2_{\tilde{\nu}_L \tilde{\nu}^*_R} = m^2_{\tilde{\nu}_R \tilde{\nu}^*_L} = -Y_{L}^1 \lambda_3 v_s v_1 + v_1 T_{L}^1.
\end{align}
\subsection{Neutralinos and charginos}
This model contains twelve neutral fermions, with their mass matrix decomposed into three independent blocks: two $2 \times 2$ blocks that describe the mixing of the $\tilde{\delta}_L / \tilde{\Delta}_L$ and $\tilde{\Phi}_{22}^0 / \tilde{\Phi}_{11}^0$ fields, respectively, and an $8 \times 8$ block involving the mixing of the other eight neutral Higgsinos and gauginos. These two $2 \times 2$ blocks are represented in the bases of $(\tilde{\delta}_L, \tilde{\Delta}_L^0)$ and $(\tilde{\Phi}_{22}^0, \tilde{\Phi}_{11}^0)$
\begin{align}
M_{\tilde{\chi}_{\delta}} =
\begin{pmatrix}
0 & \mu_L \\
\mu_L & 0
\end{pmatrix}
\quad \text{and} \quad
M_{\tilde{\chi}_{\Phi}} =
\begin{pmatrix}
0 & -\mu_{\text{eff}} \\
-\mu_{\text{eff}} & 0
\end{pmatrix},
\end{align}
while the last $8 \times 8$ block reads, in the basis of $(\tilde{\Phi}_{12}^0, \tilde{\Phi}_{21}^0, \tilde{\delta}_R^0, \tilde{\Delta}_R^0, \tilde{S}, \tilde{B}, \tilde{W}_L^0, \tilde{W}_R^0)$
\begin{align}
M_{\chi^0} =
\begin{pmatrix}
0 & -\mu_{\text{eff}} & 0 & 0 & -\mu_2 & 0 & \frac{g_L v_1}{\sqrt{2}} & -\frac{g_R v_1}{\sqrt{2}} \\
-\mu_{\text{eff}} & 0 & 0 & 0 & -\mu_1 & -\frac{g_L v_2}{\sqrt{2}} & \frac{g_R v_2}{\sqrt{2}} & 0 \\
0 & 0 & 0 & \mu_R & \frac{\lambda_R v_{R1}}{\sqrt{2}} & g_R v_{R1} & 0 & -g_R v_{R1} \\
0 & 0 & \mu_R & 0 & \frac{\lambda_R v_{R2}}{\sqrt{2}} & -g_R v_{R2} & 0 & -g_R v_{R2} \\
-\mu_2 & -\mu_1 & \frac{\lambda_R v_{R1}}{\sqrt{2}} & \frac{\lambda_R v_{R2}}{\sqrt{2}} & \mu_S & 0 & 0 & 0 \\
0 & -\frac{g_L v_2}{\sqrt{2}} & g_R v_{R1} & -g_R v_{R2} & 0 & M_1 & 0 & 0 \\
\frac{g_L v_1}{\sqrt{2}} & \frac{g_L v_2}{\sqrt{2}} & 0 & 0 & 0 & 0 & M_{2L} & 0 \\
-\frac{g_R v_1}{\sqrt{2}} & \frac{g_R v_2}{\sqrt{2}} & -g_R v_{R1} & -g_R v_{R2} & 0 & 0 & 0 & M_{2R}
\end{pmatrix}.
\end{align}

We have defined $\mu_{\text{eff}} = \lambda_3 \frac{v_s}{\sqrt{2}},\mu_S = \lambda_S \frac{v_s}{\sqrt{2}},\mu_{L,R} = \lambda_{L,R} \frac{v_s}{\sqrt{2}}$ and $\mu_{1,2} = \lambda_3 \frac{v_{1,2}}{\sqrt{2}}$. In the charged sector, the model has six singly-charged
charginos whose mass matrix can be written, in the bases of $(\tilde{\delta}_L^+, \tilde{\delta}_R^+, \tilde{\Phi}_{11}^+,\tilde{\Phi}_{21}^+, \tilde{W}_L^+, \tilde{W}_R^+) $ and $ (\tilde{\Delta}_L^-, \tilde{\Delta}_R^-, \tilde{\Phi}_{12}^-, \tilde{\Phi}_{22}^-, \tilde{W}_L^-, \tilde{W}_R^-)$
\begin{align}
M_{\tilde{\chi}^\pm} =
\begin{pmatrix}
\lambda_L v_s / \sqrt{2} & 0 & 0 & 0 & 0  & 0 \\
0 & \lambda_R v_s / \sqrt{2} & 0 & 0 & 0  & -g_R v_{R1} \\
0 & 0 & 0 & \mu_{\text{eff}} & g_L v_1 / \sqrt{2} & 0 \\
0 & 0 & \mu_{\text{eff}} & 0 & 0  & -g_R v_2 / \sqrt{2} \\
0 & 0 & 0 & g_L v_2 / \sqrt{2} & M_{2L}  & 0 \\
0 & g_R v_{R2} & -g_R v_1 / \sqrt{2} & 0 & 0  & M_{2R}
\end{pmatrix}.
\end{align}
\section{The expressions for Higgs decays in the LRSSM\label{app3}}
The concrete expressions that are present in the LRSSM are specifically discussed in the following:

1. CP-even Higgs-W boson-W boson contribution
\begin{align}
g_{h_i W_{L\mu}^+ W_{L\nu}^-} &= \frac{i}{2} g_{\mu \nu} g_L^2 \left( v_1 Z^H_{i1} + v_2 Z^H_{i2} \right), \nonumber \\
g_{h_i W_{R\mu}^+ W_{R\nu}^-} &= \frac{i}{2} g_{\mu \nu} g_R^2 \left( v_1 Z^H_{i1} + v_2 Z^H_{i2} + 2 v_{R1} Z^H_{i3} + 2 v_{R2} Z^H_{i3} \right).
\end{align}

2. CP-even Higgs-Z boson-Z boson contribution
\begin{align}
g_{h_i Z_{L\mu}Z_{L\nu}} &=\frac{i}{2}g_{\mu \nu} \frac{g_L^2}{4\cos\theta_W}(v_1 Z^H_{i1} + v_2 Z^H_{i2}), \nonumber \\
g_{h_i Z_{R\mu}Z_{R\nu}} &=\frac{i}{2}g_{\mu \nu}(g_R^2+g_B^2)\left( 4\cos\theta_R(v_1 Z^H_{i1} + v_2 Z^H_{i2})+4 v_{R1} Z^H_{i3} + 4 v_{R2} Z^H_{i3} \right).
\end{align}

3. CP-even Higgs-charged Higgs-charged Higgs contribution
\begin{align}
g_{h_ih^+_jh^-_k}=&\frac{1}{8}( g_{L}^2 v_1 Z^H_{i1} + g_{R}^2 v_1 Z^H_{i1} + g_{L}^2 v_2 Z^H_{i2} + g_{R}^2 v_2 Z^H_{i2} - 2 g_{R}^2 v_{R1} Z^H_{i3} + 2 g_{R}^2 v_{R2} Z^H_{i4} \nonumber \\
&+ 4 v_{S} Z^H_{i5} \lambda_{3}^2+ 4 v_{S} Z^H_{i5} \lambda_{4}^2 ) Z^{+}_{j1} Z^{+}_{k1}+\frac{1}{8}( g_{L}^2 v_1 Z^H_{i1} + g_{R}^2 v_1 Z^H_{i1} + g_{L}^2 v_2 Z^H_{i2} \nonumber \\
&+ g_{R}^2 v_2 Z^H_{i2} + 2 g_{R}^2 v_{R1} Z^H_{i3} - 2 g_{R}^2 v_{R2} Z^H_{i4} + 4 v_{S} Z^H_{i5} \lambda_{3}^2 + 4 v_{S} Z^H_{i5} \lambda_{4}^2 ) Z^{+}_{j2} Z^{+}_{k2} \nonumber \\
&+\frac{1}{2} \left( g_{R}^2 v_{R1} Z^H_{i3} + g_{B}^2 \left( v_{R1} Z^H_{i3} - v_{R2} Z^H_{i4} \right) + v_{S} Z^H_{i5} \lambda_{R}^2 \right) Z^{+}_{j3} Z^{+}_{k3}+\frac{1}{2} ( g_{R}^2 v_{R2} Z^H_{i4} \nonumber \\
&+ g_{B}^2 ( - v_{R1} Z^H_{i3} + v_{R2} Z^H_{i4} ) + v_{S} Z^H_{i5} \lambda_{R}^2 ) Z^{+}_{j4} Z^{+}_{k4} +\frac{1}{4} ( g_{L}^2 (v_1 Z^H_{i2}+ v_2 Z^H_{i1}) \nonumber \\
& + g_{R}^2 (v_1 Z^H_{i2} + v_2 Z^H_{i1}) + 2 \sqrt{2} T_3 Z^H_{i5} - 2 (v_1 Z^H_{i2} + v_2 Z^H_{i1}) \lambda_{3}^2 + 2 (v_{R1} Z^H_{i4} + v_{R2} Z^H_{i3}) \lambda_{3} \lambda_{R} \nonumber \\
& + 6 v_S Z^H_{i5} \lambda_{3} \lambda_{S} ) Z^{+}_{j1} Z^{+}_{k2}-\frac{1}{2 \sqrt{2}}g_{R}^2 (v_1 Z^H_{i3} + v_{R1} Z^H_{i1}) Z^{+}_{j1} Z^{+}_{k3} \nonumber \\
&-\frac{1}{2 \sqrt{2}}g_{R}^2 (v_1 Z^H_{i4} + v_{R2} Z^H_{i1}) Z^{+}_{j1} Z^{+}_{k4}-\frac{1}{2 \sqrt{2}}g_{R}^2 (v_2 Z^H_{i3} + v_{R1} Z^H_{i2}) Z^{+}_{j2} Z^{+}_{k3} \nonumber \\
&-\frac{1}{2 \sqrt{2}}g_{R}^2 (v_2 Z^H_{i4} + v_{R2} Z^H_{i2}) Z^{+}_{j2} Z^{+}_{k4}+\frac{1}{2} \Big( -g_{R}^2 (v_{R1} Z^H_{i4} + v_{R2} Z^H_{i3}) + \sqrt{2} \, T_R Z^H_{i5}  \nonumber \\
&+ \lambda_{R} \big( -(v_1 Z^H_{i2} + v_2 Z^H_{i1}) \lambda_{3} + (v_{R1} Z^H_{i4} + v_{R2} Z^H_{i3}) \lambda_{R} + 3 v_S Z^H_{i5} \lambda_{S} \big) \Big) Z^{+}_{j3} Z^{+}_{k4}
\end{align}

4. CP-even Higgs-squark-up squark contribution
\begin{align}
g_{h_i\widetilde{u}_j\widetilde{u}_k}=&-\frac{i}{12}\Big( \sum_{a=1}^{3} Z_{j3+a}^{U} Z_{k3+a}^{U}
\Big( 2 \Big( -3 g_{R}^2 + g_{B}^2 \Big) (v_{1R} Z_{i3}^{H} - v_{2R} Z_{i4}^{H})
+ 3 g_{R}^2 v_1 Z_{i1}^{H} \nonumber \\
&- 3 g_{R}^2 v_2 Z_{i2}^{H} \Big)+ \sum_{a=1}^{3} Z_{ja}^{U} Z_{ka}^{U} \Big( 2 g_{B}^2 (-v_{1R} Z_{i3}^{H} + v_{2R} Z_{i4}^{H})
+ 3 g_{L}^2 v_1 Z_{i1}^{H} - 3 g_{L}^2 v_2 Z_{i2}^{H} \Big) \nonumber \\
&- 6 \Big( 2 \sqrt{2} m_{\Phi_1} \sum_{b=1}^{3} Z_{kb}^{U} \sum_{a=1}^{3} Y_{Q1,ab}^* Z_{j3+a}^{U} Z_{i1}^{H}
+ 2 \sqrt{2} \mu_4^* \sum_{b=1}^{3} Z_{jb}^{U} \sum_{a=1}^{3} Z_{k3+a}^{U} Y_{Q1,ab} Z_{i1}^{H} \Big) \nonumber \\
&+ \lambda_3 v_S \sum_{b=1}^{3} Z_{jb}^{U} \sum_{a=1}^{3} Z_{k3+a}^{U} Y_{Q2,ab} Z_{i1}^{H} - \sqrt{2} \sum_{b=1}^{3} Z_{jb}^{U}
\sum_{c=1}^{3} Z_{kc}^{U} \sum_{a=1}^{3} Z_{j3+a}^{U} T_{Q2,ac} Z_{i2}^{H}  \nonumber \\
&- 2 v_2 \sum_{b=1}^{3} Z_{jb}^{U} \sum_{c=1}^{3} Z_{kc}^{U} \sum_{a=1}^{3} Z_{j3+a}^{U} Y_{Q2,ac}^* Z_{i2}^{H} + \lambda_3 v_S \sum_{b=1}^{3}
Z_{jb}^{U} \sum_{a=1}^{3} Z_{k3+a}^{U} Y_{Q2,ab} Z_{i5}^{H} \Big) Z_{j3}^{D} Z_{k3}^{D}.
\end{align}

5. CP-even Higgs-down squark-down squark contribution
\begin{align}
g_{h_i\widetilde{d}_j\widetilde{d}_k}=&\frac{i}{12}  \Big( \sum_{a=1}^{3} Z_{ja}^{D} Z_{ka}^{D}
\Big( 2 g_{B}^2 (v_{R1} Z_{i3}^{H} - v_{R2} Z_{i4}^{H}) + 3 g_{L}^2 v_1 Z_{i1}^{H} - 3 g_{L}^2 v_2 Z_{i2}^{H} \Big) \nonumber \\
&+ \sum_{a=1}^{3} Z_{j3+a}^{D} Z_{k3+a}^{D} \Big( -2 \Big( 3 g_{R}^2 + g_{B}^2 \Big) (v_{R1} Z_{i3}^{H} - v_{R2} Z_{i4}^{H})
+ 3 g_{R}^2 v_1 Z_{i1}^{H} - 3 g_{R}^2 v_2 Z_{i2}^{H} \Big) \nonumber \\
&- 6 \Big( \sqrt{2} \sum_{b=1}^{3} Z_{kb}^{D} \sum_{a=1}^{3} T_{Q1,ab}^* Z_{j3+a}^{D} Z_{i1}^{H}
+ \sqrt{2} \sum_{b=1}^{3} Z_{jb}^{D} \sum_{a=1}^{3} Z_{k3+a}^{D} T_{Q1,ab} Z_{i1}^{H} \Big) \nonumber \\
&+ 2 v_1 \sum_{c=1}^{3} Z_{kc}^{D} \sum_{b=1}^{3} Z_{jb}^{D} \sum_{a=1}^{3} Y_{Q1,ac}^* Z_{j3+a}^{D} Z_{i1}^{H}
+ 2 v_1 \sum_{c=1}^{3} Z_{jc}^{D} \sum_{b=1}^{3} Z_{k3+b}^{D} \sum_{a=1}^{3} Y_{Q1,ca} Y_{Q1,ba} Z_{i1}^{H} \nonumber \\
&- \lambda_3 v_S \sum_{b=1}^{3} Z_{jb}^{D} \sum_{a=1}^{3} Z_{k3+a}^{D} Y_{Q1,ab} Z_{i2}^{H}
- 2 \sqrt{2} m_{\Phi_2} \sum_{b=1}^{3} Z_{kb}^{D} \sum_{a=1}^{3} Y_{Q2,ab} Z_{j3+a}^{D} Z_{i2}^{H} \nonumber \\
&- \lambda_3 v_S \sum_{b=1}^{3} Z_{jb}^{D} \sum_{a=1}^{3} Z_{k3+a}^{D} Y_{Q2,ab} Z_{i5}^{H}
- 2 v_2 \sum_{b=1}^{3} Z_{jb}^{D} \sum_{a=1}^{3} Z_{k3+a}^{D} Y_{Q1,ab} Z_{i5}^{H} \Big).
\end{align}

6. CP-even Higgs-slepton-slepton contribution
\begin{align}
g_{h_i\widetilde{l}_j\widetilde{l}_k}=&\frac{i}{4} \Big( \sum_{a=1}^{3} Z_{j3+a}^{E} Z_{k3+a}^{E} \Big( 2 \Big( - g_{R}^2 + g_{B}^2 \Big)
(v_{R1} Z_{i3}^{H} - v_{R2} Z_{i4}^{H}) + g_{R}^2 v_1 Z_{i1}^{H} - g_{R}^2 v_2 Z_{i2}^{H} \Big) \nonumber \\
&+ \sum_{a=1}^{3} Z_{ja}^{E} Z_{ka}^{E} \Big( 2 g_{B}^2 (- v_{R1} Z_{i3}^{H} + v_{R2} Z_{i4}^{H})
+ g_{L}^2 v_1 Z_{i1}^{H} - g_{L}^2 v_2 Z_{i2}^{H} \Big) \nonumber \\
&+ 2 \Big( -\sqrt{2} \sum_{b=1}^{3} Z_{kb}^{E} \sum_{a=1}^{3} T_{L1,ab}^* Z_{j3+a}^{E} Z_{i1}^{H}
- \sqrt{2} \sum_{b=1}^{3} Z_{jb}^{E} \sum_{a=1}^{3} Z_{k3+a}^{E} T_{L1,ab} Z_{i1}^{H} \Big) \nonumber \\
&- 2 v_1 \sum_{c=1}^{3} Z_{kc}^{E} \sum_{b=1}^{3} Z_{jb}^{E} \sum_{a=1}^{3} Y_{L1,ac}^* Z_{j3+a}^{E} Z_{i1}^{H}
- 2 v_1 \sum_{c=1}^{3} Z_{jc}^{E} \sum_{b=1}^{3} Z_{k3+b}^{E} \sum_{a=1}^{3} Y_{L1,ca} Y_{L1,ba} Z_{i1}^{H} \nonumber \\
&+ \lambda_3 v_S \sum_{b=1}^{3} Z_{jb}^{E} \sum_{a=1}^{3} Z_{k3+a}^{E} Y_{L1,ab} Z_{i2}^{H}
+ \lambda_3 v_S \sum_{b=1}^{3} Z_{jb}^{E} \sum_{a=1}^{3} Z_{k3+a}^{E} Y_{L1,ab} Z_{i2}^{H} \nonumber \\
&+ 2 \sqrt{2} m_{\Phi_2} \sum_{b=1}^{3} Z_{kb}^{E} \sum_{a=1}^{3} Y_{L2,ab} Z_{j3+a}^{E} Z_{i2}^{H}
+ 2 \sqrt{2} m_{\Phi_2}^* \sum_{b=1}^{3} Z_{jb}^{E} \sum_{a=1}^{3} Z_{k3+a}^{E} Y_{L2,ab} Z_{i2}^{H} \nonumber \\
&+ 3 \lambda_3 v_2 \sum_{b=1}^{3} Z_{jb}^{E} \sum_{a=1}^{3} Z_{k3+a}^{E} Y_{L1,ab} Z_{i5}^{H} \Big).
\end{align}

7. CP-even Higgs-double charged Higgs-double charged Higgs contribution
\begin{align}
g_{h_iH_{R,j}^{++}H_{R,k}^{--}}=&-\frac{i}{2} \Big( U_{j2}^{--} \Big( U_{k2}^{--} \Big( -g_{R}^2 v_1 Z_{i1}^{H} + g_{R}^2 v_2 Z_{i2}^{H}
+ 2 g_{B}^2 v_{R1} Z_{i3}^{H} + 2 g_{R}^2 v_{R1} Z_{i3}^{H} + 2 g_{B}^2 v_{R2} Z_{i4}^{H} \nonumber \\
&- 2 g_{R}^2 v_{R2} Z_{i4}^{H} \Big)
+ 2 \lambda_{R}^2 v_S Z_{i5}^{H} \Big)
+ U_{k1}^{--} \Big( - \lambda_3 \lambda_{R} v_2 Z_{i1}^{H} - \lambda_3 \lambda_{R} v_1 Z_{i2}^{H}
+ \lambda_{R}^2 v_{R2} Z_{i3}^{H}  \nonumber \\
&+ \lambda_{R}^2 v_{R1} Z_{i4}^{H} + 2 \lambda_{R} v_S Z_{i5}^{H}
+ \sqrt{2} T_{R} Z_{i5}^{H} \Big) + U_{j1}^{--} \Big( g_{R}^2 v_1 Z_{i1}^{H} - g_{R}^2 v_2 Z_{i2}^{H} + 2 g_{B}^2 v_{R1} Z_{i3}^{H} \nonumber \\
&- 2 g_{R}^2 v_{R1} Z_{i3}^{H}- 2 g_{R}^2 v_{R2} Z_{i4}^{H} + 2 g_{R}^2 v_{R2} Z_{i4}^{H}
+ 2 \lambda_{R}^2 v_S Z_{i5}^{H} \Big)+ U_{k2}^{--} \Big( - \lambda_3 \lambda_{R} v_2 Z_{i1}^{H} \nonumber \\
&- \lambda_3 \lambda_{R} v_1 Z_{i2}^{H}
+ \lambda_{R}^2 v_{R2} Z_{i3}^{H} + \lambda_{R}^2 v_{R1} Z_{i4}^{H} + 2 \lambda_{R} v_S Z_{i5}^{H}
+ \sqrt{2} T_{R} Z_{i5}^{H} \Big) \Big),  \\
g_{h_iH_{L,j}^{++}H_{L,k}^{--}}=&- \frac{i}{2} \Big( U^{--}_{L,j2} \Big( U^{--}_{L,k2} \Big( -2g^2_{B} v_{R1} Z^H_{k3} + 2g^2_{B} v_{R2} Z^H_{k4} + 2\lambda^2_{L} v_{S} Z^H_{k5} + g^2_{L} v_{1} Z^H_{k1} - g^2_{L} v_{2} Z^H_{k2} \Big) \nonumber \\
&+ U^{--}_{L,k1} \Big( -\lambda_3 \lambda_{L} v_{2} Z^H_{k1} - \lambda_3 \lambda_{L} v_{1} Z^H_{k2} + \lambda_{L} \lambda_{R} v_{R2} Z^H_{k3} + \lambda_{L} \lambda_{R} v_{R1} Z^H_{k4} + 2 \lambda_{L} \lambda_{S} v_{S} Z^H_{k5} \Big) \nonumber \\
&+ \sqrt{2} T_{L} Z^H_{k5} \Big) \Big)+ U^{--}_{L,j1} \Big( U^{--}_{L,k1} \Big( 2g^2_{B} v_{R1} Z^H_{k3} - 2g^2_{B} v_{R2} Z^H_{k4} + 2\lambda^2_{L} v_{S} Z^H_{k5} - g^2_{L} v_{1} Z^H_{k1} \nonumber \\
&+ g^2_{L} v_{2} Z^H_{k2} \Big) + U^{--}_{L,k2} \Big( -\lambda_3 \lambda_{L} v_{2} Z^H_{k1} - \lambda_3 \lambda_{L} v_{1} Z^H_{k2} + \lambda_{L} \lambda_{R} v_{R2} Z^H_{k3} + \lambda_{L} \lambda_{R} v_{R1} Z^H_{k4} \nonumber \\
&+ 2 \lambda_{L} \lambda_{S} v_{S} Z^H_{k5} \Big)+ \sqrt{2} T_{L} Z^H_{k5} \Big) \Big).
\end{align}

8. CP-even Higgs chargino chargino contribution
\begin{align}
g_{h_i\chi_j^+\chi_k^-}=&- \frac{i}{2} \Big( \sqrt{2} g_L U^+_{k_1} U^{-,*}_{j_4} Z^H_{i_1} + \sqrt{2} g_L U^+_{k_3} U^{-,*}_{j_1} Z^H_{i_2} - \sqrt{2} g_R U^+_{k_2} U^{-,*}_{j_3} Z^H_{i_2} \Big) \nonumber \\
&- 2 g_R U^+_{k_2} U^{-,*}_{j_6} Z^H_{i_3} + 2 g_R U^+_{k_4} U^{-,*}_{j_2} Z^H_{i_4} + \sqrt{2} \lambda_L U^+_{k_5} U^{-,*}_{j_3} Z^H_{i_5} \nonumber \\
&+ \sqrt{2} \lambda_R U^+_{k_6} U^{-,*}_{j_6} Z^H_{i_5} + \sqrt{2} U^+_{k_4} \Big( - g_R U^{-,*}_{j_2} Z^H_{i_1} + \lambda_3 U^{-,*}_{j_3} Z^H_{i_5} \Big) \Big( \frac{1 - \gamma_5}{2} \Big) \nonumber \\
&+ \frac{- i}{2} \Big( \sqrt{2} g_L U^+_{j_1} U^-_{k_4} Z^H_{i_1} + \sqrt{2} g_L U^+_{j_3} U^-_{k_1} Z^H_{i_2} - \sqrt{2} g_R U^+_{j_2} U^-_{k_2} Z^H_{i_2} \Big) \nonumber \\
&+ 2 g_R U^+_{j_6} U^-_{k_2} Z^H_{i_4} + \sqrt{2} \lambda_3 U^+_{j_4} U^-_{k_3} Z^H_{i_5} + \sqrt{2} \lambda_R U^+_{j_6} U^-_{k_5} Z^H_{i_5} \nonumber \\
&+ \sqrt{2} U^+_{j_4} \Big( - g_R U^-_{k_2} Z^H_{i_2} + \lambda_3 U^-_{k_3} Z^H_{i_5} \Big) \Big( \frac{1 + \gamma_5}{2} \Big).
\end{align}

9. CP-even-Higgs-neutralino-neutralino contribution
\begin{align}
g_{h_i\chi_j^0\chi_k^0}=&- \frac{i}{2} \Big( g_L Z^{0,*}_{k_1} Z^{0,*}_{j_6} Z^H_{i_1} - g_R Z^{0,*}_{k_2} Z^{0,*}_{j_6} Z^H_{i_1} - \sqrt{2} \lambda_3 Z^{0,*}_{k_5} Z^{0,*}_{j_12} Z^H_{i_1} \Big) \nonumber \\
&- g_L Z^{0,*}_{k_5} Z^{0,*}_{j_1} Z^H_{i_2} + g_R Z^{0,*}_{k_2} Z^{0,*}_{j_7} Z^H_{i_2} - g_L Z^{0,*}_{k_11} Z^{0,*}_{j_5} Z^H_{i_2} + g_R Z^{0,*}_{k_6} Z^{0,*}_{j_5} Z^H_{i_2} \nonumber \\
&+ 2g_R Z^{0,*}_{k_1} Z^{0,*}_{j_2} Z^H_{i_3} - 2g_B Z^{0,*}_{k_3} Z^{0,*}_{j_4} Z^H_{i_3} + 2g_R Z^{0,*}_{k_5} Z^{0,*}_{j_2} Z^H_{i_4} - 2g_B Z^{0,*}_{k_3} Z^{0,*}_{j_9} Z^H_{i_4} \nonumber \\
&+ \sqrt{2} \lambda_R Z^{0,*}_{k_9} Z^{0,*}_{j_8} Z^H_{i_3} - 2g_R Z^{0,*}_{k_10} Z^{0,*}_{j_6} Z^H_{i_4} + 2g_B Z^{0,*}_{k_11} Z^{0,*}_{j_9} Z^H_{i_4} - 2g_R Z^{0,*}_{k_12} Z^{0,*}_{j_10} Z^H_{i_5} \nonumber \\
&+ 2g_B Z^{0,*}_{k_13} Z^{0,*}_{j_6} Z^H_{i_5} + \sqrt{2} \lambda_3 Z^{0,*}_{k_4} Z^{0,*}_{j_2} Z^H_{i_4} - \sqrt{2} \lambda_3 Z^{0,*}_{k_3} Z^{0,*}_{j_4} Z^H_{i_4} + \sqrt{2} \lambda_R Z^{0,*}_{k_8} Z^{0,*}_{j_5} Z^H_{i_5} \nonumber \\
&- \sqrt{2} \lambda_3 Z^{0,*}_{k_7} Z^{0,*}_{j_6} Z^H_{i_4} + \sqrt{2} \lambda_R Z^{0,*}_{k_12} Z^{0,*}_{j_10} Z^H_{i_5} + \sqrt{2} \lambda_R Z^{0,*}_{k_13} Z^{0,*}_{j_10} Z^H_{i_6} + \sqrt{2} \lambda_3 Z^{0,*}_{k_4} Z^{0,*}_{j_3} Z^H_{i_6} \nonumber \\
&+ \sqrt{2} \lambda_R Z^{0,*}_{k_8} Z^{0,*}_{j_9} Z^H_{i_6} + \sqrt{2} Z^{0,*}_{k_9} Z^{0,*}_{j_8} Z^H_{i_3} + \sqrt{2} \lambda_3 Z^{0,*}_{k_7} Z^{0,*}_{j_6} Z^H_{i_4} \Big( \frac{1 - \gamma_5}{2} \Big) \nonumber \\
&+ \frac{-i}{2} \Big( g_L Z^{0}_{k_1} Z^{0}_{j_6} Z^H_{i_1} - g_R Z^{0}_{k_2} Z^{0}_{j_6} Z^H_{i_1} - \sqrt{2} \lambda_3 Z^{0}_{k_5} Z^{0}_{j_{12}} Z^H_{i_1} - g_L Z^{0}_{k_5} Z^{0}_{j_1} Z^H_{i_2} \Big) \nonumber \\
&+ g_R Z^{0}_{k_5} Z^{0}_{j_2} Z^H_{i_2} - g_L Z^{0}_{k_1} Z^{0}_{j_5} Z^H_{i_2} + g_R Z^{0}_{k_{11}} Z^{0}_{j_2} Z^H_{i_3} + 2g_R Z^{0}_{k_1} Z^{0}_{j_2} Z^H_{i_3} \nonumber \\
&- 2g_B Z^{0}_{k_3} Z^{0}_{j_4} Z^H_{i_3} + 2g_R Z^{0}_{k_5} Z^{0}_{j_2} Z^H_{i_4} - 2g_B Z^{0}_{k_3} Z^{0}_{j_9} Z^H_{i_4} + \sqrt{2} \lambda_R Z^{0}_{k_9} Z^{0}_{j_{12}} Z^H_{i_3} \nonumber \\
&- 2g_R Z^{0}_{k_9} Z^{0}_{j_2} Z^H_{i_4} + 2g_B Z^{0}_{k_3} Z^{0}_{j_9} Z^H_{i_4} - 2g_R Z^{0}_{k_{12}} Z^{0}_{j_{10}} Z^H_{i_5} + 2g_B Z^{0}_{k_{13}} Z^{0}_{j_6} Z^H_{i_5} \nonumber \\
&+ \sqrt{2} \lambda_R Z^{0}_{k_{11}} Z^{0}_{j_{12}} Z^H_{i_4} - \sqrt{2} \lambda_3 Z^{0}_{k_7} Z^{0}_{j_4} Z^H_{i_5} - \sqrt{2} \lambda_3 Z^{0}_{k_5} Z^{0}_{j_6} Z^H_{i_5} + \sqrt{2} \lambda_L Z^{0}_{k_{10}} Z^{0}_{j_8} Z^H_{i_5} \nonumber \\
&+ \sqrt{2} \lambda_R Z^{0}_{k_{11}} Z^{0}_{j_9} Z^H_{i_5} + \sqrt{2} \lambda_L Z^{0}_{k_{10}} Z^{0}_{j_{11}} Z^H_{i_5} + \sqrt{2} \lambda_R Z^{0}_{k_{11}} Z^{0}_{j_9} Z^H_{i_5} \nonumber \\
&+ \sqrt{2} Z^{0}_{k_{12}} ( 2\lambda_S Z^{0}_{j_{12}} Z^H_{i_5} - \lambda_3 Z^{0}_{j_5} Z^H_{i_2} - \lambda_3 Z^{0}_{j_6} Z^H_{i_4} + \lambda_R Z^{0}_{j_{11}} Z^H_{i_4} + \lambda_R Z^{0}_{j_9} Z^H_{i_3} ) \nonumber \\
&+ Z^{0}_{k_6} ( g_L Z^{0}_{j_1} Z^H_{i_1} - g_R Z^{0}_{j_2} Z^H_{i_1} - \sqrt{2} \lambda_3 ( Z^{0}_{j_{12}} Z^H_{i_2} + Z^{0}_{j_5} Z^H_{i_5} ) ) \Big( \frac{1 + \gamma_5}{2} \Big).
\end{align}

10. CP-even Higgs-up quark-up quark contribution
\begin{align}
g_{h_i\bar{u}u}=&- i \frac{1}{\sqrt{2}}  \sum_{b=1}^{3} U^{u,*}_{L,jb} \sum_{a=1}^{3} U^{u,*}_{R,ka} Y_{Q2,ab} Z^H_{i_2} \Big( \frac{1 - \gamma_5}{2} \Big) \nonumber \\
&- i \frac{1}{\sqrt{2}}  \sum_{b=1}^{3} \sum_{a=1}^{3} Y^{*}_{Q2,ab} U^{u}_{R,ja} U^{u}_{L,kb} Z^H_{i_2} \Big( \frac{1 + \gamma_5}{2} \Big).
\end{align}

11. CP-even Higgs-down quark-down quark contribution
\begin{align}
g_{h_i\bar{d}d}=&- i \frac{1}{\sqrt{2}}  \sum_{b=1}^{3} \sum_{a=1}^{3} U^{d,*}_{L,jb} U^{d,*}_{R,ka} Y_{Q1,ab} Z^H_{i_1} \Big( \frac{1 - \gamma_5}{2} \Big) \nonumber \\
& - i \frac{1}{\sqrt{2}}  \sum_{b=1}^{3} \sum_{a=1}^{3} Y^{*}_{Q1,ab} U^{d}_{R,ja} U^{d}_{L,kb} Z^H_{i_1} \Big( \frac{1 + \gamma_5}{2} \Big).
\end{align}

12. CP-even Higgs-lepton-lepton contribution
\begin{align}
g_{h_i\bar{l}l}=&- i \frac{1}{\sqrt{2}}  \sum_{b=1}^{3} \sum_{a=1}^{3} U^{e,*}_{L,jb} U^{e,*}_{R,ka} Y_{L1,ab} Z^H_{i_1} \Big( \frac{1 - \gamma_5}{2} \Big) \nonumber \\
&+ - i \frac{1}{\sqrt{2}}  \sum_{b=1}^{3} \sum_{a=1}^{3} Y^{*}_{L1,ab} U^{e}_{R,ja} U^{e}_{L,kb} Z^H_{i_1} \Big( \frac{1 + \gamma_5}{2} \Big).
\end{align}
\section{The radiative corrections to the CP-even neutral
 Higgs mass matrix\label{app4}}
In the LRSSM, the forms of the mass squared matrices for the stop quark and sbottom quark are, respectively, as follows
\begin{align}
m^2_{\tilde{t}} =
\begin{pmatrix}
m^2_{\tilde{t}_L} & m^2_{\tilde{t}_{12}} \\
m^2_{\tilde{t}_{21}} & m^2_{\tilde{t}_R}
\end{pmatrix}, \quad
m^2_{\tilde{b}} =
\begin{pmatrix}
m^2_{\tilde{b}_L} & m^2_{\tilde{b}_{12}} \\
m^2_{\tilde{b}_{21}} & m^2_{\tilde{b}_R}
\end{pmatrix},
\end{align}
where
\begin{align}
&m^2_{\tilde{b}_L}=\frac{1}{3} g_{B}^2 \left(\frac{v_{R2}^2}{2}-\frac{v_{R1}^2}{2}\right)+\left(\frac{1}{4} g_{L}^2 \left(\frac{v_{2}^2}{2}-\frac{v_{1}^2}{2}\right)+m^2_{Q_L}+\frac{v_{1}^2 (Y_{Q}^1)^2}{2}\right), \nonumber \\
&m^2_{\tilde{b}_R }=-\frac{1}{3} g_{B}^2 \left(\frac{v_{R2}^2}{2}-\frac{v_{R1}^2}{2}\right)+\left(\frac{1}{4} g_{R}^2 \left(-\frac{v_{1}^2}{2}+\frac{v_{2}^2}{2}+v_{R1}^2-v_{R2}^2\right)+m^2_{Q_R}+\frac{v_{1}^2
(Y^1_{Q})^2}{2}\right), \nonumber \\
&m^2_{\tilde{b}_{12}}=m^2_{\tilde{b}_{21}}=\left( -\frac{T^1_{Q} v_{1}}{\sqrt{2}}-\frac{1}{2} \lambda_{3} v_{2} vs Y_{Q}^1\right), \nonumber \\
&m^2_{\tilde{t}_L}=\frac{1}{3} g_B^2 \left( -\frac{v_{R1}^2}{2} + \frac{v_{R2}^2}{2} \right)+\left(m^2_{Q_L} + \frac{1}{4} g_L^2 \left( \frac{v_1^2}{2} - \frac{v_2^2}{2} \right)  + \frac{v_2^2 (Y^2_{Q})^2}{2}\right), \nonumber \\
&m^2_{\tilde{t}_R }= - \frac{1}{3} g_B^2 \left( -\frac{v_{R1}^2}{2} + \frac{v_{R2}^2}{2} \right)+\left(m^2_{Q_R} - \frac{1}{4} g_R^2 \left( -\frac{v_1^2}{2} + \frac{v_2^2}{2} + v_{R1}^2 - v_{R2}^2 \right)+ \frac{v_2^2 (Y^2_{Q})^2}{2}\right), \nonumber \\
&m^2_{\tilde{t}_{12}}=m^2_{\tilde{t}_{21}}=\left(\frac{T^2_{Q} v_2}{\sqrt{2}} + \frac{1}{2} v_1 v_s Y^2_{Q} \lambda_3\right).
\end{align}
Then, the mass eigenvalues of stop and sbottom quarks will be given by
\begin{align}
m_{\tilde{t}_{1,2}}^2 &= \frac{1}{2} \left( m_{\tilde{t}_L}^2 + m_{\tilde{t}_R}^2 \right)
\pm \frac{1}{2} \left[ \left( m_{\tilde{t}_L}^2 - m_{\tilde{t}_R}^2 \right)^2
+ 4 m_{\tilde{t}_{12}}^4 \right]^{\frac{1}{2}}, \\
m_{\tilde{b}_{1,2}}^2 &= \frac{1}{2} \left( m_{\tilde{b}_L}^2 + m_{\tilde{b}_R}^2 \right)
\pm \frac{1}{2} \left[ \left( m_{\tilde{b}_L}^2 - m_{\tilde{b}_R}^2 \right)^2
+ 4 m_{\tilde{b}_{12}}^4\right]^{\frac{1}{2}}.
\end{align}
Therefore, the radiative corrections $\Delta \Pi^1$ from the one-loop zero temperature effective potential will be deduced as
\begin{align}
\Delta \Pi^1_{ii}&=[-\frac{1}{\phi_i}\frac{\partial V_1}{\partial\phi_i}+\frac{\partial^2 V_1}{\partial\phi_i^2}]_{\phi_i=v_i}, \\ \nonumber
\Delta \Pi^1_{ij}&=[\frac{\partial^2 V_1}{\partial\phi_i\partial\phi_j}]_{\phi_i=v_i},
\end{align}

where
\begin{align}
\Delta\Pi^{(1)}_{ij}
&=\sum_{\ell}\frac{n_{\ell}}{32\pi^{2}}\!\left[
\ln\frac{m_{\ell}^{2}}{Q^{2}}\,
\frac{\partial m_{\ell}^{2}}{\partial \phi_{i}}\,
\frac{\partial m_{\ell}^{2}}{\partial \phi_{j}}
+m_{\ell}^{2}\!\left(\ln\frac{m_{\ell}^{2}}{Q^{2}}-1\right)
\frac{\partial^{2} m_{\ell}^{2}}{\partial \phi_{i}\,\partial \phi_{j}}
\right]
\qquad (i\neq j) \\ \nonumber
\Delta\Pi^{(1)}_{ii}
&=\sum_{\ell}\frac{n_{\ell}}{32\pi^{2}}\!\left[
\ln\frac{m_{\ell}^{2}}{Q^{2}}
\left(\frac{\partial m_{\ell}^{2}}{\partial \phi_{i}}\right)^{2}
+m_{\ell}^{2}\!\left(\ln\frac{m_{\ell}^{2}}{Q^{2}}-1\right)
\frac{\partial^{2} m_{\ell}^{2}}{\partial \phi_{i}^{2}}
\right] \\ \nonumber
&-\frac{1}{v_{i}}\sum_{\ell}\frac{n_{\ell}}{32\pi^{2}}
\left[
m_{\ell}^{2}\!\left(\ln\frac{m_{\ell}^{2}}{Q^{2}}-1\right)
\frac{\partial m_{\ell}^{2}}{\partial \phi_{i}}
\right]
\,.
\end{align}
Here, $i=1,2,R1,R2,S$ and $\ell=t,b,\tau,\nu_R,\tilde{u_i},\tilde{d_i},\tilde{e_i},\tilde{\nu_i},W_R,W_L,Z_R,Z_L,\chi^0_i,H_{i}^{\pm},\chi_{i}^0$, Q is the new physics scale and we take $Q=m_{W}$ in the following numerical discussion. In the one-loop effective potential, the numerically dominant contributions come from
the top/stop and bottom/sbottom sectors, together with the heavy gauge and right---handed neutrino states---namely $V_t$, $V_{\tilde t_i}$ $(i=1,2)$, $V_b$, $V_{\tilde b_i}$, $V_{W_R}$, $V_{Z_R}$, and $V_{\nu_R}$. All remaining terms are subleading. Below we present the explicit expressions for the derivatives of the field-dependent masses with respect to $\phi_i$ that enter these dominant contributions.

The two-loop effective potential corrections to the CP-even neutral Higgs mass matrix is given by
\begin{align}
\Delta \Pi^2_{ii}&=[-\frac{1}{\phi_i}\frac{\partial V_2}{\partial\phi_i}+\frac{\partial^2 V_2}{\partial\phi_i^2}]_{\phi_i=v_i}, \\ \nonumber
\Delta \Pi^2_{ij}&=[\frac{\partial^2 V_2}{\partial\phi_i\partial\phi_j}]_{\phi_i=v_i},
\end{align}
where

\begin{align}
\frac{\partial V_2}{\partial\phi_i}=&\frac{\partial V_2}{\partial m^2_{t}}\frac{\partial m^2_{t}}{\partial \phi_i}+\frac{\partial V_2}{\partial m^2_{\tilde{t}_1}}\frac{\partial m^2_{\tilde{t}_1}}{\partial \phi_i}+\frac{\partial V_2}{\partial m^2_{\tilde{t}_2}}\frac{\partial m^2_{\tilde{t}_2}}{\partial \phi_i}+\frac{\partial V_2}{\partial c^2_{2\theta_t}}\frac{\partial c^2_{2\theta_t}}{\partial \phi_i},
\\
\frac{\partial^2 V_2}{\partial\phi_i \partial\phi_j}=&\Bigg[(
\frac{\partial^2V_2}{\partial m_t^2 \partial m_t^2}\frac{\partial m_t^2}{\partial\phi_j}
+\frac{\partial^2V_2}{\partial m_t^2 \partial m^2_{\tilde{t}_1}}\frac{\partial m^2_{\tilde{t}_1}}{\partial\phi_j}
+\frac{\partial^2V_2}{\partial m_t^2 \partial m^2_{\tilde{t}_2}}\frac{\partial m^2_{\tilde{t}_2}}{\partial\phi_j}
+\frac{\partial^2V_2}{\partial m_t^2 \partial c^2_{2\theta_t}}\frac{\partial c^2_{2\theta_t}}{\partial\phi_j})\frac{\partial m_t^2}{\partial \phi_i}
\nonumber \\
&+(\frac{\partial^2V_2}{\partial m^2_{\tilde{t}_1} \partial m_t^2}\frac{\partial m_t^2}{\partial\phi_j}
+\frac{\partial^2V_2}{\partial m^2_{\tilde{t}_1} \partial m^2_{\tilde{t}_1}}\frac{\partial m^2_{\tilde{t}_1}}{\partial\phi_j}
+\frac{\partial^2V_2}{\partial m^2_{\tilde{t}_1} \partial m^2_{\tilde{t}_2}}\frac{\partial m^2_{\tilde{t}_2}}{\partial\phi_j}
+\frac{\partial^2V_2}{\partial m^2_{\tilde{t}_1} \partial c^2_{2\theta_t}}\frac{\partial c^2_{2\theta_t}}{\partial\phi_j})\frac{\partial m_{\tilde{t}_1}^2}{\partial \phi_i}
  \nonumber \\
&+(\frac{\partial^2V_2}{\partial m^2_{\tilde{t}_2} \partial m_t^2}\frac{\partial m_t^2}{\partial\phi_j}
+\frac{\partial^2V_2}{\partial m^2_{\tilde{t}_2} \partial m^2_{\tilde{t}_1}}\frac{\partial m^2_{\tilde{t}_1}}{\partial\phi_j}
+\frac{\partial^2V_2}{\partial m^2_{\tilde{t}_2} \partial m^2_{\tilde{t}_2}}\frac{\partial m^2_{\tilde{t}_2}}{\partial\phi_j}
+\frac{\partial^2V_2}{\partial m^2_{\tilde{t}_2} \partial c^2_{2\theta_t}}\frac{\partial c^2_{2\theta_t}}{\partial\phi_j})\frac{\partial m_{\tilde{t}_2}^2}{\partial \phi_i}
 \nonumber \\
&+(\frac{\partial^2V_2}{\partial c^2_{2\theta_t} \partial m_t^2}\frac{\partial m_t^2}{\partial\phi_j}
+\frac{\partial^2V_2}{\partial c^2_{2\theta_t} \partial m^2_{\tilde{t}_1}}\frac{\partial m^2_{\tilde{t}_1}}{\partial\phi_j}
+\frac{\partial^2V_2}{\partial c^2_{2\theta_t} \partial m^2_{\tilde{t}_2}}\frac{\partial m^2_{\tilde{t}_2}}{\partial\phi_j}
+\frac{\partial^2V_2}{\partial c^2_{2\theta_t} \partial c^2_{2\theta_t}}\frac{\partial c^2_{2\theta_t}}{\partial\phi_j})\frac{\partial c^2_{2\theta_t}}{\partial \phi_i}
 \nonumber \\
&+\frac{\partial V_2}{\partial m^2_t}\frac{\partial^2m_t^2}{\partial \phi_i \partial \phi_j}+\frac{\partial V_2}{\partial m^2_{\tilde{t}_1}}\frac{\partial^2m_t^2}{\partial \phi_i \partial \phi_j}+\frac{\partial V_2}{\partial m^2_{\tilde{t}_2}}\frac{\partial^2m_t^2}{\partial \phi_i \partial \phi_j}+\frac{\partial V_2}{\partial c^2_{2\theta_t}}\frac{\partial^2m_t^2}{\partial \phi_i \partial \phi_j} \Bigg],
\end{align}
We have
\begin{align}
c_{2\theta_t}^2 = 1 - \frac{4 |m_{\tilde{t}_{12}}^2|^2}{(m_{\tilde{t}_1}^2 - m_{\tilde{t}_2}^2)^2},
\end{align}
$\text{and the results of } \frac{\partial V^2}{\partial M^2}, \frac{\partial^2 V^2}{\partial M^2 \partial M^2} \text{ with } M = m_t, m_{\tilde{t}_1}, m_{\tilde{t}_2} \text{ can be found in the Appendix of Ref. \cite{Degrassi:2009yq}} \\
\text{The derivatives of } m^2_t, m^2_{\tilde{t}_1}, m^2_{\tilde{t}_2}, c^2_{2\tilde{\theta}_t} \text{ over } \phi_{i(i = 1,2,R1,R2,S)} \text{ can be written as}$
\begin{align}
\frac{\partial c_{2\theta_t}^2}{\partial \phi_i}=&\frac{2m_{\tilde{t}_{12}}^2}{(m_{\tilde{t}_1}^2 - m_{\tilde{t}_2}^2)^2}\frac{\partial m_{\tilde{t}_{12}}^2 }{\partial \phi_i}+
\frac{-2m_{\tilde{t}_{12}}^4}{(m_{\tilde{t}_1}^2 - m_{\tilde{t}_2}^2)^3}\left( \frac{\partial m_{\tilde{t}_1}^2}{\partial \phi_i}-\frac{\partial m_{\tilde{t}_2}^2}{\partial \phi_i}\right) \\
\frac{\partial^2 c_{2\theta_t}^2}{\partial \phi_i \partial \phi_j}=&\Bigg[\frac{2}{(m_{\tilde{t}_1}^2 - m_{\tilde{t}_2}^2)^2}\frac{\partial m^2_{\tilde{t}_{12}}}{\partial \phi_j}+\frac{2m_{\tilde{t}_{12}}^2}{(m_{\tilde{t}_1}^2 - m_{\tilde{t}_2}^2)^3}\left( \frac{\partial m_{\tilde{t}_1}^2}{\partial \phi_j}-\frac{\partial m_{\tilde{t}_2}^2}{\partial \phi_j}\right)\Bigg]\frac{\partial m_{\tilde{t}_{12}}^2}{\partial \phi_i} \nonumber \\
&+\frac{2m_{\tilde{t}_{12}}^2}{(m_{\tilde{t}_1}^2 - m_{\tilde{t}_2}^2)^2}\frac{\partial^2 m_{\tilde{t}_{12}}^2 }{\partial \phi_i \partial \phi_j}+\Bigg[\frac{-2m_{\tilde{t}_{12}}^2}{(m_{\tilde{t}_1}^2 - m_{\tilde{t}_2}^2)^3}\frac{\partial m_{\tilde{t}_{12}}^2}{\partial \phi_j}
\nonumber \\
&+\frac{-6m_{\tilde{t}_{12}}^4}{(m_{\tilde{t}_1}^2-m_{\tilde{t}_2}^2)^4}\left(
\frac{\partial m_{\tilde{t}_1}^2}{\partial \phi_j}-\frac{\partial m_{\tilde{t}_2}^2}{\partial \phi_j}\right)\Bigg]\left(\frac{\partial m_{\tilde{t}_1}^2}{\partial \phi_i}-\frac{\partial m_{\tilde{t}_2}^2}{\partial \phi_i}\right)
\nonumber \\
&+\frac{-2m_{\tilde{t}_{12}}^4}{(m_{\tilde{t}_1}^2 - m_{\tilde{t}_2}^2)^3}\left(\frac{\partial^2 m_{\tilde{t}_{1}}^2 }{\partial \phi_i \partial \phi_j}-\frac{\partial^2 m_{\tilde{t}_{2}}^2 }{\partial \phi_i \partial \phi_j}\right)
\end{align}

\begin{align}
\frac{\partial m_k^2}{\partial \phi_i} \bigg|_{k = \tilde{t}_1, \tilde{t}_2, \tilde{b}_1, \tilde{b}_2} =&
\frac{1}{2}
\left(
\frac{\partial m_{k_L}^2}{\partial \phi_i} + \frac{\partial m_{k_R}^2}{\partial \phi_i}
\right)
\pm \frac{1}{4}
 \Bigg[
2 \left(
\frac{\partial m_{k_L}^2}{\partial \phi_i}
- \frac{\partial m_{k_R}^2}{\partial \phi_i}
\right)  \\ \nonumber
&+ 8  m_{k_{12}}^2 \frac{\partial m_{k_{12}}^2}{\partial \phi_i}
\Bigg]
\Bigg[ \left( m_{\tilde{k}_L}^2 - m_{\tilde{k}_R}^2 \right)^2
+ 4 m_{\tilde{k}_{12}}^4 \Bigg]^{-\frac{1}{2}}
\\ \nonumber
\frac{\partial^2 m_k^2}{\partial \phi_i \partial \phi_j} \bigg|_{k=\tilde{t}_1, \tilde{t}_2, \tilde{b}_1, \tilde{b}_2} &= \frac{1}{2} \left( \frac{\partial^2 m_{k_L}^2}{\partial \phi_j \partial \phi_i} + \frac{\partial^2 m_{k_R}^2}{\partial \phi_j \partial \phi_i} \right) \\ \nonumber
&\quad \pm \frac{1}{4} \left[ (m_{k_L}^2 - m_{k_R}^2)^2 + 4m_{k_{12}}^4 \right]^{-\frac{1}{2}} \\ \nonumber
&\quad \times \Bigg\{ 2 \left[ \frac{\partial}{\partial \phi_j}(m_{k_L}^2 - m_{k_R}^2) \right] \left[ \frac{\partial}{\partial \phi_i}(m_{k_L}^2 - m_{k_R}^2) \right] \\ \nonumber
&\quad + 2 (m_{k_L}^2 - m_{k_R}^2) \left( \frac{\partial^2 m_{k_L}^2}{\partial \phi_j \partial \phi_i} - \frac{\partial^2 m_{k_R}^2}{\partial \phi_j \partial \phi_i} \right) \\ \nonumber
&\quad + 8 \frac{\partial m_{k_{12}}^2}{\partial \phi_i} \frac{\partial m_{k_{12}}^2}{\partial \phi_j} + 8 m_{k_{12}}^2 \frac{\partial^2 m_{k_{12}}^2}{\partial \phi_j \partial \phi_i} \Bigg\} \\ \nonumber
&\quad \mp \frac{1}{8} \left[ (m_{k_L}^2 - m_{k_R}^2)^2 + 4m_{k_{12}}^4 \right]^{-\frac{3}{2}} \\ \nonumber
&\quad \times \left[ 2(m_{k_L}^2 - m_{k_R}^2)\frac{\partial(m_{k_L}^2 - m_{k_R}^2)}{\partial \phi_i} + 8m_{k_{12}}^2 \frac{\partial m_{k_{12}}^2}{\partial \phi_i} \right] \\
&\quad \times \left[ 2(m_{k_L}^2 - m_{k_R}^2)\frac{\partial(m_{k_L}^2 - m_{k_R}^2)}{\partial \phi_j} + 8m_{k_{12}}^2 \frac{\partial m_{k_{12}}^2}{\partial \phi_j} \right]
\end{align}
In the preceding expressions, the derivatives of the matrix elements with $m^2_{\tilde{t}_L}$,$m^2_{\tilde{t}_R}$,$m^2_{\tilde{t}_12}$,$m^2_{\tilde{b}_L}$,$m^2_{\tilde{b}_R}$ and $m^2_{\tilde{b}_12}$ with respect to $\phi_{i(i = 1,2,R1,R2,S)}$ are given below

\begin{align}
&\frac{\partial m^2_{\tilde{t}_R}}{\partial \phi_1}=-\frac{g_R^2}{4}\phi_1,
&&\frac{\partial m^2_{\tilde{t}_R}}{\partial \phi_2}=\frac{g_R^2}{4}\phi_2+(Y_{Q}^2)^2\phi_2,
&&\frac{\partial m^2_{\tilde{t}_R}}{\partial \phi_{R1}}=(\frac{g_B^2}{3}-\frac{g_R^2}{2})\phi_{R1},
\nonumber \\
&\frac{\partial m^2_{\tilde{t}_R}}{\partial \phi_{R2}}=(\frac{g_R^2}{2}-\frac{g_B^2}{3})\phi_{R2},\
&&\frac{\partial m^2_{\tilde{t}_L}}{\partial \phi_1}=\frac{g_L^2}{4}\phi_1,
&&\frac{\partial m^2_{\tilde{t}_L}}{\partial \phi_2}=-\frac{g_L^2}{4}\phi_2+(Y_{Q}^2)^2\phi_2,
\nonumber \\
&\frac{\partial m^2_{\tilde{t}_L}}{\partial \phi_{R1}}=-\frac{g_B^2}{3}\phi_{R1},
&&\frac{\partial m^2_{\tilde{t}_L}}{\partial \phi_{R2}}=\frac{g_B^2}{3}\phi_{R2},
&&\frac{\partial m^2_{\tilde{t}_{12}}}{\partial \phi_1}=\frac{\lambda_3 Y^2_Q }{2} \phi_S ,
\nonumber \\
&\frac{\partial m^2_{\tilde{t}_{12}}}{\partial \phi_2}=\frac{T_Q^2}{\sqrt{2}},
&&\frac{\partial m^2_{\tilde{t}_{12}}}{\partial \phi_S}=\frac{\lambda_3 Y^2_Q }{2} \phi_1,
\nonumber \\
&\frac{\partial m^2_{\tilde{t}_L}}{\partial \phi_{S}}=\frac{\partial m^2_{\tilde{t}_R}}{\partial \phi_{S}}=\frac{\partial m^2_{\tilde{t}_{12}}}{\partial \phi_{R1}}=&&\frac{\partial m^2_{\tilde{t}_{12}}}{\partial \phi_{R2}}=0
\\
&\frac{\partial m^2_{\tilde{b}_R}}{\partial \phi_1}=(-\frac{g_R^2}{4}+(Y_{Q}^1)^2)\phi_1,
&&\frac{\partial m^2_{\tilde{b}_R}}{\partial \phi_2}=\frac{g_R^2}{4}\phi_2,
&&\frac{\partial m^2_{\tilde{b}_R}}{\partial \phi_{R1}}=(\frac{g_B^2}{2}+\frac{g_R^2}{2})\phi_{R1},
\nonumber \\
&\frac{\partial m^2_{\tilde{b}_R}}{\partial \phi_{R2}}=(\frac{g_R^2}{2}-\frac{g_B^2}{3})\phi_{R2}
&&\frac{\partial m^2_{\tilde{b}_L}}{\partial \phi_1}=(-\frac{g_L^2}{4}+(Y_{Q}^1)^2)\phi_1,
&&\frac{\partial m^2_{\tilde{b}_L}}{\partial \phi_2}=\frac{g_L^2}{4}\phi_2,
\nonumber \\
&\frac{\partial m^2_{\tilde{b}_L}}{\partial \phi_{R1}}=-\frac{g_B^2}{3}\phi_{R1},
&&\frac{\partial m^2_{\tilde{b}_L}}{\partial \phi_{R2}}=\frac{g_B^2}{3}\phi_{R2}
&&\frac{\partial m^2_{\tilde{b}_{12}}}{\partial \phi_1}=-\frac{T_{Q}^1}{\sqrt{2}},
\nonumber \\
&\frac{\partial m^2_{\tilde{b}_{12}}}{\partial \phi_2}=-\frac{Y_{Q}^1 \lambda_3}{2}\phi_S,
&&\frac{\partial m^2_{\tilde{b}_{12}}}{\partial \phi_S}=-\frac{Y_{Q}^1 \lambda_3}{2}\phi_2, \nonumber \\
&\frac{\partial m^2_{\tilde{b}_L}}{\partial \phi_{S}}=\frac{\partial m^2_{\tilde{b}_R}}{\partial \phi_{S}}=\frac{\partial m^2_{\tilde{b}_{12}}}{\partial \phi_{R1}}=&&\frac{\partial m^2_{\tilde{b}_{12}}}{\partial \phi_{R2}}=0
\end{align}
\begin{align}
&\frac{\partial^2 m_{\tilde{t}_L}^2}{\partial \phi_1 \partial \phi_1} = \frac{g_L^2}{4},\ \ \
\frac{\partial^2 m_{\tilde{t}_L}^2}{\partial \phi_1 \partial \phi_2} = \frac{\partial^2 m_{\tilde{t}_L}^2}{\partial \phi_1 \partial \phi_{R1}} = \frac{\partial^2 m_{\tilde{t}_L}^2}{\partial \phi_1 \partial \phi_{R2}} = \frac{\partial^2 m_{\tilde{t}_L}^2}{\partial \phi_1 \partial \phi_S} = 0,
 \nonumber \\
&\frac{\partial^2 m_{\tilde{t}_L}^2}{\partial \phi_2 \partial \phi_2} = -\frac{g_L^2}{4} + (Y_Q^2)^2, \ \ \
\frac{\partial^2 m_{\tilde{t}_L}^2}{\partial \phi_2 \partial \phi_1} = \frac{\partial^2 m_{\tilde{t}_L}^2}{\partial \phi_2 \partial \phi_{R1}} = \frac{\partial^2 m_{\tilde{t}_L}^2}{\partial \phi_2 \partial \phi_{R2}} = \frac{\partial^2 m_{\tilde{t}_L}^2}{\partial \phi_2 \partial \phi_S} = 0,
\nonumber \\
&\frac{\partial^2 m_{\tilde{t}_L}^2}{\partial \phi_{R1}\partial \phi_{R1}} = -\frac{g_B^2}{3}
,\ \ \
\frac{\partial^2 m_{\tilde{t}_L}^2}{\partial \phi_{R1} \partial \phi_1} = \frac{\partial^2 m_{\tilde{t}_L}^2}{\partial \phi_{R1} \partial \phi_{R1}} = \frac{\partial^2 m_{\tilde{t}_L}^2}{\partial \phi_{R1} \partial \phi_{R2}} = \frac{\partial^2 m_{\tilde{t}_L}^2}{\partial \phi_{R1} \partial \phi_S} = 0,
\nonumber \\
&\frac{\partial^2 m_{\tilde{t}_L}^2}{\partial \phi_{R2}\phi_{R2}} = -\frac{g_B^2}{3},
\ \ \
\frac{\partial^2 m_{\tilde{t}_L}^2}{\partial \phi_{R2} \partial \phi_{R1}} = \frac{\partial^2 m_{\tilde{t}_L}^2}{\partial \phi_{R2} \partial \phi_{R1}} = \frac{\partial^2 m_{\tilde{t}_L}^2}{\partial \phi_{R2} \partial \phi_{R2}} = \frac{\partial^2 m_{\tilde{t}_L}^2}{\partial \phi_{R2} \partial \phi_S} = 0,
\nonumber \\
&\frac{\partial^2 m_{\tilde{t}_L}^2}{\partial \phi_S\phi_S} = \frac{\partial^2 m_{\tilde{t}_L}^2}{\partial \phi_S \partial \phi_S}=
\frac{\partial^2 m_{\tilde{t}_L}^2}{\partial \phi_S \partial \phi_{R1}} = \frac{\partial^2 m_{\tilde{t}_L}^2}{\partial \phi_S \partial \phi_{R2}} = 0,
\nonumber \\
&\frac{\partial^2 m_{\tilde{t}_R}^2}{\partial \phi_1 \partial \phi_1} = -\frac{g_R^2}{4}
,\ \ \
\frac{\partial^2 m_{\tilde{t}_R}^2}{\partial \phi_1 \partial \phi_2} =
\frac{\partial^2 m_{\tilde{t}_R}^2}{\partial \phi_1 \partial \phi_{R1}} =
\frac{\partial^2 m_{\tilde{t}_R}^2}{\partial \phi_1 \partial \phi_{R2}} =
\frac{\partial^2 m_{\tilde{t}_R}^2}{\partial \phi_1 \partial \phi_S} = 0,
  \nonumber \\
&\frac{\partial^2 m_{\tilde{t}_R}^2}{\partial \phi_2 \partial \phi_2} = -\frac{g_R^2}{4} + (Y_Q^2)^2
,\ \ \
\frac{\partial^2 m_{\tilde{t}_R}^2}{\partial \phi_2 \partial \phi_1} =
\frac{\partial^2 m_{\tilde{t}_R}^2}{\partial \phi_2 \partial \phi_{R1}} =
\frac{\partial^2 m_{\tilde{t}_R}^2}{\partial \phi_2 \partial \phi_{R2}} =
\frac{\partial^2 m_{\tilde{t}_R}^2}{\partial \phi_2 \partial \phi_S} = 0,
\nonumber \\
&\frac{\partial^2 m_{\tilde{t}_R}^2}{\partial \phi_{R1} \partial \phi_{R1}} = -\frac{g_B^2}{3} - \frac{g_R^2}{2},\ \ \
\frac{\partial^2 m_{\tilde{t}_R}^2}{\partial \phi_{R1} \partial \phi_1} =
\frac{\partial^2 m_{\tilde{t}_R}^2}{\partial \phi_{R1} \partial \phi_{R1}} =
\frac{\partial^2 m_{\tilde{t}_R}^2}{\partial \phi_{R1} \partial \phi_{R2}} =
\frac{\partial^2 m_{\tilde{t}_R}^2}{\partial \phi_{R1} \partial \phi_S} = 0,
\nonumber \\
&\frac{\partial^2 m_{\tilde{t}_R}^2}{\partial \phi_{R2} \partial \phi_{R2}} = -\frac{g_B^2}{3} + \frac{g_R^2}{2},\ \ \
\frac{\partial^2 m_{\tilde{t}_R}^2}{\partial \phi_{R2} \partial \phi_{R1}} =
\frac{\partial^2 m_{\tilde{t}_R}^2}{\partial \phi_{R2} \partial \phi_{R1}} =
\frac{\partial^2 m_{\tilde{t}_R}^2}{\partial \phi_{R2} \partial \phi_{R2}} =
\frac{\partial^2 m_{\tilde{t}_R}^2}{\partial \phi_{R2} \partial \phi_S} = 0,
\nonumber \\
&\frac{\partial^2 m_{\tilde{t}_R}^2}{\partial \phi_S \partial \phi_S} =
\frac{\partial^2 m_{\tilde{t}_R}^2}{\partial \phi_S \partial \phi_S} =
\frac{\partial^2 m_{\tilde{t}_R}^2}{\partial \phi_S \partial \phi_{R1}} =
\frac{\partial^2 m_{\tilde{t}_R}^2}{\partial \phi_S \partial \phi_{R2}} = 0,
\nonumber \\
&\frac{\partial^2 m_{\tilde{t}_{12}}^2}{\partial \phi_1 \partial \phi_S} = \frac{Y^2_Q \lambda_3}{2},\ \ \
\frac{\partial^2 m_{\tilde{t}_{12}}^2}{\partial \phi_1 \partial \phi_1} =
\frac{\partial^2 m_{\tilde{t}_{12}}^2}{\partial \phi_1 \partial \phi_2} -
\frac{\partial^2 m_{\tilde{t}_{12}}^2}{\partial \phi_1 \partial \phi_{R1}} -
\frac{\partial^2 m_{\tilde{t}_{12}}^2}{\partial \phi_1 \partial \phi_{R2}} = 0, \nonumber \\
&\frac{\partial^2 m_{\tilde{t}_{12}}^2}{\partial \phi_2 \partial \phi_2} =
\frac{\partial^2 m_{\tilde{t}_{12}}^2}{\partial \phi_2 \partial \phi_{R1}} -
\frac{\partial^2 m_{\tilde{t}_{12}}^2}{\partial \phi_2 \partial \phi_{R2}} -
\frac{\partial^2 m_{\tilde{t}_{12}}^2}{\partial \phi_2 \partial \phi_S} = 0,
\nonumber \\
&\frac{\partial^2 m_{\tilde{t}_{12}}^2}{\partial \phi_{R1} \partial \phi_{R1}} =
\frac{\partial^2 m_{\tilde{t}_{12}}^2}{\partial \phi_{R1} \partial \phi_{R1}} -
\frac{\partial^2 m_{\tilde{t}_{12}}^2}{\partial \phi_{R1} \partial \phi_{R2}} -
\frac{\partial^2 m_{\tilde{t}_{12}}^2}{\partial \phi_{R1} \partial \phi_S} = 0, \nonumber \\
&\frac{\partial^2 m_{\tilde{t}_{12}}^2}{\partial \phi_S \partial \phi_1} = \frac{Y^2_Q\lambda_3}{2}, \ \ \
\frac{\partial^2 m_{\tilde{t}_{12}}^2}{\partial \phi_{R2} \partial \phi_{R2}} =
\frac{\partial^2 m_{\tilde{t}_{12}}^2}{\partial \phi_{R2} \partial \phi_{R1}} -
\frac{\partial^2 m_{\tilde{t}_{12}}^2}{\partial \phi_{R2} \partial \phi_{R2}} -
\frac{\partial^2 m_{\tilde{t}_{12}}^2}{\partial \phi_{R2} \partial \phi_S} = 0,
\nonumber \\
&\frac{\partial^2 m_{\tilde{t}_{12}}^2}{\partial \phi_S \partial \phi_S} =
\frac{\partial^2 m_{\tilde{t}_{12}}^2}{\partial \phi_S \partial \phi_1} -
\frac{\partial^2 m_{\tilde{t}_{12}}^2}{\partial \phi_S \partial \phi_{R1}} -
\frac{\partial^2 m_{\tilde{t}_{12}}^2}{\partial \phi_S \partial \phi_{R2}} = 0.
\end{align}
\begin{align}
&\frac{\partial^2 m_{\tilde{b}_L}^2}{\partial \phi_1 \partial \phi_1} = -\frac{g_L^2}{4} + (Y_Q^1)^2, \ \ \
\frac{\partial^2 m_{\tilde{b}_L}^2}{\partial \phi_1 \partial \phi_2} =
\frac{\partial^2 m_{\tilde{b}_L}^2}{\partial \phi_1 \partial \phi_{R1}} =
\frac{\partial^2 m_{\tilde{b}_L}^2}{\partial \phi_1 \partial \phi_{R2}} =
\frac{\partial^2 m_{\tilde{b}_L}^2}{\partial \phi_1 \partial \phi_S} = 0,
 \nonumber \\
&\frac{\partial^2 m_{\tilde{b}_L}^2}{\partial \phi_2 \partial \phi_2} = \frac{g_L^2}{4},\ \ \
\frac{\partial^2 m_{\tilde{b}_L}^2}{\partial \phi_2 \partial \phi_1} =
\frac{\partial^2 m_{\tilde{b}_L}^2}{\partial \phi_2 \partial \phi_{R1}} =
\frac{\partial^2 m_{\tilde{b}_L}^2}{\partial \phi_2 \partial \phi_{R2}} =
\frac{\partial^2 m_{\tilde{b}_L}^2}{\partial \phi_2 \partial \phi_S} = 0,
\nonumber \\
&\frac{\partial^2 m_{\tilde{b}_L}^2}{\partial \phi_{R1} \partial \phi_{R1}} = -\frac{g_B^2}{3},\ \ \
\frac{\partial^2 m_{\tilde{b}_L}^2}{\partial \phi_{R1} \partial \phi_1} =
\frac{\partial^2 m_{\tilde{b}_L}^2}{\partial \phi_{R1} \partial \phi_{R1}} =
\frac{\partial^2 m_{\tilde{b}_L}^2}{\partial \phi_{R1} \partial \phi_{R2}} =
\frac{\partial^2 m_{\tilde{b}_L}^2}{\partial \phi_{R1} \partial \phi_S} = 0,
\nonumber \\
&\frac{\partial^2 m_{\tilde{b}_L}^2}{\partial \phi_{R2} \partial \phi_{R2}} = \frac{g_B^2}{3},\ \ \
\frac{\partial^2 m_{\tilde{b}_L}^2}{\partial \phi_{R2} \partial \phi_{R1}} =
\frac{\partial^2 m_{\tilde{b}_L}^2}{\partial \phi_{R2} \partial \phi_{R1}} =
\frac{\partial^2 m_{\tilde{b}_L}^2}{\partial \phi_{R2} \partial \phi_{R2}} =
\frac{\partial^2 m_{\tilde{b}_L}^2}{\partial \phi_{R2} \partial \phi_S} = 0,
\nonumber \\
&\frac{\partial^2 m_{\tilde{b}_L}^2}{\partial \phi_S \partial \phi_S} =
\frac{\partial^2 m_{\tilde{b}_L}^2}{\partial \phi_S \partial \phi_S} =
\frac{\partial^2 m_{\tilde{b}_L}^2}{\partial \phi_S \partial \phi_{R1}} =
\frac{\partial^2 m_{\tilde{b}_L}^2}{\partial \phi_S \partial \phi_{R2}} = 0,
\nonumber \\
&\frac{\partial^2 m_{\tilde{b}_R}^2}{\partial \phi_1 \partial \phi_1} = -\frac{g_R^2}{4} + (Y^1_Q)^2, \ \ \
\frac{\partial^2 m_{\tilde{b}_R}^2}{\partial \phi_1 \partial \phi_2} =
\frac{\partial^2 m_{\tilde{b}_R}^2}{\partial \phi_1 \partial \phi_{R1}} -
\frac{\partial^2 m_{\tilde{b}_R}^2}{\partial \phi_1 \partial \phi_{R2}} -
\frac{\partial^2 m_{\tilde{b}_R}^2}{\partial \phi_1 \partial \phi_S} = 0,
\nonumber \\
&\frac{\partial^2 m_{\tilde{b}_R}^2}{\partial \phi_2 \partial \phi_2} = \frac{g_R^2}{4}, \ \ \
\frac{\partial^2 m_{\tilde{b}_R}^2}{\partial \phi_2 \partial \phi_1} =
\frac{\partial^2 m_{\tilde{b}_R}^2}{\partial \phi_2 \partial \phi_{R1}} -
\frac{\partial^2 m_{\tilde{b}_R}^2}{\partial \phi_2 \partial \phi_{R2}} -
\frac{\partial^2 m_{\tilde{b}_R}^2}{\partial \phi_2 \partial \phi_S} = 0,
\nonumber \\
&\frac{\partial^2 m_{\tilde{b}_R}^2}{\partial \phi_{R1} \partial \phi_{R1}} = -\frac{g_B^2}{3} - \frac{g_R^2}{2},\ \ \
\frac{\partial^2 m_{\tilde{b}_R}^2}{\partial \phi_{R1} \partial \phi_1} =
\frac{\partial^2 m_{\tilde{b}_R}^2}{\partial \phi_{R1} \partial \phi_{R1}} -
\frac{\partial^2 m_{\tilde{b}_R}^2}{\partial \phi_{R1} \partial \phi_{R2}} -
\frac{\partial^2 m_{\tilde{b}_R}^2}{\partial \phi_{R1} \partial \phi_S} = 0,
\nonumber \\
&\frac{\partial^2 m_{\tilde{b}_R}^2}{\partial \phi_{R2} \partial \phi_{R2}} = -\frac{g_B^2}{3} + \frac{g_R^2}{2},\ \ \
\frac{\partial^2 m_{\tilde{b}_R}^2}{\partial \phi_{R2} \partial \phi_{R1}} =
\frac{\partial^2 m_{\tilde{b}_R}^2}{\partial \phi_{R2} \partial \phi_{R1}} -
\frac{\partial^2 m_{\tilde{b}_R}^2}{\partial \phi_{R2} \partial \phi_{R2}} -
\frac{\partial^2 m_{\tilde{b}_R}^2}{\partial \phi_{R2} \partial \phi_S} = 0,
\nonumber \\
&\frac{\partial^2 m_{\tilde{b}_R}^2}{\partial \phi_S \partial \phi_S} =
\frac{\partial^2 m_{\tilde{b}_R}^2}{\partial \phi_S \partial \phi_S} -
\frac{\partial^2 m_{\tilde{b}_R}^2}{\partial \phi_S \partial \phi_{R1}} -
\frac{\partial^2 m_{\tilde{b}_R}^2}{\partial \phi_S \partial \phi_{R2}} = 0,
\nonumber \\
&\frac{\partial^2 m_{\tilde{b}_{12}}^2}{\partial \phi_S \partial \phi_1} =
\frac{\partial^2 m_{\tilde{b}_{12}}^2}{\partial \phi_S \partial \phi_{R1}} =
\frac{\partial^2 m_{\tilde{b}_{12}}^2}{\partial \phi_S \partial \phi_{R2}} =
\frac{\partial^2 m_{\tilde{b}_{12}}^2}{\partial \phi_S \partial \phi_2} = 0,
\nonumber \\
&\frac{\partial^2 m_{\tilde{b}_{12}}^2}{\partial \phi_2 \partial \phi_1} =
-\frac{Y^1_Q \lambda_3}{2},
\frac{\partial^2 m_{\tilde{b}_{12}}^2}{\partial \phi_2 \partial \phi_2} =
\frac{\partial^2 m_{\tilde{b}_{12}}^2}{\partial \phi_2 \partial \phi_{R1}} =
\frac{\partial^2 m_{\tilde{b}_{12}}^2}{\partial \phi_2 \partial \phi_{R2}} =
\frac{\partial^2 m_{\tilde{b}_{12}}^2}{\partial \phi_2 \partial \phi_S} = 0,
\nonumber \\
&\frac{\partial^2 m_{\tilde{b}_{12}}^2}{\partial \phi_{R1} \partial \phi_{R1}} =
\frac{\partial^2 m_{\tilde{b}_{12}}^2}{\partial \phi_{R1} \partial \phi_{R1}} =
\frac{\partial^2 m_{\tilde{b}_{12}}^2}{\partial \phi_{R1} \partial \phi_{R2}} =
\frac{\partial^2 m_{\tilde{b}_{12}}^2}{\partial \phi_{R1} \partial \phi_S} = 0, \nonumber \\
&\frac{\partial^2 m_{\tilde{b}_{12}}^2}{\partial \phi_{R2} \partial \phi_{R2}} =
\frac{\partial^2 m_{\tilde{b}_{12}}^2}{\partial \phi_{R2} \partial \phi_{R1}} =
\frac{\partial^2 m_{\tilde{b}_{12}}^2}{\partial \phi_{R2} \partial \phi_{R2}} =
\frac{\partial^2 m_{\tilde{b}_{12}}^2}{\partial \phi_{R2} \partial \phi_S} = 0, \nonumber \\
&\frac{\partial^2 m_{\tilde{b}_{12}}^2}{\partial \phi_S \partial \phi_2} =
-\frac{Y^1_Q \lambda_3}{2},
\frac{\partial^2 m_{\tilde{b}_{12}}^2}{\partial \phi_S \partial \phi_S} =
\frac{\partial^2 m_{\tilde{b}_{12}}^2}{\partial \phi_S \partial \phi_1} =
\frac{\partial^2 m_{\tilde{b}_{12}}^2}{\partial \phi_S \partial \phi_{R1}} =
\frac{\partial^2 m_{\tilde{b}_{12}}^2}{\partial \phi_S \partial \phi_{R2}} = 0, \quad
\end{align}

In our setup, the field-dependent masses satisfy
$m_t = \frac{Y_t}{\sqrt{2}}v_1$ and $m_{\nu_R} = \frac{h_{RR}}{\sqrt{2}}v_{R1}$; that is, the top-quark mass depends only on $v_1$, while the right-handed neutrino mass depends only on $v_{R1}$. The derivatives with respect to $\phi_i$ are provided below.
\begin{align}
\frac{\partial m_{t}^2}{\partial \phi_1}=Y^2_{t}v_1,\frac{\partial^2 m_{t}^2}{\partial \phi_1\partial \phi_1}=Y^2_{t},\frac{\partial m_{\nu_R}^2}{\partial \phi_{R1}}=h_{RR}^2v_{R1},\frac{\partial^2 m_{\nu_R}^2}{\partial \phi_{R1}\partial \phi_{R1}}=h_{RR}^2
\end{align}
For the right-handed gauge bosons, $W_R$ and $Z_R$, the derivatives with respect
to the $\phi_i$ are given below.
\begin{align}
&\frac{\partial m_{W_R}^2}{\partial \phi_1}=g_R^{2}\,v_1,
&&\frac{\partial m_{W_R}^2}{\partial \phi_2}=g_R^{2}\,v_2,
&&\frac{\partial m_{W_R}^2}{\partial \phi_{R1}}=2g_R^{2}\,v_{R1},\nonumber \\
&\frac{\partial m_{W_R}^2}{\partial \phi_{R2}}=2g_R^{2}\,v_{R2},
&&\frac{\partial^2 m_{W_R}^2}{\partial \phi_1 \partial \phi_1}=g_R^{2},
&&\frac{\partial^2 m_{W_R}^2}{\partial \phi_1 \partial \phi_1}=g_R^{2},\nonumber \\
&\frac{\partial^2 m_{W_R}^2}{\partial \phi_{R1} \partial \phi_{R1}}=2g_R^{2},
&&\frac{\partial^2 m_{W_R}^2}{\partial \phi_{R2} \partial \phi_{R2}}=2g_R^{2},
&&\frac{\partial^2 m_{W_R}^2}{\partial \phi_{S}}=0, \nonumber \\
&\frac{\partial^2 m_{W_R}^2}{\partial \phi_{i} \partial \phi_{j}}=0(i\neq j),\nonumber\\
&\frac{\partial m_{Z_R}^2}{\partial \phi_1}
=\frac{g_R^{4}}{\,g_R^{2}+g_B^{2}\,}\,v_1,
&&\frac{\partial m_{Z_R}^2}{\partial \phi_2}
=\frac{g_R^{4}}{\,g_R^{2}+g_B^{2}\,}\,v_2,
&&\frac{\partial m_{Z_R}^2}{\partial \phi_{R1}}
=4\!\left(g_R^{2}+g_B^{2}\right)\!\left(v_{R1}+v_{R2}\right),
\nonumber \\
&\frac{\partial m_{Z_R}^2}{\partial \phi_{R2}}
=4\!\left(g_R^{2}+g_B^{2}\right)\!\left(v_{R1}+v_{R2}\right),
&&\frac{\partial^{2} m_{Z_R}^2}{\partial \phi_1 \partial \phi_1}
=\frac{g_R^{4}}{\,g_R^{2}+g_B^{2}\,},
&&\frac{\partial^{2} m_{Z_R}^2}{\partial \phi_2 \partial \phi_2}
=\frac{g_R^{4}}{\,g_R^{2}+g_B^{2}\,},
\nonumber \\
&\frac{\partial^{2} m_{Z_R}^2}{\partial \phi_{R1} \partial \phi_{R1}}
=4\!\left(g_R^{2}+g_B^{2}\right),
&&\frac{\partial^{2} m_{Z_R}^2}{\partial \phi_{R2} \partial \phi_{R2}}
=4\!\left(g_R^{2}+g_B^{2}\right),
&&\frac{\partial^{2} m_{Z_R}^2}{\partial \phi_{R1} \partial \phi_{R2}}
=4\!\left(g_R^{2}+g_B^{2}\right),
\nonumber \\
&\frac{\partial^{2} m_{Z_R}^2}{\partial \phi_{S} \partial \phi_{S}}=0,
&&\frac{\partial^{2} m_{Z_R}^2}{\partial \phi_{i} \partial \phi_{j}}=0|_ {(i\neq j\ and\ i,j=1,2)}.
\end{align}

\begin{figure}
\setlength{\unitlength}{5.0mm}
\centering
\includegraphics[width=2.5in]{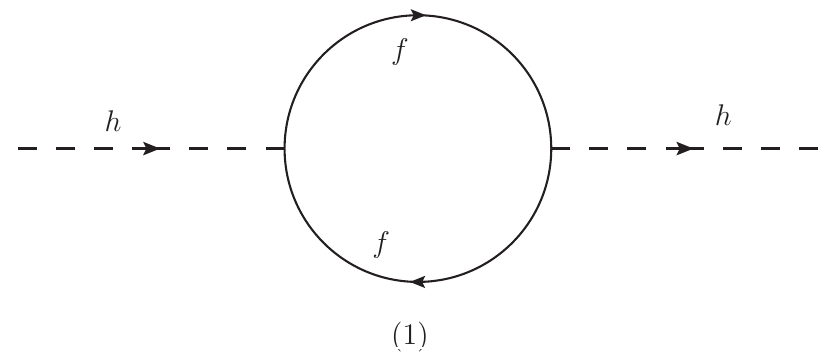}
\vspace{0.2cm}
\setlength{\unitlength}{5.0mm}
\centering
\includegraphics[width=2.5in]{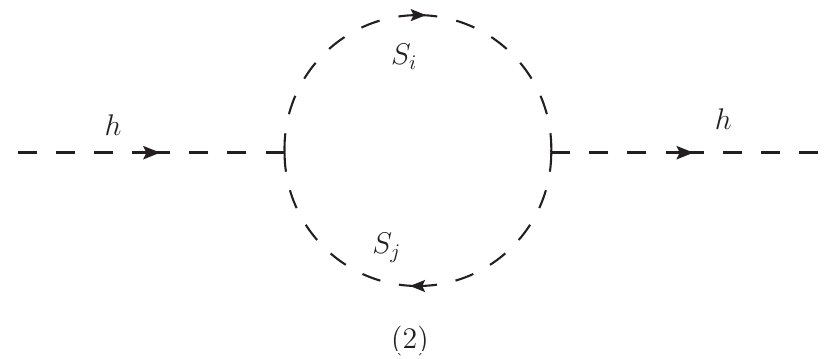}
\vspace{0.2cm}
\caption{ One-loop diagrams which induce the radiative corrections to the lightest Higgs boson mass,
 where $f = t, b,\tilde{\chi}^{0}_i$ and $S_i=\tilde{t}_{1,2},\tilde{b}_{1,2}$.}\label{fig:pm}
\end{figure}

 In order to get the pole mass of the lightest Higgs consistently, we write the explicit expressions for the self-energy diagrams. It is well known that the dominant corrections to the lightest Higgs mass originate from bottom, sbottom, top and stop. The relevant diagrams are plotted in Fig. \ref{fig:pm}, and the corresponding corrections are given by Eq. (\ref{pmc}).
\begin{align}
\Delta \Pi_{hh}^{(1)} &= \frac{-6C_{hff}^2}{(4\pi)^2}
\left[
(4m_f^2 - m_h^2) B(m_f, m_f, m_h) - 4m_f^2 \ln \frac{m_f^2}{Q^2}
\right],
\nonumber \\
\Delta \Pi_{hh}^{(2)} &= \frac{3C_{hs_i,s_j}^2}{(4\pi)^2}
\left[
B(m_{s_i}, m_{s_j}, m_h) - B(m_{s_i}, m_{s_j}, 0)
\right], \label{pmc}
\end{align}

 The function $B$ is defined as
\begin{align}
B(m_1, m_2, m_3) = \int_0^1 dx \ln \frac{(1 - x)m_1^2 + x m_2^2 + x(x - 1)m_3^2}{Q^2}.
\end{align}

\section{Approximate Results for $m_{h_1}$, $m_{h_2}$, $m_{H_1^{\pm}}$, and $m_{H^{\pm\pm}}$\label{app5}}

We performed an approximate analytical study of the lightest doubly charged Higgs and the lightest singly charged Higgs under the condition $\tan\beta_R = 1$, in order to more clearly reveal their relationship with $\tan\beta_R$. Since the mass matrix of the doubly charged Higgs is a 2$\times$2 matrix, the mass of the lightest doubly charged Higgs can be expressed as:
\begin{align}
M_{H^{\pm\pm}_1}^2=\frac{1}{2}\bigg[m^2_{\Delta_R^{++}\Delta_R^{--}}+m^2_{\delta_R^{++}\delta_R^{--}}-
\sqrt{(m^2_{\Delta_R^{++}\Delta_R^{--}}-m^2_{\delta_R^{++}\delta_R^{--}})^2+4m^4_{\Delta_R^{++}\delta_R^{--}}}\bigg].
\end{align}
We begin by performing an approximate expansion around $\tan\beta_R = 1$. For this purpose, we define $\tan\beta_R = 1 + \delta$ and $v_1 / v_{R1} = \sigma$. The squared mass of the lightest doubly charged Higgs, $M_{H^{\pm\pm}1}^2$, can thus be regarded as a function of $\delta$ and $\sigma$, denoted as $M_{H^{\pm\pm}_1}^2(\delta, \sigma)$. Expanding this function as a Taylor series around the point $\delta = 0$, $\sigma = 0$.
\begin{align}
m_{H^{\pm\pm}_1}^2=&-\frac{4g_{R}^2v_{R1}(-2g_{R}^2v_{R1}^2+\sqrt{2}T_{R}v_{S}+\lambda_{R}(\lambda_{R}v_{R1}^2
+3\lambda_Sv_{S}^2+2\xi_F))}{\sqrt{2}T_{R}v_{S}+\lambda_{R}(\lambda_{R}v_{R1}^2
+3\lambda_Sv_{S}^2+2\xi_F)}\delta^2.\label{e1}
\end{align}
The calculation results show that the mass of the lightest doubly charged Higgs is strongly correlated with $\delta^2$. Since the mass matrix of the lightest singly charged Higgs boson is $4 \times 4$, its approximation procedure is more complicated. To facilitate the calculation, we first divide the singly charged mass matrix $M_{H^{\pm}}^2$ by $v_{R1}$ and then substitute $\tan\beta_R = 1 + \delta$ and $v_1 / v_{R1} = \sigma$. The resulting form of the matrix is
\begin{align}
\frac{1}{v_{R1}^2}M_{H^{\pm}}^2=\begin{pmatrix}
M_{1(2\times2)}^2 &  M_{3(2\times2)}^2 \\
M_{3(2\times2)}^{\dagger2} &  M_{2(2\times2)}^2
\end{pmatrix},
\end{align}
where
\begin{align}
M_{3(2\times2)}^2=\begin{pmatrix}
-\frac{g_R^2\sigma}{2\sqrt{2}} & \frac{g_R^2(1+\delta)\sigma}{2\sqrt{2}} \\
-\frac{g_R^2t_{\beta}\sigma}{2\sqrt{2}} & \frac{g_R^2t_{\beta}(1+\delta)\sigma}{2\sqrt{2}}
\end{pmatrix}.
\end{align}
It is evident that all elements of the matrix $M_{3(2\times2)}^2$ depend on $\delta$ and $\sigma$. Therefore, we treat $M_{3(2\times2)}^2$ as a perturbation and perform an approximate treatment on $M_{1(2\times2)}^2$. We have
\begin{align}
M_{1(2\times2)}^{'2}=M_{1(2\times2)}^2+M_{3(2\times2)}^{2T}(M_{1(2\times2)}^2-M_{2(2\times2)}^2)^{-1}M_{3(2\times2)}^2.
\end{align}
We then similarly perform a Taylor expansion on the eigenvalues of $M_{1(2\times2)}^{'2}$ as described above, yielding
\begin{align}
m^2_{H^{\pm}_1}=\begin{cases}
-\frac{g_R^2(t_{\beta}-1)(1+t\beta)v^2_{R1}}{1+t_{\beta}^2}\delta  \ \ \text{if}\ \
\sqrt{2}T_3+\lambda_3(\lambda_R v_{R1}^2+3 \lambda_S^2 v_S^2+2\xi_F)>0 \\
\frac{g_R^2(t_{\beta}^2-1)v_{R1}^2\delta}{1+t_{\beta}}+
\frac{\lambda_3\lambda_Rv_{R1}^2\delta}{2t_{\beta}}\ \ \text{if}\ \
\sqrt{2}T_3+\lambda_3(\lambda_R v_{R1}^2+3 \lambda_S^2 v_S^2+2\xi_F)<0
\end{cases}.
\end{align}
The calculation shows that only terms proportional to $\delta$ remain in the above expression, while all other terms vanish exactly. It is evident from the expressions for $m^2_{H^{\pm\pm}_1}$ and $m^2_{H^{\pm}_1}$ that both approach zero as $\tan\beta_R$ tends toward 1. We also performed an approximate analytical study of the dependence of the Higgs boson masses (\( m_{h_1} \) and \(m_{h_2} \)) on \( \tan\beta \). Since the mass matrix of the CP-even Higgs bosons is a $5 \times 5$ matrix, the approximation process becomes significantly more complex. For simplicity, we define \(\tan\beta_R = 1\) and \(v_1 / v_{R1} = \sigma\). We first perform a transformation on the CP-even mass matrix ($m_{H_3}^2$)
\begin{align}
M_{H_{t1}}^2=U_1^{\dagger} m_{H_3}^2 U_1,
\end{align}
where
\begin{align}
U_1=\begin{pmatrix}
\frac{1}{\sqrt{1+\tan^2\beta}}  &  \frac{-\tan\beta}{\sqrt{1+\tan^2\beta}^2} & 0 & 0 & 0 \\
\frac{\tan\beta}{\sqrt{1+\tan^2\beta}}  &  \frac{1}{\sqrt{1+\tan^2\beta}} & 0 & 0 & 0 \\
0&0&\frac{1}{\sqrt{2}}  & \frac{-1}{\sqrt{2}}  &0  \\
0&0&\frac{1}{\sqrt{2}}  & \frac{1}{\sqrt{2}}  &0  \\
0&0&0&0&1
\end{pmatrix}.
\end{align}
This formulation is based on the following set of basis vectors
\begin{align}
U_1^\dagger \begin{pmatrix}
H_{12}^0 \\
H_{21}^2 \\
H_{\Delta_R}^0\\
H_{\delta_R}^0\\
H_{S}^0
\end{pmatrix}=
\begin{pmatrix}
\frac{1}{\sqrt{1+\tan^2\beta}} (H_{12}^0+\tan\beta H_{21}^2) \\
\frac{1}{\sqrt{1+\tan^2\beta}} (H_{21}^0-\tan\beta H_{12}^2) \\
\frac{1}{\sqrt{2}}(H_{\Delta_R}^0+H_{\delta_R}^0)\\
\frac{1}{\sqrt{2}}(H_{\delta_R}^0-H_{\Delta_R}^0)\\
H_{S}^0
\end{pmatrix}.
\end{align}
The form of is $M_{H_{t1}}^2$ given by:
\begin{align}
M_{H_{t1}}^2=\begin{pmatrix}
m^2_{11} & m^2_{12} & m^2_{13} & m^2_{14} & m^2_{15}\\
m^2_{21} & m^2_{22} & m^2_{23} & m^2_{24} & m^2_{25}\\
m^2_{31} & m^2_{32} & m^2_{33} & m^2_{34} & m^2_{35}\\
m^2_{41} & m^2_{42} & m^2_{43} & m^2_{44} & m^2_{45}\\
m^2_{51} & m^2_{52} & m^2_{53} & m^2_{54} & m^2_{55}\\
\end{pmatrix}.
\end{align}
The matrix entries are given by
\begin{align}
m_{11}^2&=\frac{v_1^2 \left( g_L^2 \left( -1 + t_{\beta}^2 \right)^2 + g_R^2 \left( -1 + t_{\beta}^2 \right)^2 + 8 t_{\beta}^2 \lambda_3^2 \right)}{4 \left( 1 + t_{\beta}^2 \right)} ,
m_{12}^2=m_{21}^2=\frac{t_{\beta} \left( -1 + t_{\beta}^2 \right) v_1^2 \left( g_L^2 + g_R^2 - 2 \lambda_3^2 \right)}{2 \left( 1 + t_{\beta}^2 \right)}, \nonumber \\
m_{13}^2&=m_{31}^2=-\frac{\sqrt{2} \, t_{\beta} \, v_1 \, v_{R1} \, \lambda_3 \, \lambda_R}{\sqrt{1 + t_{\beta}^2}} ,
m_{14}^2=m_{41}^2=\frac{g_R^2 \left( -1 + t_{\beta}^2 \right) v_1 v_{R1}}{\sqrt{2} \sqrt{1 + t_{\beta}^2}}, \nonumber \\
m_{15}^2&=m_{51}^2=\frac{v_1 \left( -\sqrt{2} \, T_3 \, t_{\beta} + v_S \, \lambda_3 \left( \lambda_3 + t_{\beta}^2 \, \lambda_3 - 6 \, t_{\beta} \, \lambda_S \right) \right)}{\sqrt{1 + t_{\beta}^2}}, \nonumber \\
m_{22}^2&=\frac{1}{2 \left( t_{\beta}^3 + t_{\beta} \right)}(2 g_L^2 t_{\beta}^3 v_1^2 + 2 g_R^2 t_{\beta}^3 v_1^2 + 2 \lambda_3 \xi_F  + \sqrt{2} T_3 t_{\beta}^4 v_S + 2 \sqrt{2} T_3 t_{\beta}^2 v_S + \sqrt{2} T_3 v_S \nonumber \\
&+2 \lambda_3 \xi_F t_{\beta}^4 + 4 \lambda_3 \xi_F t_{\beta}^2 - 4 \lambda_3^2 t_{\beta}^3 v_1^2 + \lambda_3 \lambda_R t_{\beta}^4 v_{R1}^2 + 2 \lambda_3 \lambda_R t_{\beta}^2 v_{R1}^2 + 3 \lambda_3 \lambda_S t_{\beta}^4 v_S^2 \nonumber \\
&+ 6 \lambda_3 \lambda_S t_{\beta}^2 v_S^2 + \lambda_3 \lambda_R v_{R1}^2 + 3 \lambda_3 \lambda_S v_S^2), \nonumber \\
m_{23}^2&=m_{32}^2=\frac{\lambda_3 \lambda_R \left(t_{\beta}^2 - 1\right) v_1 v_{R1}}{\sqrt{2} \sqrt{t_{\beta}^2 + 1}} ,
m_{24}^2=m_{42}^2=\frac{\sqrt{2} g_R^2 t_{\beta} v_1 v_{R1}}{\sqrt{t_{\beta}^2 + 1}}, \nonumber \\
m_{25}^2&=m_{52}^2=\frac{\left(t_{\beta}^2 - 1\right) v_1 \left(\sqrt{2} T_3 + 6 \lambda_3 \lambda_s v_s\right)}{2 \sqrt{t_{\beta}^2 + 1}}, m_{33}^2=\lambda_{R}^2v_{R1}^2,m_{34}^2=m_{43}^2=0, \nonumber \\
m_{35}^2&=m_{53}^2=v_{R1} \left(T_R + \sqrt{2} \lambda_R v_s \left(\lambda_R + 3 \lambda_s\right)\right), m_{45}^2=m_{54}^2=0, \nonumber \\
m_{44}^2&=2 g_B^2 v_{R1}^2 + 2 g_R^2 v_{R1}^2 - 2 \lambda_R \xi_F + \lambda_3 \lambda_R t_{\beta} v_1^2 - \sqrt{2} T_R v_s - \lambda_R^2 v_{R1}^2 - 3 \lambda_R \lambda_s v_s^2, \nonumber \\
m_{55}^2&=\frac{-2 \sqrt{2} \xi_s + \sqrt{2} T_3 t_{\beta} v_1^2 - \sqrt{2} T_R v_{R1}^2 + 3 \sqrt{2} T_S v_s^2 + 36 \lambda_s^2 v_s^3}{2 v_s}.
\end{align}
It is straightforward to observe that when $\tan\beta_R = 1$, the Higgs boson with the basis state $\frac{1}{\sqrt{1+t_\beta^2}} (H_{21}^0 - t_\beta H_{12}^0)$ exhibits negligible mixing with the other Higgs bosons. Therefore, we decouple this particular Higgs state from the system. As a result, the original $5\times5$ mass matrix $M_{H_{t1}}^2$ is effectively reduced to a $4\times4$ matrix. We define the decoupled $4\times4$ matrix as $M_{H_{t2}}^2$, whose explicit form is given by:
\begin{align}
M_{H_{t2}}^2=\begin{pmatrix}
(M_{t11}^2)_{2\times2} & (M_{t12}^2)_{2\times2} \\
(M_{t21}^2)_{2\times2} & (M_{t22}^2)_{2\times2}
\end{pmatrix},
\end{align}
where
\begin{align}
M_{t11}^2=\begin{pmatrix}
m_{11}^2 & m_{13}^2 \\
m_{31}^2 & m_{33}^2
\end{pmatrix},M_{t12}^2=\begin{pmatrix}
m_{14}^2 & m_{15}^2 \\
m_{24}^2 & m_{25}^2
\end{pmatrix},  \nonumber \\
M_{t21}^2=\begin{pmatrix}
m_{41}^2 & m_{43}^2 \\
m_{51}^2 & m_{53}^2
\end{pmatrix},M_{t22}^2=\begin{pmatrix}
m_{44}^2 & m_{45}^2 \\
m_{54}^2 & m_{55}^2
\end{pmatrix},
\end{align}
where the matrix elements $m_{ij}^2$ are the entries of $M_{H_{t1}}^2$. By substituting \( v_1 / v_{R1} = \sigma \) into \( M_{H_{t2}}^2 \), it is evident that both \( M_{t21}^2 \) and \( M_{t22}^2 \) exhibit strong dependence on \( \sigma \). We therefore treat these terms as perturbations and apply a perturbative approximation to \( M_{t11}^2 \)
\begin{align}
M_{\alpha}^2=M_{t11}^2+M_{t12}^2 (M_{t11}^2-M_{t22}^2)^{-1} M_{t21}^2.
\end{align}
In the approximated matrix $M_\alpha$, the lightest and next-to-lightest Higgs bosons from the original $5\times5$ matrix are preserved. Due to the exceedingly complex form of the expression, we provide only their general form here.
\begin{align}
M_{\alpha}^2=\begin{pmatrix}
m_{\alpha11}^2 & m_{\alpha12}^2 \\
m_{\alpha21}^2 & m_{\alpha22}^2 \\
\end{pmatrix}.
\end{align}
The expressions for the lightest and next-to-lightest Higgs bosons are provided below
\begin{align}
F_{h_1}^2(\sigma)&=\frac{1}{2}(m_{\alpha11}^2+m_{\alpha22}^2-\sqrt{(m_{\alpha11}^2-m_{\alpha22}^2)^2+4m_{\alpha12}^2m_{\alpha2 1}^2}), \nonumber \\
F_{h_2}^2(\sigma)&=\frac{1}{2}(m_{\alpha11}^2+m_{\alpha22}^2+\sqrt{(m_{\alpha11}^2-m_{\alpha22}^2)^2+4m_{\alpha12}^2m_{\alpha2 1}^2}).
\end{align}
Finally, we perform a Taylor expansion around $\sigma=0$ to derive approximate solutions for the masses of the lightest and next-to-lightest Higgs bosons
\begin{align}
m_{h_1}^2&=\frac{F_{h_1}^{(2)}(\sigma)}{2!}\sigma^2+\frac{F_{h_1}^{(4)}(\sigma)}{4!}\sigma^4+\frac{F_{h_1}^{(6)}(\sigma)}{6!}\sigma^6, \nonumber \\
m_{h_2}^2&=F_{h_2}^2(0)+\frac{F_{h_2}^{(2)}(\sigma)}{2!}\sigma^2+\frac{F_{h_2}^{(4)}(\sigma)}{4!}\sigma^4+\frac{F_{h_2}^{(6)}(\sigma)}{6!}\sigma^6.
\end{align}
The form of $F_{h_2}^2(0)$ is given by
\begin{align}
F_{h_2}^2(0)=v_{R1}^2(\lambda_R^2-\frac{2v_{S}(T_R+\sqrt{2}v_S\lambda_R(\lambda_R+3\lambda_S))^2}
{-\sqrt{2}T_Rv_{R1}^2+3\sqrt{2}T_Sv_S^2+36\lambda_S^2v_S^2-2\sqrt{2}\xi_S}),
\end{align}
where $f^{(n)}_i(\sigma)$ is $\frac{\partial^{(n)} F_{h_i}^2(\sigma)}{(\partial \sigma)^n}|_{\sigma=0,i=1,2}$.
Due to the complexity of the explicit analytical expressions, we refrain from providing the detailed forms of \( f^{(n)}_i(\sigma) \). Under the parameter choices considered in this analysis, the zeroth-order term in the expansion of \( m^2_{h_1} \) around \( \sigma = 0 \) is zero, whereas the zeroth-order term for \( m^2_{h_2} \) is nonzero and constitutes the dominant contribution. Additionally, \( f^{(4)}_1(\sigma) \) and \( f^{(6)}_1(\sigma) \) are both negative, while \( f^{(2)}_1(\sigma) \), \( f^{(2)}_2(\sigma) \), \( f^{(4)}_2(\sigma) \), and \( f^{(6)}_2(\sigma) \) are all positive. The parameter $\sigma$ is inversely proportional to $\tan\beta$. As $\tan\beta$ increases, the mass $m_{h_1}$ exhibits an initial rise followed by a gradual leveling-off, while $m_{h_2}$ shows an initial decrease and then asymptotically stabilizes. This behavior is consistent with the analysis presented in the text.

\section{Neutrino Masses and Mixing in LRSSM\label{app6}}

In the paper, the right-handed neutrinos can make important contributions to the mass correction of the 95 GeV Higgs particle, while the effect on the 125 GeV Higgs particle is minimal. Therefore, we present simple analysis of the neutrino sector in the LRSSM. The form of the neutrino mass matrix in the LRSSM is given by
\begin{align}
M_{\nu}=\begin{pmatrix}
0 & m_{\nu_L \nu_R} \\
m_{\nu_R \nu_L} & m_{\nu_R \nu_R}
\end{pmatrix}
\end{align}
where
\begin{align}
m_{\nu_L \nu_R}=\frac{Y_{\nu}}{\sqrt{2}}v_1,m_{\nu_R \nu_L}=\frac{Y_{\nu}^{\dagger}}{\sqrt{2}}v_1,m_{\nu_R \nu_R}=\frac{h_{RR}}{\sqrt{2}}v_{R1}
\end{align}
The lightest 95 GeV Higgs is primarily composed of $\Delta_R$ and $\delta_R$. In this case, in addition to the contributions from top quark and stop quark particles, the contributions from right-handed bosons and right-handed neutrinos also become important. It is straightforward to obtain the masses of the left-handed and right-handed neutrinos as follows
\begin{align}
M_{\nu_L}&\approx-\frac{v_1^2}{2}Y_{\nu}^{\dagger} m_{\nu_R \nu_R}^{-1}Y_{\nu} \nonumber \\
M_{\nu_R}&\approx\frac{h_{RR}}{\sqrt{2}}v_{R1}
\end{align}
The form of $Y_{\nu}$ is given by
\begin{align}
Y_{\nu}=\sqrt{2}\frac{1}{v_1}\sqrt{M_{\nu_R}^{dig}} R \sqrt{M_{\nu_L}^{dig}} U^{\dagger}_{PMNS}.
\end{align}
The form of $U_{PMNS}$ is given by
\begin{align}
U_{PMNS} = \left(
\begin{array}{ccc}
 c_{12}c_{13} & s_{12}c_{13} & s_{13}e^{-i\delta} \\
 -s_{12}c_{23} - c_{12}s_{23}s_{13}e^{i\delta} & c_{12}c_{23} - s_{12}s_{23}s_{13}e^{i\delta} & s_{23}c_{13} \\
 s_{12}s_{23} - c_{12}c_{23}s_{13}e^{i\delta} & -c_{12}s_{23} - s_{12}c_{23}s_{13}e^{i\delta} & c_{23}c_{13}
\end{array}
\right),
\end{align}
where $s_{ij}=\sin(\theta_{ij})_{i,j=1,2,3},
c_{ij}=\cos(\theta_{ij})_{i,j=1,2,3}$ and
\begin{align}
&\sin^2(\theta_{12}) = 0.307 \pm 0.012,\nonumber \\
&\sin^2(\theta_{23}) = 0.537 \pm 0.020 \quad (\text{Inverted order}), \nonumber \\
&\sin^2(\theta_{23}) = 0.534^{+0.015}_{-0.019} \quad (\text{Normal order}) , \nonumber \\
&\sin^2(\theta_{13}) = (2.16 \pm 0.06) \times 10^{-2}.
\end{align}
The neutrino mass differences are provided
\begin{align}
&\Delta m_{21}^2 = (7.50 \pm 0.19) \times 10^{-5} \, \text{eV}^2, \nonumber \\
&\Delta m_{32}^2 = (-2.527 \pm 0.034) \times 10^{-3} \, \text{eV}^2 \quad (\text{Inverted order}), \nonumber \\
&\Delta m_{32}^2 = (2.451 \pm 0.026) \times 10^{-3} \, \text{eV}^2 \quad (\text{Normal order}), \nonumber \\
&\delta, \, \text{CP violating phase} = 1.21^{+0.19}_{-0.22} \, \pi \, \text{rad} \quad (S = 1.2) \nonumber, \\
&\langle \Delta m_{21}^2 - \Delta m_{21}^2 \rangle < 1.1 \times 10^{-4} \, \text{eV}^2, \, \text{CL} = 99.7\% ,\nonumber \\
&\langle \Delta m_{32}^2 - \Delta m_{32}^2 \rangle = (-0.12 \pm 0.25) \times 10^{-3} \, \text{eV}^2.
\end{align}
We have calculated \( Y_{\text{NO}} \) and \( Y_{\text{IO}} \) for two scenarios: one with the lightest Higgs boson mass at 125 GeV, and the other with the lightest Higgs boson mass at 95 GeV and the next-to-lightest at 125 GeV, respectively. In both cases, $\delta$ is taken as 1.21$\pi$.
In the case of IO (Inverted order), we set the mass of the lightest neutrino $m_{v_1}$ to be 0.01 eV. In the case of NO (Normal order), we set the mass of the lightest neutrino $m_{v_1}$ to be 0.0509 eV. In the scenario where the lightest Higgs boson mass is 125 GeV, the parameter values we used are $\tan \beta = 20$, $\tan \beta_R = 1.1$, $v_{R1} = 2.5 \, \text{TeV}$, and $h_{RR} = 2.0$. The results for $Y_{\nu(NO)}^{h_{125}}$ and $Y_{\nu(IO)}^{h_{125}}$ are given by
\begin{align}
\scriptsize
&Y_{\nu(NO)}^{h_{125}}=
\scriptsize
\begin{pmatrix}
1.262 \times 10^{-5} & -4.473 \times 10^{-6} - 7.918 \times 10^{-7}i & 7.372 \times 10^{-6} - 7.046 \times 10^{-7}i \\
9.302 \times 10^{-6} & 1.026 \times 10^{-5} - 5.844 \times 10^{-7}i & -9.791 \times 10^{-6} - 5.214 \times 10^{-7}i \\
-4.214 \times 10^{-6} - 2.864 \times 10^{-6}i & 2.544 \times 10^{-5} & 2.638 \times 10^{-5}
\end{pmatrix}, \\
&Y_{\nu(IO)}^{h_{125}}=
\scriptsize
\begin{pmatrix}
2.843 \times 10^{-4} & -1.092 \times 10^{-4} - 1.786 \times 10^{-6}i & 1.663 \times 10^{-4} - 1.590 \times 10^{-5}i \\
1.905 \times 10^{-4} & 2.104 \times 10^{-4} - 1.970 \times 10^{-6}i & -2.005 \times 10^{-4} - 1.065 \times 10^{-6}i \\
-1.873 \times 10^{-6} - 1.273 \times 10^{-6}i & 1.308 \times 10^{-5} & 1.006 \times 10^{-4}
\end{pmatrix}.
\end{align}
In the scenario where the lightest Higgs boson mass is 95 GeV and the next-to-lightest is 125 GeV, we adopt the following parameter values: $\tan\beta = 20$, $\tan\beta_R = 0.99$, $v_{R1} = 4$ TeV, and $h_{RR} = 2.0$. The resulting values for $Y_{\nu(NO)}^{h_{95}}$ and $Y_{\nu(IO)}^{h_{95}}$ are
\begin{align}
&Y_{\nu(NO)}^{h_{95}}=
\scriptsize
\begin{pmatrix}
2.267 \times 10^{-5} & -8.109 \times 10^{-6} - 1.417 \times 10^{-6}i & 1.321 \times 10^{-5} - 1.274 \times 10^{-6}i \\
1.674 \times 10^{-5} & 1.852 \times 10^{-5} - 1.046 \times 10^{-6}i & -1.759 \times 10^{-5} - 9.404 \times 10^{-7}i \\
-7.577 \times 10^{-6} - 5.149 \times 10^{-6}i & 4.553 \times 10^{-5} & 4.093 \times 10^{-5}
\end{pmatrix}, \\
&Y_{\nu(IO)}^{h_{95}}=
\scriptsize
\begin{pmatrix}
5.111 \times 10^{-5} & -1.814 \times 10^{-5} - 3.211 \times 10^{-6}i & 2.990 \times 10^{-5} - 2.858 \times 10^{-6}i \\
3.425 \times 10^{-5} & 3.776 \times 10^{-5} - 2.152 \times 10^{-6}i & -3.605 \times 10^{-5} - 1.915 \times 10^{-6}i \\
-3.368 \times 10^{-6} - 2.289 \times 10^{-6}i & 2.033 \times 10^{-5} & 1.809 \times 10^{-5}
\end{pmatrix}.
\end{align}
In summary, while our primary goal is to account for the $\sim95$\,GeV hints in
the diphoton and $b$-jet channels under the Higgs and collider constraints, we have now made explicit the realization of the neutrino sector consistent with our benchmarks.

\end{document}